\documentclass{imammb}

\jno{dqnxxx}
\usepackage{epsfig}
\usepackage{graphicx}
\usepackage{epstopdf}
\usepackage{amssymb,amsmath,amsthm,amscd}
\usepackage{latexsym}
\usepackage{mathrsfs}
\usepackage{verbatim}
\usepackage{color}
\usepackage{float}
\usepackage{enumerate}
\usepackage[round, sort, authoryear]{natbib}
\usepackage{subcaption}

\usepackage{color}

\usepackage{bm}

 \newtheoremstyle{theorem}{6pt}{6pt}{\rm}{}{\sffamily}{ }{ }{}
 \theoremstyle{theorem}

 \newtheoremstyle{algorithm}{6pt}{6pt}{\rm}{}{\sffamily}{ }{ }{}
 \theoremstyle{algorithm}

 \newtheoremstyle{lemma}{6pt}{6pt}{\rm}{}{\sffamily}{ }{ }{}
 \theoremstyle{lemma}

\newtheoremstyle{case}{6pt}{6pt}{\rm}{}{\sffamily}{. }{ }{}
 \theoremstyle{case}

 \newtheoremstyle{statement}{6pt}{6pt}{\rm}{}{\sffamily}{ }{ }{}
\theoremstyle{statement}

 \newtheoremstyle{corollary}{6pt}{6pt}{\rm}{}{\sffamily}{ }{ }{}
 \theoremstyle{corollary}

  \newtheoremstyle{definition}{6pt}{6pt}{\rm}{}{\sffamily}{ }{ }{}
 \theoremstyle{definition}

\newtheoremstyle{example}{6pt}{6pt}{\rm}{}{\sffamily}{ }{ }{}
\theoremstyle{example}

\newtheoremstyle{remark}{6pt}{6pt}{\rm}{}{\sffamily}{ }{ }{}
\theoremstyle{remark}

\newtheoremstyle{approximation}{6pt}{6pt}{\rm}{}{\sffamily}{ }{ }{}
\theoremstyle{approximation}

\newtheoremstyle{scheme}{6pt}{6pt}{\rm}{}{\sffamily}{ }{ }{}
\theoremstyle{scheme}

\newtheoremstyle{Algorithm}{6pt}{6pt}{\rm}{}{\sffamily}{ }{ }{}
\theoremstyle{Algorithm}

\newtheoremstyle{Assumption}{6pt}{6pt}{\rm}{}{\sffamily}{ }{ }{}
\theoremstyle{Assumption}

\newtheoremstyle{proposition}{6pt}{6pt}{\rm}{}{\sffamily}{ }{ }{}
\theoremstyle{proposition}

\newtheoremstyle{hypo}{6pt}{6pt}{\rm}{}{\sffamily}{ }{ }{}
 \theoremstyle{hypo}

  \newtheoremstyle{Step}{6pt}{6pt}{\rm}{}{}{ }{ }{}
 \theoremstyle{Step}

\renewcommand{\theequation}{\thesection.\arabic{equation}}
\numberwithin{equation}{section}

\newcommand{\be}{\begin{equation}}
\newcommand{\ee}{\end{equation}}
\newcommand{\bes}{\begin{equation*}}
\newcommand{\ees}{\end{equation*}}

\newcommand\Tstrut{\rule{0pt}{2.6ex}}    % = `top' strut
\newcommand\Bstrut{\rule[-0.9ex]{0pt}{0pt}}

\newcommand{\abs}[1]{\left |#1\right |}
\newcommand{\norm}[1]{\|#1\|}

  \newcommand{\h}{*}

\begin{document}

\title{Lattice and Continuum Modelling of a Bioactive Porous Tissue Scaffold}
\author{ {\sc Andrew L. Krause, Dmitry Beliaev, Robert A. Van Gorder, Sarah L. Waters\thanks{Corresponding author, waters@maths.ox.ac.uk}}\\[2pt]
Mathematical Institute, Andrew Wiles Building, University of Oxford \\
 Radcliffe Observatory Quarter, Woodstock Rd, OX2 6GG, UK.\\[6pt]
{\rm [Received on XX XXX XXXX]}\vspace*{6pt}}
\pagestyle{headings}
\markboth{A. L. KRAUSE ET AL.}{\rm LATTICE AND CONTINUUM MODELLING OF BIOACTIVE SCAFFOLD}
\maketitle

\begin{abstract}
{A contemporary procedure to grow artificial tissue is to seed cells onto a porous biomaterial scaffold and culture it within a perfusion bioreactor to facilitate the transport of nutrients to growing cells. Typical models of cell growth for tissue engineering applications make use of spatially homogeneous or spatially continuous equations to model cell growth, flow of culture medium, nutrient transport, and their interactions. The network structure of the physical porous scaffold is often incorporated through parameters in these models, either phenomenologically or through techniques like mathematical homogenization. We derive a model on a square grid lattice to demonstrate the importance of explicitly modelling the network structure of the porous scaffold, and compare results from this model with those from a modified continuum model from the literature. We capture two-way coupling between cell growth and fluid flow by allowing cells to block pores, and by allowing the shear stress of the fluid to affect cell growth and death. We explore a range of parameters for both models, and demonstrate quantitative and qualitative differences between predictions from each of these approaches, including spatial pattern formation and local oscillations in cell density present only in the lattice model. These differences suggest that for some parameter regimes, corresponding to specific cell types and scaffold geometries, the lattice model gives qualitatively different model predictions than typical continuum models. Our results inform model selection for bioactive porous tissue scaffolds, aiding in the development of successful tissue engineering experiments and eventually clinically successful technologies.}%We demonstrate quantitative differences between lattice and continuum models in terms of the mean cell density reached after some period of time, and the time scales for which spatially uniform growth remains uniform. We display qualitative differences between these models in terms of spatial pattern formation that depends on the density of pores in the lattice model, and show that there is less spatial heterogeneity exhibited in the continuum formulation. Finally, we demonstrate the existence of a dynamical effect present in the lattice model, small oscillations of cell density, which is not present in the spatially continuous equations. We argue why these oscillations occur in the finite discrete system, and why they may be important for some parameter regimes of relevance to tissue engineering experiments.
{tissue engineering, bioactive porous media, lattice and continuum models, model selection.}
\end{abstract}
\section{Introduction}
Tissue engineering is a rapidly growing field seeking to apply techniques from a variety of disciplines to create tissues and organs. There is increased need for these technologies as patient transplant lists continue to grow, especially due to increased longevity \citep{beard_global_2013}.  Despite significant progress in recent years, there is still a need to better understand the basic biological and physical aspects underpinning the engineering of artificially constructed tissues and organs in order to achieve widespread clinical success \citep{van_blitterswijk_tissue_2008}.

A contemporary strategy for \emph{in vitro} tissue engineering is to seed cells onto a porous biomaterial scaffold which is placed inside a bioreactor where it is perfused with nutrient-rich culture medium \citep{glowacki_perfusion_1998, kim_perfusion_2007, cimetta_enhancement_2007}. There is a large variation in cell types, scaffolds, and bioreactor geometries used to grow artificial tissues. Some cell types are mechano-sensitive and can be induced to proliferate, differentiate, or die in response to the local cellular mechanical environment, e.g. due to the fluid shear stress \citep{sankar2011culturing, duan2008shear, riha_roles_2005, iskratsch_appreciating_2014, bakker2004shear}. Some cells produce an extracellular matrix (ECM) that is sufficiently dense to substantially affect the flow of culture medium through individual pores \citep{hossain_computational_2015}. Understanding these fluid flow and cell growth interactions, and what roles they have in a tissue engineering experiment, is an important challenge.

Mathematical modelling plays a key role in underpinning experimental design, optimizing operating regimes, and predicting experimental outcomes \citep{odea_continuum_2012}. Mathematical models have also been used in lieu of costly and time-consuming experimental trials. There is a growing literature of models used to understand the interactions between fluid mechanical forces exerted on mechano-sensitive cells, and the effect that cell growth has on the flow via blocking of scaffold pores. The simplest of these are spatially homogeneous models where ODEs are used to model the evolution of each phase, such as in \cite{lemon_mathematical_2009}. However, the spatial structure of a scaffold can have profound effects on the growth of the tissue, due to the transport of cells and nutrients, as well as the variation of mechanical stress \citep{osborne2010influence, lanza_principles_2014}. Determining the spatial distribution of cells within a porous tissue scaffold, and the nature of their local biomechano-chemical environment, is a key challenge that mathematical modelling can address \citep{melchels2011influence, thevenot2008method}. Understanding the time-dependent coupling between cell growth and fluid flow is currently an active area of modelling and experimental investigation in tissue engineering and related fields \citep{geris_computational_2013}.

To describe the spatial properties of artificial tissue growth, many macroscale continuum models for bioactive porous media have been proposed \citep{odea_continuum_2012}. These approaches are computationally much cheaper than a full simulation of flow within the pores and do not require detailed knowledge of the microstructure. Due to the complexity of growing tissue \emph{in vitro}, and the wealth of biochemical and biophysical processes involved over a range of spatial and temporal scales, many different theoretical models exist in the literature \citep{german_applications_2016}. These models result in partial differential equations for the dependent variables, such as cell density and fluid velocity, with constitutive assumptions describing the interactions between these variables.

Multiphase models account for multiple phases explicitly (e.g. fluid, scaffold, cells, etc). The governing equations are derived from conservation of mass and momentum for each phase, and interactions between the phases are captured via the specification of appropriate constitutive laws \citep{pearson_multiphase_2013,odea_multiphase_2010}. Alternatively, many models include the effect of cell growth on the fluid flow implicitly, by considering a cell density equation based on conservation of mass, and then prescribing the scaffold porosity to be a function of cell density \citep{coletti_mathematical_2006,shakeel_continuum_2013, pohlmeyer_cyclic_2013,pohlmeyer_mathematical_2013}. We will refer to these as implicit models, in comparison to multiphase models which explicitly account for interactions between phases in terms of physical forces. The flow of fluid through the porous scaffold is then governed by Darcy's law. In all of these macroscale approaches, constitutive assumptions relate microscale processes,  such as cell growth, to macroscale parameters. In both multiphase and implicit models, equations have been derived that account for the influence of shear stress on cell growth. Additionally, homogenization techniques have been proposed to precisely capture the relationship between macroscopic parameters in these models, such as porosity, and growth and flow processes at the pore scale \citep{shipley_design_2009, chen2017, odea_multiscale_2015}. We briefly give examples of each of these kinds of approaches below.
% \cite{}  \cite{}  \citep{}.  \cite{irons2017microstructural}  \cite{penta_effective_2014} \cite{collis2017effective} 

%Incorporating a scaffold phase
\cite{pearson_multiphase_2013,pearson_multiphase_2016,pearson_dispersion-enhanced_2016,pearson_multiphase_2015} employed a multiphase approach to model culture medium fluid, scaffold, and cell phases within a hollow fibre bioreactor. This bioreactor design consists of parallel hollow permeable tubes which facilitate nutrient transport to the cells seeded in the extracapillary space surrounding the fibres. The small aspect ratio of the bioreactor was used to simplify these models, and the reduced systems of governing equations were then solved numerically to gain insight into how to stimulate uniform cell growth throughout the scaffold. %Mechanotransduction 
\cite{odea_multiphase_2010} developed a model of cells, culture medium, and tissue scaffold with the aim of understanding the role of cell-cell and cell-scaffold interactions on tissue growth in a perfusion bioreactor. The focus was on mechanotransduction effects, such as the influence of fluid shear stress on the cell growth. Their analytical and simulation results gave insight into using experimental data to determine the dominant mechanical regulatory mechanisms within a cell population. There is also a very large literature on the use of multiphase models in cancer biology and wound healing, but for brevity we do not review it here; see \cite{byrne2003modelling, byrne2003two} as examples and \cite{roose2007mathematical, preziosi2003cancer, alarcon2009modelling, powathil2015systems, rieger2016physics} for reviews of vascular and avascular tumour modelling. Multiphase models also appear in other areas of bioactive porous material, such as in the development and growth of biofilms \citep{anguige2006multi, klapper2010mathematical, fagerlind2012dynamic}.

We now discuss the class of implicit models. \cite{coletti_mathematical_2006} proposed a model of nutrient transport, fluid flow, and nutrient-limited cell growth in a tissue-engineering bioreactor which was used to investigate the effect of experimental protocols employed to overcome diffusion-limited transport, and to study adverse conditions that are observed experimentally, such as flow channelling along bioreactor walls. Accounting for both cell density-dependent permeability and mechanotransduction of shear stress by cells, \cite{shakeel_continuum_2013} investigated a model of nutrient transport and cell growth in a porous scaffold within a perfusion bioreactor. Cell proliferation and ECM deposition were assumed to decrease the local porosity, and hence permeability of the scaffold, while moderate values of the shear stress were assumed to enhance both cell growth and uptake of a generic nutrient by the cells. Although it was suggested that cell death would occur at higher values of shear stress (corresponding to higher flow rates into the scaffold), this was not explored. \cite{pohlmeyer_cyclic_2013} and  \cite{pohlmeyer_mathematical_2013} used a similar approach to \cite{shakeel_continuum_2013} to model other aspects of cell growth in perfusion bioreactors, including growth-factor driven haptotaxis. Using comparable implicit modelling approaches, \cite{nava_multiphysics_2013} considered a nutrient and shear stress-dependent moving boundary problem to model cell growth in a multi-compartment computational model. Their results included development of computational approaches for coupling sub-models with 2 and 3 dimensional spatial regions. \cite{hossain_computational_2015} considered the coupling of fluid shear stress, cell growth, and the fluid flow in a scaffold-free perfusion bioreactor.

Mathematical homogenization approaches have been proposed to upscale microscale features of the scaffold. \cite{shipley_design_2009} considered a model for fluid, glucose, and lactate transport in a microstructured porous medium. In this study, the variation of the effective permeability of the scaffold due to cell growth was not considered. \cite{odea_multiscale_2015} considered the advection of a generic diffusible nutrient through a porous medium and derived macroscale equations accounting for microscale accretion of biomass at the pore surface due to nutrient uptake. Homogenisation has also been used to upscale fluid flow and elastic deformation in a fibre-reinforced hydrogel scaffold for cartilage tissue engineering \citep{chen2017}. Outside of tissue engineering, several other homogenisation approaches have been developed to upscale tissue growth processes. \cite{irons2017microstructural} developed a model of biomass accretion in a porous medium aimed at understanding vascular tumour growth. \cite{penta_effective_2014} derived a macroscopic model of a porous linearly elastic medium that accounted for mass exchange at the microscale between the phases due to accretion of the solid phase at the pore surface. \cite{collis2017effective} determined effective equations for a poroelastic medium where microscale elastic stresses and deformations were homogenized alongside growth and solute transport. All of the above homogenization results are typically valid for specific parameter regimes, such as when the elastic and growth time scales are strongly separated.

In contrast to the spatially continuous models described above, network models have been used increasingly to describe porous media \citep{sahimi_applications_1994, sahimi_flow_1993}. These are `mesoscopic' models, whereby the porous medium is discretized as a graph with nodes and edges linking them. Fluid flow can then be described by assigning a pressure to each node, and determining the flow rate between nodes, including any sources or sinks of fluid into the network. Network models of flow through porous media explicitly describing the pore structure have not, to the best of our knowledge, been used in the context of tissue engineering scaffolds. However, network approaches have been used to model flow in geological and geochemical porous media, such as the growth of biofilms in soils or mass accretion in pores due to solid forming reactions, as well as in angiogenesis and its applications to cancer modelling and wound healing \citep{thullner_computational_2008, tsimpanogiannis_fluid_2012, anderson1998continuous, mcdougall_mathematical_2002}. We now review some of these models as motivation for our network modelling approach. We also mention that \cite{mely_double_2012, barbotteau_modelling_2003} used percolation theory to model random cell growth in a lattice to determine effective properties of bone tissue engineering scaffolds, although we note that the effect of flow on the growth process was not explicitly modelled, and biomechanical effects such as shear stress were therefore not included.

Biofilms growing in porous media have been modelled using a dynamic (e.g. time-dependent) pore network with coupling between the growth of the biofilm and flow throughout the network. \cite{thullner_computational_2008} proposed a lattice model of a porous medium, where growing biofilms clogged pores. Nutrients were carried by the fluid, affecting the biofilm growth throughout the porous medium. Different constitutive assumptions relating pressure differences to flow rates in pore throats were explored and resulted in global changes in hydraulic conductivity throughout the entire network. \cite{gharasoo2012reactive} and \cite{rosenzweig2014modeling} extended this approach to consider the influence of pore heterogeneity, as well as different kinds of reactions, within the network. \cite{tsimpanogiannis_fluid_2012} also considered a reactive transport model utilizing networks in a geochemical setting with solid formation leading to pore blocking, which resulted in changes in the flow throughout the network.

Network modelling has been used extensively in the study of angiogenesis, which is the formation of new blood vessels from existing ones. Alongside vascular network remodelling, angiogenesis plays a major role in cancer biology, wound healing, and developmental biology; see \citep{otrock2007understanding, yadav2015tumour, scianna_review_2013} for contemporary reviews of these fields. The discrete structure of microvessels has motivated a long history of using network models to understand their formation, and associated fluid and mass transport. \cite{anderson1998continuous} considered continuum and discrete models of endothelial-cell (tip cell) led vessel sprouting, and specifically derived the (probabilistic) movement of discrete cells via a finite-difference discretization of the corresponding continuum description. Capillaries then form behind these tip cells, growing the vascular network. Extending this approach, \cite{mcdougall_mathematical_2002} developed a network model for angiogenesis induced by a nearby tumour, and investigated fluid flow through the resulting network structure. Their results suggested that topological properties of the network, such as the number of anastomoses (loops), were key factors in determining the flow through the resulting network, which impacts the effectiveness of drug delivery via the vasculature. Reviews of some of the earlier theoretical literature utilizing these approaches can be found in \cite{chaplain2004mathematical, chaplain2006mathematical}. More recent network remodelling frameworks have been developed and applied to vascular tumours \citep{secomb_angiogenesis:_2013, pries_making_2014, rieger2016physics, vilanova2017mathematical} whereby vascular remodelling occurs due to lack of flow or tumour angiogenic factors being produced by cells due to hypoxia. 

Many recent models of cancer growth and treatment, including many of the network models above, adopt an individual-cell-based modelling approach for some components of the overall tumour model, such as discrete cells. Individual-based models track the local microenvironment around a cell, composed of mechanical stresses and chemical concentrations, in order to capture biophysical interactions between cells and between cells and the surrounding environment \citep{byrne2009individual, fozard2009continuum}. These models can either treat the cells as discrete cellular-automata confined to a spatial lattice \citep{gerlee2007evolutionary}, use off-lattice models of cells as points \citep{drasdo2007center}, or use cellular-Potts models which specify intracellular and environmental interactions via a Hamiltonian \citep{glazier2007ii}. Similar discrete modelling approaches have also been employed in vasculogenesis \citep{perfahl20173d}. Simulations of many individual cells are then used to gain insight into tissue-level behaviour \citep{powathil2015systems}.

Comparisons between individual-based and continuum models of growing cell populations have suggested certain parameter regimes where the approaches agree and disagree, which has implications for applicability and computational tractability of each kind of model \citep{pillay2017multiscale, byrne2009individual,osborne2017comparing}. Hybdrid discrete and continuum models of vascular tumours have also been studied, where the vasculature itself has a discrete structure, but transport of nutrients, cell migration, and cell/ECM growth throughout the tumour is modelled as spatially continuous \citep{welter2013interstitial, vilanova2017mathematical, welter2016computer, wu2014effect,figueredo2013lattice, peng2017multiscale}. These hybdrid models offer the advantage of describing the discrete structure of the vasculature, but can be efficiently simulated computationally \citep{de2017coarse}. Other multiscale frameworks linking vasculature, cell cycle dynamics, and other effects have also been pursued in the context of infectious diseases and tissue dynamics in general \citep{cilfone2015strategies, jessica2016multi}, which combine spatially continuous and spatially discrete submodels.

Finally, we also mention analogous approaches to those used in angiogenesis applied to wound healing; these consist of similar continuum, discrete, and hybdrid modelling approaches. \cite{mcdougall2012hybrid} develops a hybdrid discrete and continuum model of angiogenesis in retinal wound healing, using many of the same ideas of discrete tip cell migration as in \cite{anderson1998continuous}. \cite{flegg2015mathematical} reviews the current state of reaction-transport models, in particular considering angiogenesis in the context of dermal wounds that heal via replenishment of granulation tissue and ECM. We particularly note the use of discrete models that characterize vasculature, as in the tumour literature discussed above, and continuum models for reaction and transport of cells and nutrients \cite{Chaplain1996, spill2015mesoscopic}.

Having reviewed some of the network and hybdrid modelling literature, we return to the discussion of a tissue engineering scaffold as an active porous medium. The pore and scaffold length scales in typical experiments with perfusion bioreactors can lead to scaffolds with relatively few pores, and it is unclear that macroscopic spatially continuous models will capture features present in small networks of pores. See Figure 6 of \cite{cox_3d_2015} for an example of a porous scaffold with very few pores, and \cite{loh_three-dimensional_2013} and \cite{vafai_porous_2010} for general discussions of pore size, scaffold geometry, and transport phenomena in porous tissue scaffolds. Motivated by concerns with these spatially continuous modelling approaches, we propose a novel lattice model of a bioactive porous medium. We compare this to a typical continuum model, modified from \cite{shakeel_continuum_2013}, to elucidate the different kinds of behaviour displayed by each modelling paradigm. 

We consider the following biological system that captures many of the features encountered when engineering artificial tissues. We model a cell phase, which incorporates the ECM produced by the cells, and a fluid phase modelling the culture medium, which together saturate a two-dimensional porous scaffold. We consider model behaviours on the timescale of cell proliferation, focusing in particular on the role of shear stress on cell proliferation. In \cite{shakeel_continuum_2013} and \cite{chapman_optimising_2014} it was assumed that moderate levels of shear stress would enhance cell proliferation, but the detrimental effects of large shear stress were not considered. Several studies have shown detrimental effects of high shear on renal epithelial cells \citep{duan2008shear} and pancreatic endothelial cells \citep{sankar2011culturing}. The influence of shear stress on cell adhesion and `wash-out' has been studied for human fibroblast cells \citep{lu2004microfluidic, korin2007parametric}. We are interested in the effects of high levels of shear stress not explored in previously-mentioned models \citep{odea_multiphase_2010, shakeel_continuum_2013, chapman_optimising_2014}, and assume that cells die if the local shear stress exceeds a certain threshold, either due to specific mechano-transduction mechanisms for apoptosis, or due to detachment induced by the flow which has been investigated in the tissue engineering literature \citep{mccoy_influence_2012, whited2014influence}. Under this latter mechanism, our model assumes that cells do not reattach to the scaffold downstream after detachment, which is appropriate for some cell types under high velocity viscous flows \citep{sircar2016surface}. High levels of shear-stress were assumed to induce cell death and detachment in \cite{chapman2017mathematical}, where nutrient limitations and other biomechanical stimuli were considered in a hollow-fibre bioreactor. We assume that the culture medium is pumped into the scaffold at a fixed flow rate, and that local increases in cell density decrease the local permeability of the scaffold. For the sake of simplicity, we assume the cells are in a nutrient-rich environment and do not explicitly model a nutrient phase; we instead emphasize qualitative differences between lattice and continuum models with cell growth mediated only by fluid shear stress. Focusing on a single stimulus allows us to elucidate paradigmatic differences between spatially continuous and spatially discrete models in a simple setting.

We note that the lattice model we derive has a similar discrete network structure to those found in the angiogenesis literature, much of which originated in \cite{anderson1998continuous}, but our model differs in two key ways from the majority of models discussed above. Firstly, the underlying lattice topology is intended to represent the pore space of the tissue scaffold as a static network, rather than evolving a dynamic network. Secondly, we do not track individual cells with random motility, but instead consider nodal cell densities which diffuse via a continuous-time deterministic process. The active medium in our models consists of cells changing the local pore size; cells do not grow or remodel the pore network as is considered in the angiogenesis literature.

In Section \ref{ModellingSec} we present our continuum and lattice modelling approaches, and discuss the physical and biological parameters. In Section \ref{ResultsDiscussion} we compare results from these two models, primarily using numerical simulations. Finally in Section \ref{Conclusions_M} we discuss the implications of our study for the successful modelling of porous media in tissue engineering applications.

\section{Motivation and Modelling Frameworks}\label{ModellingSec}
\begin{figure}
\centering
\includegraphics[width=0.6\textwidth]{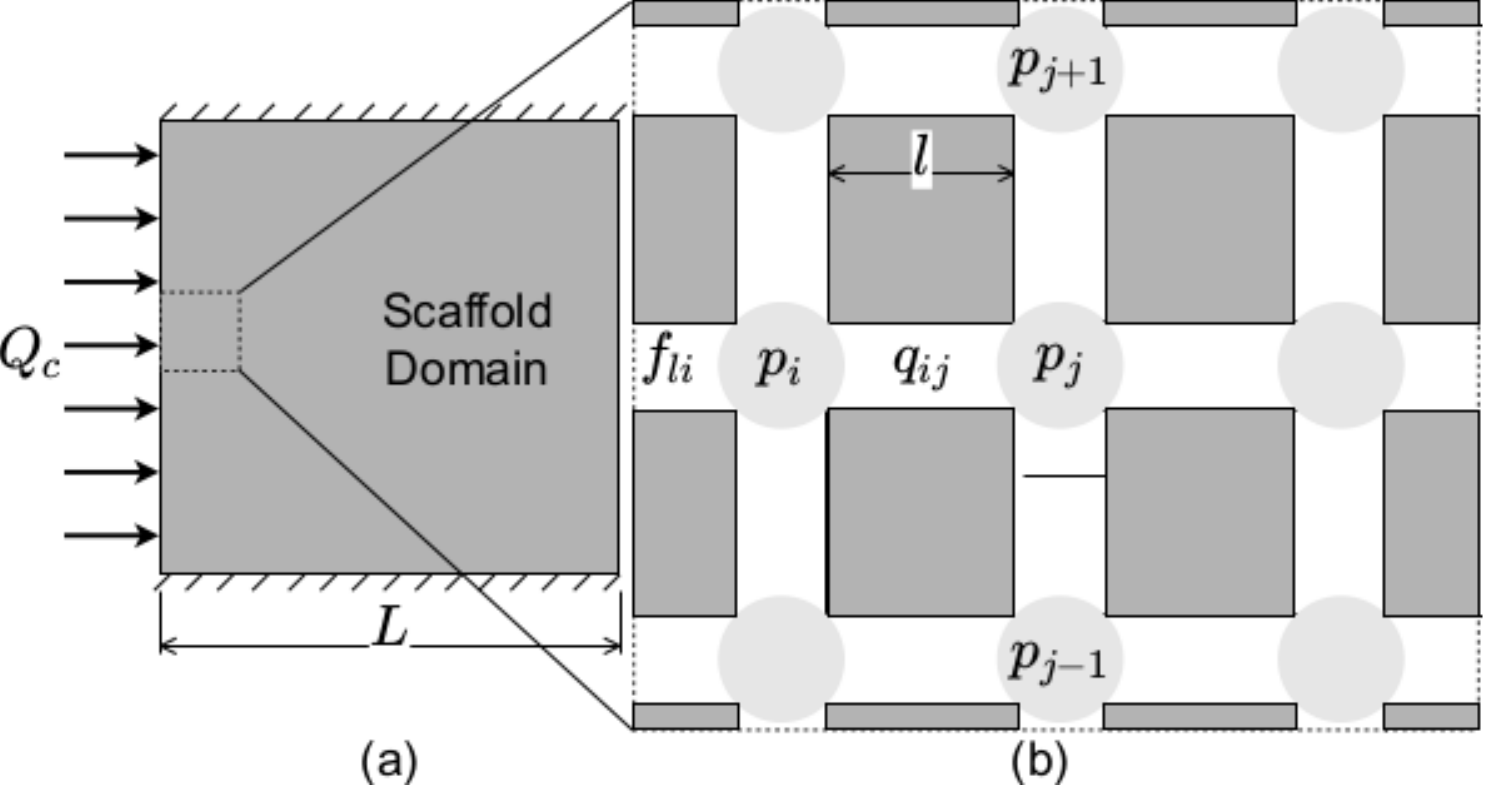}
\caption{A diagram of the scaffold domain (a) where fluid enters at a prescribed flow rate, $Q_c$, from the left and exits on the right side of the scaffold. The horizontal boundaries are impermeable. Enlarged (b) is the lattice structure of the scaffold, with fluid entering node $i$ from the left boundary at a volumetric flow rate $f_{li}$. $L$ denotes the size of the scaffold, $l$ the length of an internal `pipe,' $p_i$ the pressure at node $i$, and $q_{ij}$ the volumetric flow rate from node $i$ to node $j$. Fluid is present in the white and light gray regions (denoting nodes in the lattice model), and the dark gray regions are impermeable.}
\label{FlowDiagram}
\end{figure}

We consider a two dimensional rigid porous medium as a model of the tissue scaffold. For simplicity, we consider a square domain with side length $L$. The culture medium is an incompressible viscous Newtonian fluid, with dynamic viscosity $\mu$. We assume that fluid is pumped at a constant flow rate $Q_c$ ($\mathrm{m}^2\mathrm{s}^{-1}$) into the left side of the domain, and exits through the right side. We assume no fluid enters or leaves through the horizontal boundaries of the domain. See Figure \ref{FlowDiagram}(a) for a visualization of the domain. We assume that cells grow logistically, diffuse within the scaffold, and cannot leave the domain. We assume that cells will stop growing and will die if the shear stress exceeds a threshold value. The timescale of interest is that of cell proliferation. Finally, we neglect advection of the cells by the fluid, as we assume cells which are detached from the scaffold do not reattach downstream. 
\subsection{Continuum Model of a Bioactive Porous Medium}\label{PDEModel}
We describe the square domain as spatially continuous with Cartesian coordinates $\bm{x}^\h = (x^\h,y^\h) \in [0,L]^2$ (we use asterisks throughout to denote dimensional variables). We assume that the pore Reynolds number is small and neglect fluid inertia. We use Darcy's Law to model flow through the scaffold \citep{bear_dynamics_1972}, so that
\begin{equation}\label{darcy_law}
\bm{u}^\h = -\frac{k^\h(x^\h,y^\h,N^\h)}{\mu}\nabla^\h p^\h \quad \text{and }\quad \nabla^\h \cdot \bm{u}^\h = 0,
\end{equation}
where $\bm{u}^\h$ is the Darcy velocity of the fluid, $p^\h$ is the fluid pressure, and $k^\h(x^\h,y^\h,N^\h)$ is the permeability of the scaffold which depends on the cell density $N^\h$. 

For ease of computation, we prescribe a pressure drop across the domain so that, along with no flux conditions at the horizontal boundaries, we impose
\begin{equation}\label{Fluid_Boundaries}
\bm{n}\cdot\bm{u}^\h = 0 \enspace \text{at} \enspace y^\h= 0,L, \enspace 0\leq x^\h \leq L,
\end{equation}
\begin{equation}
p^\h = p_0 \enspace \text{at} \enspace x^\h = 0, \enspace 0\leq y^\h \leq L, \quad \text{and} \quad
p^\h = p_1 \enspace \text{at} \enspace x^\h = L, \enspace 0\leq y^\h \leq L,
\end{equation}
where $p_0$ is the upstream pressure, $p_1$ is the downstream pressure, and $\bm{n}$ is the outward unit normal.

We use the linearity between pressure and the Darcy velocity to rescale the fluid variables to match the prescribed fluid flow rate through the scaffold. The total flow rate through the boundary at $x^\h=0$ is
\begin{equation}
Q_0^\h = \int_0^L-\frac{k^\h(0,y^\h,N^\h)}{\mu}\frac{\partial p^\h}{\partial x^\h}dy^\h.
\end{equation}
We define the rescaled Darcy velocity to be,
\begin{equation}\label{uscale}
\bm{u_r}^\h = \frac{Q_c}{Q_0^\h}\bm{u}^\h.
\end{equation}

The cell density $N^\h$ is governed by the reaction-diffusion equation,
\begin{equation}\label{continuum_cells_full}
\frac{\partial N^\h}{\partial t^\h} = \beta \left(F_1(\sigma^\h) N^\h \left(1-\frac{N^\h}{N_c}\right)-F_2(\sigma^\h)N^\h\right)+ \nabla^{\h} \cdot (D^\h(N^\h) \nabla^\h N^\h),
\end{equation}
where $t^\h$ is time, $N_c$ is the maximum cell density, $\beta$ is the cell proliferation rate, $D^\h$ is a nonlinear cell diffusion coefficient, $\sigma^\h$ is the fluid shear stress, and the dimensionless functions $F_1(\sigma^\h)$ and $F_2(\sigma^\h)$ capture the effect of fluid shear stress on growth and death respectively. The terms on the right hand side correspond to logistic growth, cell death, and cell diffusion respectively. For simplicity we assume that $\beta$ also quantifies the death rate due to high shear, and $F_1$ and $F_2$ are dimensionless functions of the shear stress that vary between $0$ and $1$ (see \ref{pde_pressure_functions_sigma}).

The no flux conditions for the cell density are,
\begin{equation}
\bm{n}\cdot\nabla N^\h = 0 \enspace \text{for } \bm{x} \in \partial [0,L]^2,
\end{equation}
where $\partial [0,L]^2$ denotes the boundaries of the square domain. Finally we impose the following initial condition for cell density,
\begin{equation}
N^\h(x^\h,y^\h,0) = N_0^\h(x^\h,y^\h),
\end{equation}
where $N_0^\h(x^\h,y^\h)$ is the initial scaffold seeding density.

We now specify how the scaffold permeability $k^\h$ appearing in Equation \eqref{darcy_law} depends on the cell density $N^\h$. Following \cite{coletti_mathematical_2006}, we write the porosity as
\begin{equation}\label{porosity_eqn}
\hat{\phi}(N^\h) =  \phi_0(1-\nu N^\h),
\end{equation}
where $\phi_0$ is the volume fraction of pore space in the scaffold without cells, and $\nu$ is the volume of pore space occupied by an individual cell. We require that $\nu N_c \leq 1$ so that under the bounded dynamics of Equation \eqref{continuum_cells_full}, $0 \leq \hat{\phi}(N^\h) \leq \phi_0$ for all $N^\h \leq N_c$.

We follow \cite{shakeel_continuum_2013} and relate the cell-dependent porosity to the scaffold permeability via
\begin{equation}\label{permeability_eqn}
k^\h(N^\h) = k_0\hat{\phi}(N^\h)^3,
\end{equation}
where $k_0$ is the permeability of the cell-free scaffold.% in the absence of cells.

We require an expression for the shear stress $\sigma^\h$ in Equation \eqref{continuum_cells_full}. We follow \cite{whittaker_mathematical_2009} and \cite{shakeel_continuum_2013} and write,
\begin{equation}\label{shear_eqn}
\sigma^\h = \frac{4 \mu \tau}{R_0}\frac{\norm{\bm{u_r}^\h}}{\hat{\phi}(N^\h)},
\end{equation}
as the shear stress experienced by cells at the pore scale in terms of the Darcy velocity, where $R_0$ is the typical radius of a pore and $\tau$ is the tortuousity of a typical fluid path. As in \cite{whittaker_mathematical_2009} and \cite{shakeel_continuum_2013}, we neglect the influence that cells have on the tortuousity $\tau$ and treat it as a nondimensional constant corresponding to the average ratio of streamline lengths to the straight-line distance between two points. The more tortuous the scaffold, the faster the interstitial fluid has to be to travel longer paths in the same time.

\begin{figure}
\centering
\includegraphics[width=.65\textwidth, height=.40\textwidth]{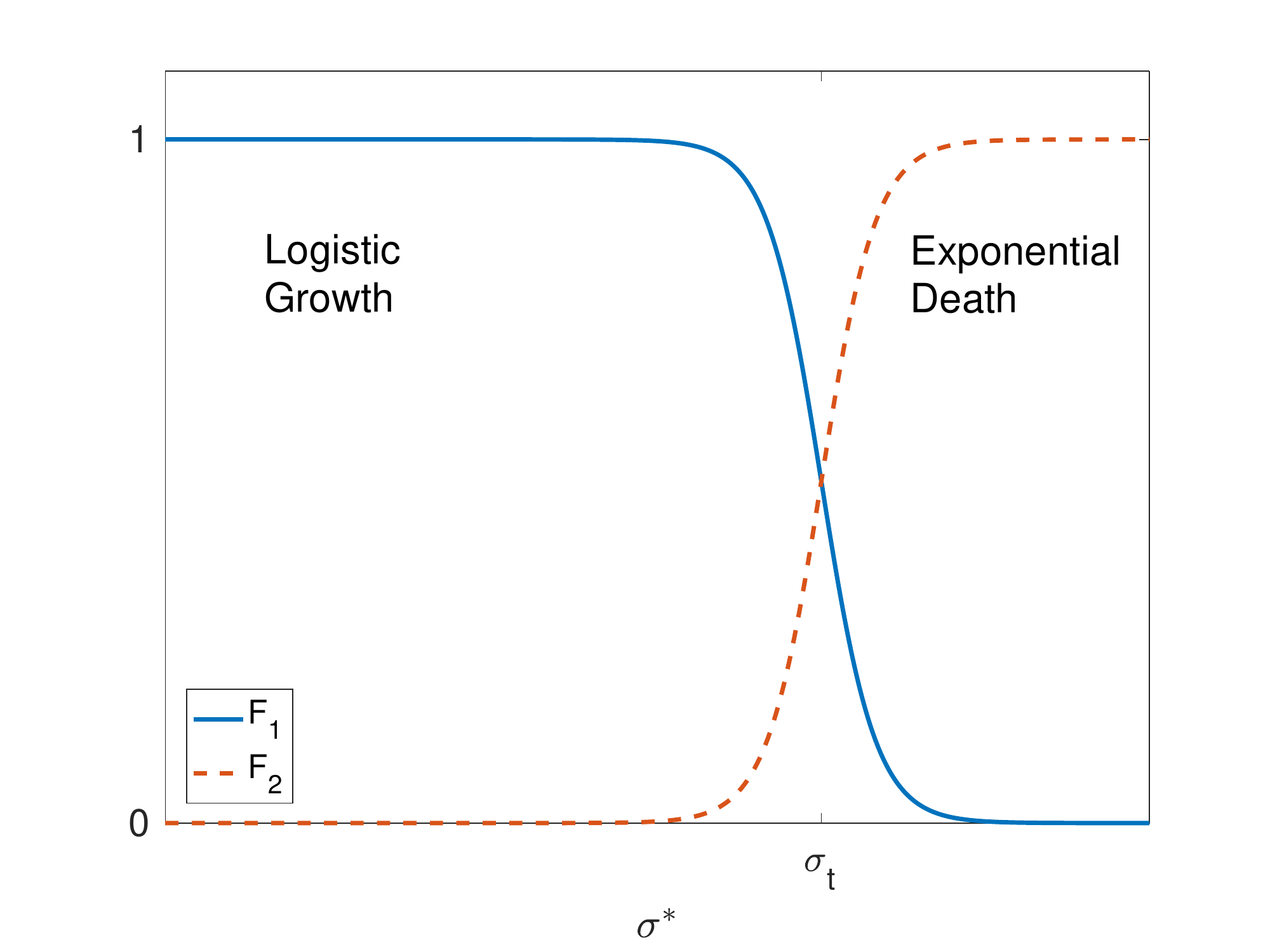}
\caption{Plots of the shear stress functions, $F_1(\sigma^\h)$ and $F_2(\sigma^\h)$. For $\sigma^\h < \sigma_t$, cells grow logistically, whereas for $\sigma^\h > \sigma_t$, cells die. These effects are local in space as the value of the shear stress depends on the position in the scaffold, and the overall profile of the flow.}\label{shear_function_plot}
\end{figure}

Finally we specify the functions $F_1(\sigma^\h)$ and $F_2(\sigma^\h)$ to be
\begin{equation}\label{pde_pressure_functions_sigma}
F_1(\sigma^\h) = 1
 - \left( \frac{1}{2} \right)(\tanh[g(\sigma^\h - \sigma_t)]+1),\enspace \text{and }
F_2(\sigma^\h) = \left( \frac{1}{2} \right)(\tanh[g(\sigma^\h - \sigma_t)]+1),
\end{equation}
where $g$ and $\sigma_t$ are sharpness and threshold parameters. Caricatures of these functions are plotted in Figure \ref{shear_function_plot}. These functions model smoothed step-function behaviour and are commonly used in the literature  \citep{coletti_mathematical_2006, shakeel_continuum_2013}. These functions capture logistic growth for small and moderate shear stress, and cell death at large values of shear stress.

We nondimensionalise as follows
\bes
(x^\h,y^\h) = L(x,y),\enspace \nabla^\h = \frac{1}{L}\nabla,\enspace (N^\h,N_0^\h) = N_c(N,N_0),\enspace \hat{\phi}(N^\h) = \phi_0 \phi(N),
\ees
\bes
k^\h(N^\h) = k_0\phi_0^3 k(N), \enspace p^\h = (p_0-p_1)p + p_1,\enspace \bm{u}^\h = \frac{k_0 \phi^3_0 (p_0-p_1)}{\mu L}\bm{u}, \enspace \bm{u_r}^\h = \frac{Q_c}{L}\bm{u_r}, 
\ees
\be
D^\h(N^\h) = D_n D(N), \enspace Q_0^\h = \frac{k_0 \phi^3_0 (p_0-p_1)}{\mu }Q_0, \enspace 
\sigma^\h = \frac{4 \mu \tau Q_c}{L R_0 \phi_0}\sigma, \enspace
t^\h = \frac{t}{\beta},\label{nondim_1}
\ee
leading to the following nondimensional system of equations,
\refstepcounter{equation} \label{pde_eqns_M}
\be
\bm{u} = -k(N)\nabla p,
\enspace \nabla \cdot \bm{u} = 0,\tag{\theequation \textit{a,b}}
\ee
\bes
\frac{\partial N}{\partial t} = F_1(\sigma)N(1-N) - F_2(\sigma)N+\delta \nabla \cdot \left ( D(N) \nabla N \right ),\tag{\theequation \textit{c}}
\ees
\bes
Q_0 = \int_0^1 -k(N(0,\hat{y},t))\frac{\partial p}{\partial x}(0,\hat{y})d\hat{y},
\enspace
\bm{u_r}=\frac{\bm{u}}{Q_0},\tag{\theequation \textit{d,e}}
\ees
\bes
\enspace \phi(N) = (1-\rho N),
\enspace k(N) = \phi(N)^3,
\enspace \sigma = \frac{\norm{\bm{u_r}}}{\phi(N)}, \tag{\theequation \textit{f,g,h}}
\ees
\bes
F_1(\sigma) = 1
 - \left( \frac{1}{2} \right)(\tanh[g_c(\sigma - \sigma_c)]+1),\enspace
F_2(\sigma) = \left( \frac{1}{2} \right)(\tanh[g_c(\sigma - \sigma_c)]+1),\tag{\theequation \textit{i,j}}
\ees
where $\rho = \nu N_c$ is the maximum available fraction of the pore space the cells can occupy, $\delta = D_n/ \beta L^2$ is the ratio of proliferation and diffusion timescales,  $g_c = (4 \mu \tau Q_c)/(L R_0 \phi_0)g$ is a sharpness parameter, and $\sigma_c = (L R_0 \phi_0)/(4 \mu \tau Q_c)\sigma_t$ is the threshold parameter. 

The nondimensional boundary conditions are
\be
\bm{u}\cdot\bm{n} = 0 \enspace \text{at} \enspace y= 0,1, \enspace 0 \leq x \leq 1,
\ee
\be
p = 1 \enspace \text{at} \enspace x = 0, \enspace 0 \leq y \leq 1,\quad \text{and} \quad
p = 0 \enspace \text{at} \enspace x = 1, \enspace 0 \leq y \leq 1,
\ee
\be
\bm{n}\cdot\nabla N = 0 \enspace \text{for } \bm{x} \in  \partial [0,1]^2,
\ee
with the initial data
\be
N(x,y,0) = N_0(x,y). \label{pde_cell_inits_M}
\ee

Equations \eqref{pde_eqns_M}-\eqref{pde_cell_inits_M} are a modified form of those used by \cite{shakeel_continuum_2013}, where we have neglected nutrient transport and nonlinear cell diffusion, but included a mechanism for cell death induced by high shear stress.
\subsection{Lattice Model}\label{Modelling}
We idealize the scaffold domain of length $L$ as an $n$ by $n$ square lattice representing the connectivity between pores. At each node we specify the local cell density together with fluid pressures which we use to define volumetric flow rates along the edges between nodes. We assume these edges represent pipes of length $l = L/n$. See Figure \ref{FlowDiagram} for a diagram of this idealization.

We prescribe a constitutive law relating the pressure $p_i$ at each node $i$ with the volumetric flow rate between the nodes, where $1 \leq i \leq n^2$. The particular form of this relationship depends on the microstructure of the scaffold under consideration, but for simplicity we assume Poiseuille flow between each node, so that
\begin{equation}\label{Poiseuille}
q_{ij}^\h= \frac{\pi (R_{ij}^{\h})^{4}}{8 l \mu}(p_i^\h-p_j^\h),
\end{equation}
where $q_{ij}^\h$ is the volumetric flow rate ($\mathrm{m}^3\mathrm{s}^{-1}$) from node $i$ to node $j$, and $R_{ij}^\h$ is the effective radius of the pipe between these nodes, which will depend on the cell density at the nearby nodes (see Equation \eqref{Cell_Radii}).

Due to incompressibility, we can write conservation of mass at each node as
\begin{equation}\label{Kirchoff}
\sum_{j=1}^{n^2} A_{ij}q_{ij}^\h=\begin{cases}
f_{li}, \enspace &1 \leq i \leq n,\\
0, \enspace &n < i \leq n^2-n,\\
-f_{ir}, \enspace &n^2-n < i \leq n^2,
\end{cases}
\end{equation}
where $A_{ij}$ is the unweighted undirected adjacency matrix of the lattice representing the connectivity of nodes, and $f_{li}$ and $f_{ir}$ are the volumetric flow rates into the nodes at the left and the right side of the scaffold respectively, which will be used to prescribe the total volumetric flow rate into the scaffold. We have that $A_{ij}=1$ if there is an edge between node $i$ and node $j$, and otherwise $A_{ij}=0$. While we only consider a square lattice here, this formulation is general and any pore network topology can be accounted for by providing a different adjacency structure.

As in Section \ref{PDEModel} we prescribe a pressure drop across the domain. We fix the pressure upstream of the left boundary as $p_0$, and the pressure downstream of the right boundary as $p_1$. Then the volumetric flow rate of fluid entering each node at the left is given by $f_{li}=(\pi R_{0}^{4}(p_0-p_i^\h))/(8 l \mu)$ for $1 \leq i \leq n$, and the volumetric flow rate of fluid leaving each node along the right by $f_{ir}=(\pi R_{0}^{4}(p_{i}^\h-p_1))/(8 l \mu)$ for $n^2-n < i \leq n^2$, where $R_{0}$ is the radius of pipes entering and exiting the scaffold. 

We combine equations \eqref{Poiseuille} and \eqref{Kirchoff} together with the expressions for the boundary volumetric flow rates $f_{li}$ and $f_{ir}$, to find that for all nodes $i$,
\begin{equation}
\sum_{j=1}^{n^2} A_{ij}(R_{ij}^{\h})^{4}(p_i^\h-p_j^\h) = \begin{cases}
R_{0}^{4}(p_0-p_i^\h), \enspace &1 \leq i \leq n,\\
0, \enspace &n < i \leq n^2-n,\\
-R_{0}^{4}(p_i^\h-p_1), \enspace &n^2-n < i \leq n^2.
\end{cases}\label{one_flow}
\end{equation}
Equations \eqref{one_flow} represent an algebraic system of $n^2$ equations for the nodal pressures, $p_i^\h$. As before, the physically relevant boundary condition is a constant volumetric flow rate into the scaffold. We exploit the linearity of the fluid problem above to rescale the fluid variables to match this condition. The total volumetric flow rate into the pipes along the left boundary is
\begin{equation}
Q^\h = \sum_{i=1}^n \frac{\pi R_{0}^{4}}{8 l \mu}(p_0-p_i^\h).
\end{equation}
We rescale the volumetric flow rates $q_{ij}^\h$ by,
\begin{equation}
q_{ij}^{r\h} = q_{ij}^\h \frac{Q_l}{Q^\h},
\end{equation}
where $Q_l$ is the volumetric flow rate into the scaffold. Note that $Q_l$ has dimensions of $\mathrm{m}^3\mathrm{s}^{-1}$, whereas for the 2-D spatially continuous model, the flow rate $Q_c$ had dimensions of $\mathrm{m}^2\mathrm{s}^{-1}$. If we assume that the mean fluid velocity (not the Darcy velocity) into the pore space of the scaffold along the left boundary is the same between the two models, we must have $Q_c/(L\phi_0) = Q_l/(n \pi R_0^2)$ where the factor of $\pi R_0^2$ is the cross-sectional area of each pipe, and $n$ accounts for the number of pipes along the leftmost boundary. Hence $Q_l = (n\pi R_0^2 Q_c)/(L \phi_0)$, and we use this relationship in the remainder of the paper.

Again assuming that cells grow logistically, die, and diffuse to neighbouring nodes, the evolution equation for the cell density $N_i^\h$ at node $i$ is
\begin{equation}\label{lattice_cells}
\frac{d N_{i}^\h}{d t^\h} = \beta \left(F_{1}^l (\sigma_i^\h)N_{i}^\h\left(1-\frac{N_{i}^\h}{N_c}\right ) - F_{2}^l(\sigma_i^\h)N^\h_i\right) +  \sum_{j=1}^{n^2}A_{ij} D^\h(N_i^\h,N_j^\h)(N_j^\h - N_i^\h), \enspace i=1\dots n^2,
\end{equation}
where $D^\h(N_i^\h,N_j^\h)$ is the local nonlinear cell diffusion rate to move between neighboring nodes, $\sigma_i^\h$ is a local average of the shear stress defined in Equation \eqref{lattice_sigma_def}, and the functions $F_{1}^l$ and $F_{2}^l$ model the effect of shear stress on cell growth and death. Equations \eqref{lattice_cells} are analogous to Equation \eqref{continuum_cells_full}. In particular, the cell proliferation rate $\beta$ and maximum cell density $N_c$ have the same meaning as in Equation \eqref{continuum_cells_full}. The cell diffusion rate between nodes can be related to cell diffusion on the length scale of the scaffold $D_n$ by $D_l =  D_n/l^2 = n^2 D_n/L^2$. We specify an initial condition at each node as
\begin{equation}
N_i^\h(0) = N_{i0}^\h.
\end{equation}

We model the effect of cell growth on the fluid flow by taking the effective radius of a pipe to depend on the nearby cell densities as
\begin{equation}\label{Cell_Radii}
R_{ij}^\h = R_0\left(1-\frac{\nu }{2}(N_i^\h+N_j^\h)\right),
\end{equation}
where again $\nu$ represents cell volume. Equation \eqref{Cell_Radii} models pipes with radii that linearly decrease as the cell density increases at nearby nodes. Note that we must have $\nu N_c  \leq 1$ for this radius to be non-negative for all feasible cell densities. This relationship is analogous to the porosity relationship given in Equation \eqref{porosity_eqn}. 

We compute the shear stress in each pipe under the assumption of Poiseuille flow made in Equation \eqref{Poiseuille}. The local velocity profile in the pipe $R_{ij}^\h$ is
\begin{equation}\label{pipe_flux}
u_{ij}^\h(r^\h) = -2\left(\frac{(R_{ij}^{\h})^{2}-(r^\h)^2}{\pi (R_{ij}^{\h})^{4}}\right)q_{ij}^{r\h},
\end{equation}
where $0 \leq r^\h  \leq R_{ij}^\h$ is the radial coordinate. The magnitude of the shear stress at the wall of each pipe is then
\begin{equation}
\sigma_{ij}^\h = \mu\abs{\frac{\partial u_{ij}^\h}{\partial r^\h}\left(R_{ij}^\h\right)} = \frac{4 \mu}{\pi (R_{ij}^{\h})^{3}}\abs{q_{ij}^{r\h}}.
\end{equation}

As cell densities are defined at the nodes, we must relate the shear stress in each idealized pipe to an average shear stress $\sigma_i^\h$ at each node. We note that adding the shear stresses from the four neighboring nodes accounts for the total volumetric flow rate twice, so we sum the shear stress in each pipe connected to node $i$ and divide by $2$ to obtain
\begin{equation}\label{lattice_sigma_def}
\sigma_i^\h = \frac{1}{2}\sum_{j=1}^{n^2}A_{ij}\sigma_{ij}^\h = \sum_{j=1}^{n^2}A_{ij} \frac{ 2\mu}{\pi (R_{ij}^{\h})^{3}}\abs{q_{ij}^{r\h}}.
\end{equation}

We model the influence of this averaged shear stress on cell proliferation and death as before by
\begin{equation}
F_{1}^l(\sigma_i^\h) = 1-\left( \frac{1}{2} \right)(\tanh[g(\sigma_i^\h - \sigma_t)]+1),\enspace
F_{2}^l(\sigma_i^\h) = \left( \frac{1}{2} \right)(\tanh[g(\sigma_i^\h - \sigma_t)]+1).
\end{equation}
These functions have the same form as those in Equations \eqref{pde_pressure_functions_sigma}; see Figure \ref{shear_function_plot} for a visualization.

For each $1 \leq i,j \leq n^2$ we nondimensionalize by taking
\bes
p_i^\h = (p_0-p_1)p_i+p_1, 
\enspace Q^\h = \frac{\pi R_{0}^{4} (p_0-p_1)}{8 l \mu }Q,
\enspace (N_i^\h,N_{i0}^\h) = N_c (N_i,N_{i0}), 
\enspace D^\h(N_i^\h,N_j^\h) = D_l D(N_i,N_j),
\ees
\be
q_{ij}^{r\h} = \frac{n\pi R_0^2 Q_c}{L\phi_0} q_{ij}^{r},
\enspace q_{ij}^\h = \frac{\pi R_{0}^{4} (p_0-p_1)}{8 l \mu } q_{ij},
\enspace \sigma_i^\h = \frac{2\mu Q_c}{L \phi_0 R_0} \sigma_i, 
\enspace t^\h = \frac{t}{\beta}, 
\enspace R_{ij}^\h = R_0 R_{ij},
\ee
from which we obtain \refstepcounter{equation} \label{lattice_fluid_flow_eqn}
\be
\enspace q_{ij}= R^4(N_i,N_j)(p_i-p_j),\tag{\theequation \textit{a}}
\ee
\bes
 \sum_{j=1}^{n^2}A_{ij}q_{ij}= 
\begin{cases}
1-p_i, \enspace &1 \leq i \leq n,\\
0, \enspace &n < i \leq n^2-n,\\
-p_i, \enspace &n^2-n < i \leq n^2,
\end{cases} \tag{\theequation \textit{b}}
\ees
\bes
\frac{d N_{i}}{d t} = F_{1}^l(\sigma_i)N_{i}(1-N_{i}) - F_{2}^l(\sigma_i)N_{i} + \delta n^2 \sum_{j=1}^{n^2} A_{ij}D(N_i,N_j)(N_j - N_i), \tag{\theequation \textit{c}}
\ees
\bes
Q = \sum_{i=1}^n 1-p_i,\enspace q_{ij}^{r} = q_{ij} \frac{1}{Q}, 
\enspace R_{ij}=1-\frac{\rho}{2}(N_i+N_j),
\enspace \sigma_i = n\sum_{j=1}^{n^2}A_{ij} R_{ij}\abs{q_{ij}^r},\tag{\theequation \textit{d,e,f,g}}
\ees
\bes
F_{1}^l(\sigma_i) = 1-\left( \frac{1}{2} \right)(\tanh[g_l(\sigma_i - \sigma_l)]+1),\enspace F_{2}^l(\sigma_i) = \left( \frac{1}{2} \right)(\tanh[g_l(\sigma_i - \sigma_l)]+1),\tag{\theequation \textit{h}}
\ees
where $i=1\dots n^2$, $\delta = (D_n)/(L^2\beta)$ is the ratio of proliferation and diffusion timescales, $\rho=\nu N_c$ is the maximum available fraction of pore radius cells can occupy, $g_l = (2\mu Q_c g)/(L R_0\phi_0)$ is a sharpness parameter, and $\sigma_l = (L R_0 \phi_0\sigma_t)/(2\mu Q_c)$ is the shear stress threshold parameter. Note that $\rho$ and $\delta$ are identical parameters between the lattice and continuum models. The parameters $g_l$ and $\sigma_l$ are analogous to $g_c$ and $\sigma_c$ but may have different values for a particular experimental system. The number of pores, $n$, does not explicitly appear in the cell density equation (except to scale diffusion) or the constitutive law for its relationship to pipe radii as we have accounted for issues of lengthscale in thinking of $N_i$ as the nondimensional density of cells, rather than a cell number. %The effects of size constraints on cell growth are hence captured by the parameters $\nu$ and $N_c$ which we have nondimensionalized into the parameter $\rho$ which will in practice vary with the scaffold size $L$, pore lengthscale $l$ and pore radius $R_0$, but we have left this relationship unspecified for generality in order to model different geometric pore structures. 
We also have the initial data 
\be\label{init_lattice}
N_i(0) = N_{i0}, \quad 1 \leq i \leq n^2.
\ee

In addition to fundamental differences between a spatial continuum and a discrete lattice, there are two significant constitutive differences between our models. The exponents in the relationships between the Darcy velocity, $\bm{u_r}$, or volumetric flow rates, $q_{ij}$, which play analogous roles, and the cell density are not the same between the two models. To see this, we write Equation (\ref{pde_eqns_M}\textit{a}) using the permeability $k$ and porosity $\phi$ in Equations (\ref{pde_eqns_M}\textit{f,g}) as
\be\label{PDE_Vel}
\bm{u} = -\left(1-\rho N\right)^3\nabla p,
\ee
and similarly rewrite (\ref{lattice_fluid_flow_eqn}\textit{a}) using the effective pipe radii from Equation (\ref{lattice_fluid_flow_eqn}\textit{f}) to find,
\be\label{lattice_Vel}
q_{ij}= \left(1-\frac{\rho}{2}\left(N_i+N_j\right)\right)^4(p_i-p_j).
\ee
These relationships show that for a given local pressure drop, the local Darcy velocity in the PDE is larger in regions of high cell density than the corresponding volumetric flow rate in the lattice. 

There are also differences in the nondimensional expressions for the shear stress in each model (see Equation (\ref{pde_eqns_M}\textit{h}) and Equation (\ref{lattice_fluid_flow_eqn}\textit{g})). To visualize how these constitutive differences lead to quantitatively different predictions, consider a uniform cell density in both models. Let $N(x,y,t) = \hat{N}(t)$ for all $(x,y)$ in the domain and $N_i(t) = \hat{N}(t)$ for $1 \leq i \leq n^2$. For a uniform cell density, the fluid flow is uniform so $\norm{\bm{u_r}}=1$ and $q_{ij}^r=1/n$ for all $i,j$. This can be seen from solving Equations (\ref{pde_eqns_M}\textit{a,b}) and (\ref{pde_eqns_M}\textit{d-h}) for the PDE, or Equations (\ref{lattice_fluid_flow_eqn}\textit{a,b}), and (\ref{lattice_fluid_flow_eqn}\textit{d-g}) for the lattice. Note that in the lattice, we have nondimensionalized such that the total volumetric flow rate into the scaffold from the left boundary is $1$, leading to the flow in each horizontal pipe being $1/n$ as there are $n$ pipes along the boundary. We can then compute the (spatially uniform) shear stress for the PDE model as,
\be\label{PDEShear}
\sigma = \frac{\norm{\bm{u_r}}}{\phi(N)} = \frac{1}{1-\rho \hat{N}},
\ee
and for the lattice model as,
\be\label{LatticeShear}
\sigma_i = n\sum_{j=1}^{n^2}A_{ij} R_{ij}\abs{q_{ij}^r} = \frac{2}{(1-\rho \hat{N})^3}.
\ee
\begin{figure}
\centering
\includegraphics[width=0.6\textwidth]{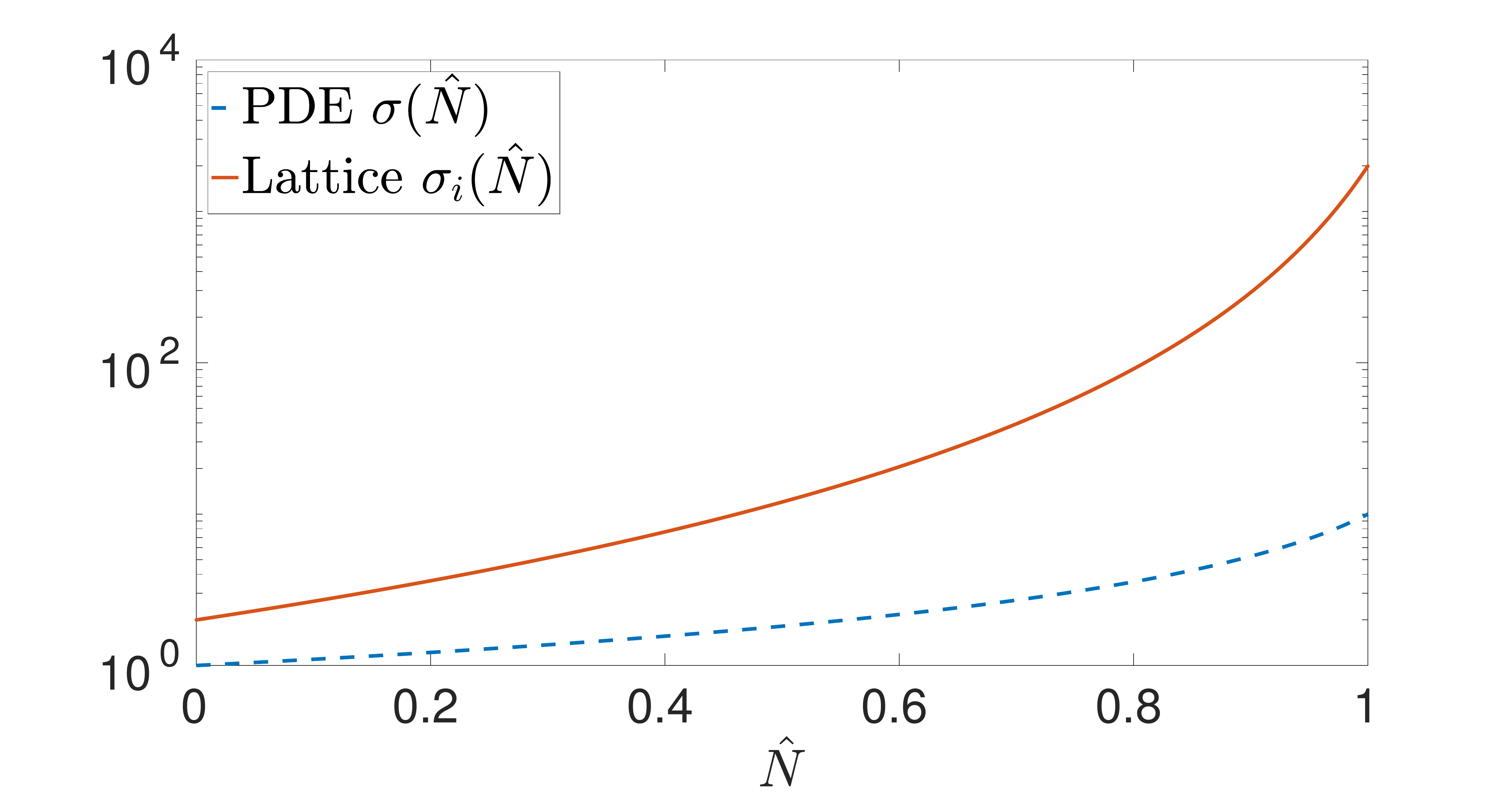}
\caption{Plots of the uniform shear stress for a uniform cell density distribution $\hat{N}$, for $\rho=0.9$. Note that the vertical axis is scaled logarithmically.}
\label{ShearNPlot}
\end{figure}

We plot these shear stresses as functions of $\hat{N}$ in the interval $[0,1]$ in Figure \ref{ShearNPlot}. For large cell densities, the shear stress experienced by the cells in the continuum and lattice models differs by two orders of magnitude. These differences between Equations \eqref{PDEShear} and \eqref{LatticeShear} prevent quantitative agreement between our models; we emphasize qualitative differences between the model predictions by exploring ranges of parameters in the following discussion.

Various forms of nonlinear cell diffusion have been used to model the movement of cells within porous media. For simplicity, we primarily consider only linear cell diffusion ($D(N)=D(N_i,N_j)=1$), so that both models have linear cell diffusion at a rate $\delta$. In Section \ref{NonlinearDiffusion} we demonstrate that the linear diffusion regime still accurately captures the main features we are trying to illustrate.
\subsection{Model Paremeters}
The continuum and lattice models contain a number of parameters that can vary significantly between tissue engineering experiments. As we are primarily interested in informing model selection, rather than constructing quantitatively accurate models, we focus on the orders of magnitude over which these parameters vary. Table \ref{parameter_table} lists typical parameter values found in the literature. Using these values we estimate the ranges of the two nondimensional parameters, $\rho$ and $\delta$, that appear in both models, and estimate the number of pores per side of the square lattice $n$. 
\begin{table}
	\centering
    \begin{tabular}{| l  l  l |}
    \hline
    Parameter & Expression & Values \\ \hline
    Cell volume$^1$ & $\nu$ & $4.2 \times 10^3-1.7 \times 10^4\mathrm{\mu m}^3/\text{cell}$ \Tstrut \\ 
    Maximum cell density $^1$ $^2$ $^3$  & $N_c$ & $10^{13}-10^{14}\text{cell}/\mathrm{m}^3$  \\ 
    Scaffold length$^1$  & $L$ & $2-10\mathrm{mm}$ \\
    Pore length scale$^4$ & $l,R_0$ & $85-325\mathrm{\mu m}$ \\
    Cell proliferation timescale$^2$ $^3$ & $1/\beta$ & $6.59\times 10^4\mathrm{s}$ \\
    Viscosity$^1$ & $\mu$ & $0.7 \mathrm{mP s}$  \\
    Cell-free scaffold porosity$^3$ $^5$ & $\phi_0$ & $40$-$97$\% \\
    Pump flow rate$^4$ $^6$ & $Q_c$ & $5.2-6.24\mathrm{mm}^2/\mathrm{s}$  \\
    Cell diffusion rate$^2$ $^3$ & $D_n$ & $10^{-12}-10^{-11}\mathrm{m}^2/\mathrm{s}$ \\
    Tortuosity$^7$ $^8$ $^{9}$ & $\tau$ & 1.15 \\
    \hline\Tstrut
    Ratio of timescales & $\delta = D_n/L^2\beta$ & $10^{-4}-10^{1}$\\
    Maximum cell volume fraction & $\rho=\nu N_c$ & $10^{-2}-1$\\    
    Pores per lattice side & $n=L/l$ & $10-10^{3}$ \Bstrut\\
    \hline
    \end{tabular}
    \caption[Table Caption]{Parameters used in the models. 
$^1$ \cite{vunjak1998dynamic},    
$^2$ \cite{vafai_porous_2010}, 
$^3$ \cite{shakeel_continuum_2013},
$^4$ \cite{mccoy_influence_2012},
$^5$ \cite{truscello_prediction_2012},
$^6$ \cite{glowacki_perfusion_1998},%%%%
$^7$ \cite{whittaker_mathematical_2009},
$^8$ \cite{koponen1996tortuous},
$^{9}$ \cite{kou2012tortuosity},
.}\label{parameter_table}
\end{table}

The shear stress threshold $\sigma_t$ and the sharpness parameter $g$ are not readily available in the literature. We treat the nondimensional thresholds $\sigma_l$ and $\sigma_c$ as model parameters and demonstrate model behaviours as they vary. Motivated by existing theoretical models in which the function $F_1$ approximated a step function \citep{shakeel_continuum_2013}, we set $g_l = g_c = 60$. In our simulations we fix $\rho = 0.9$ so that cell growth significantly affects the effective permeability of the scaffold.

The nondimensional parameters $\rho$ and $n$ can readily be determined from the properties of a specific porous scaffold microstructure and cell type. We can use the values in Table \ref{parameter_table} to estimate that the number of pores in a typical scaffold ranges from approximately $10^2$ to $10^6$. Similarly, given the values of the cell volume $\nu$ and pore radii $R_0$, we compute that the number of cells required to fill a pore can vary in the range of $10^2-10^4$. We now simulate our models in order to explore behaviours as these nondimensional parameters vary.
\subsection{Numerical Simulations}\label{ShearNumerics}
The lattice size $n$ and diffusion to proliferation rate $\delta$ are varied to demonstrate the range of qualitative behaviours displayed by these models. Specifically we take $n=25, 50,75$, and $100$ for the lattice simulations, and $\delta=10^{-4}, 10^{-3}, 10^{-2}$, and $10^{-1}$ for both models. We take values of both lattice and PDE thresholds to be $\sigma_c=\sigma_l=2.5, 5, 7.5, 10, 100$, and $1000$.

The continuum model given by Equations \eqref{pde_eqns_M}-\eqref{pde_cell_inits_M} was simulated using the finite element solver Comsol with 24,912 triangular elements. Time and space refinements were carried out to ensure that the numerical approach converged. Additionally, the results found via Comsol were consistent with those found from a finite-difference scheme implemented to solve the same model. The lattice model given by Equations \eqref{lattice_fluid_flow_eqn}-\eqref{init_lattice} was solved using an explicit adaptive Runge-Kutta method in Matlab. To ensure accuracy of our simulations with respect to the bifurcation behaviour discussed in Section \ref{Oscillations}, we constrain the maximum Runge-Kutta time step to be $10^{-3}$ \citep{christodoulou_discrete_2008}. Specific simulations were also undertaken with a fixed time step Runge-Kutta scheme with refinements in the size of the time step to ensure convergence.

We consider our initial condition to be a perturbed uniform cell density. On a lattice with $n=100$ nodes per side, we set $N_i(0) = 0.1 + h\xi_i$ where $\xi_i \sim \mathcal{N}(0,1)$ is a normally distributed noise term, and we use three values of the variance, $h=10^{-2},10^{-3},$ and $10^{-4}$. Each realization of this perturbed uniform state is then interpolated on the smaller lattices, or onto the triangular elements of the continuum model, so that all simulations have approximately consistent initial conditions. We compute mean cell densities as,
\be\label{lattice_mean}
\hat{N}(t) = \left(\frac{\sum_{i=1}^{n^2}N_i(t)}{n^2}\right),
\ee 
for the lattice and 
\be\label{PDE_mean}
\hat{N}(t) = \int_0^1\int_0^1 N(x,y,t)dxdy,
\ee
for the PDE (computed numerically using the triangular elements from the finite element scheme). For each set of parameters simulated, and each value of the variance $h$, we repeat the simulation 100 times using different random seeds in order to investigate how robust each result is to different initial cell densities.
\section{Results and Discussion}\label{ResultsDiscussion}
In Section \ref{Patterning} we discuss how the cell density changes from uniform logistic growth to death in local regions of the scaffold, leading to spatial heterogeneity in cell density which is observed over long timescales. In Section \ref{instability_time} we discuss the onset time of this spatial patterning. In Section \ref{Oscillations} we discuss oscillations in cell density that are observed only in simulations of the lattice model. In Section \ref{Discussions}, we compare the model predictions for the final cell densities. Finally, in Section \ref{NonlinearDiffusion}, we compare the results of a linear cell diffusion model with those of a model with nonlinear cell diffusion.

\subsection{Spatial Heterogeneity}\label{Patterning}
Over long timescales (nondimensional times of $t> 10$) all of our simulations exhibit significant spatial heterogeneity in cell density and shear stress. In some regions of the scaffold the cell density is close to the nondimensional carrying capacity, and there is reduced fluid flow and low shear stress in these areas. There are also regions of high fluid flow and associated shear, and low cell density. The spatial structure of the cell density distributions depends heavily on the parameter $\delta$.

\begin{figure}
\begin{subfigure}{0.24\textwidth}
\includegraphics[width=1\textwidth]{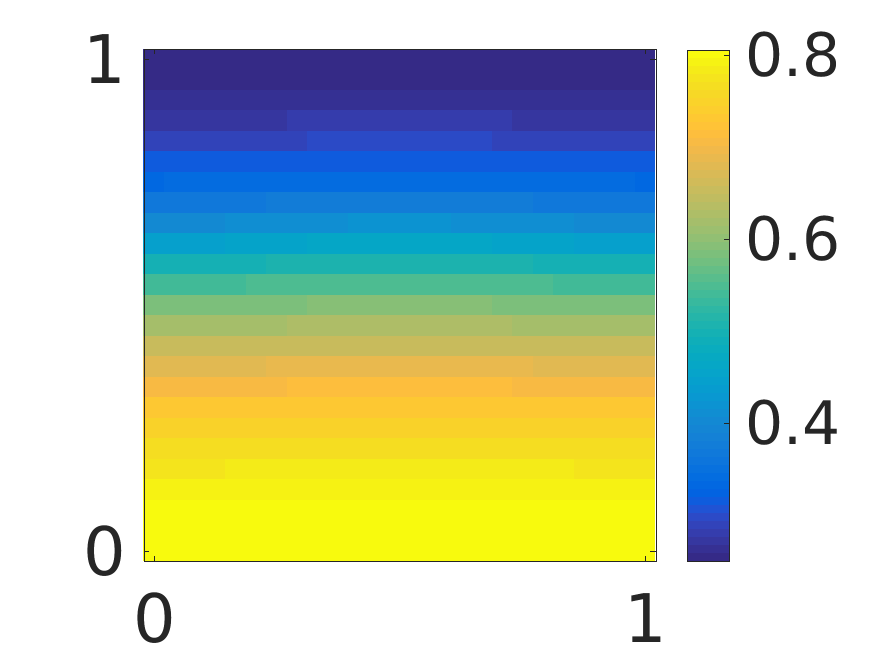} 
\caption{$n=25$, $\delta = 10^{-1}$}
\end{subfigure}\hfill%
\begin{subfigure}{0.24\textwidth}
\includegraphics[width=1\textwidth]{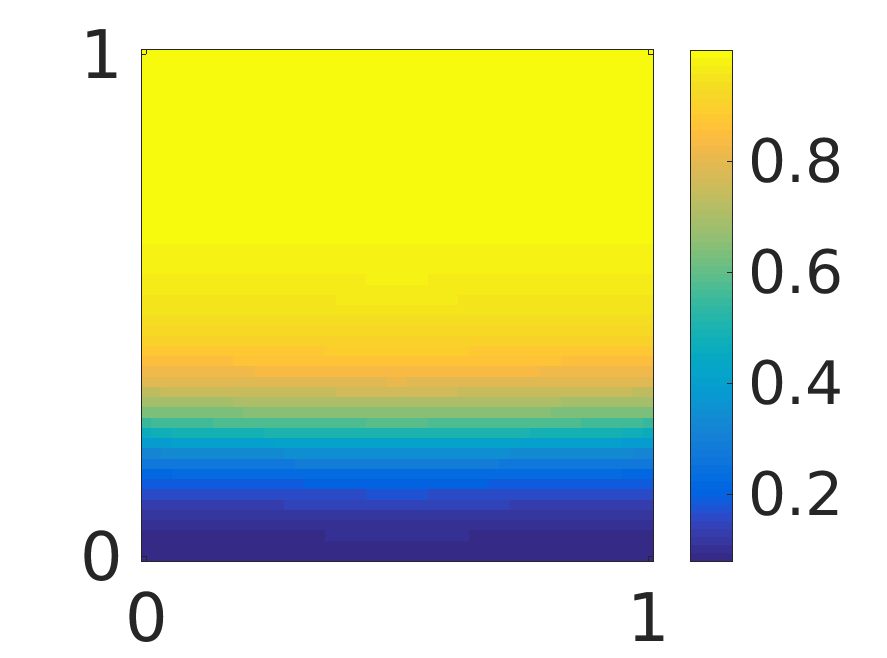}
\caption{$n=50$, $\delta = 10^{-2}$}
\end{subfigure}\hfill%
\begin{subfigure}{0.24\textwidth}
\includegraphics[width=1\textwidth]{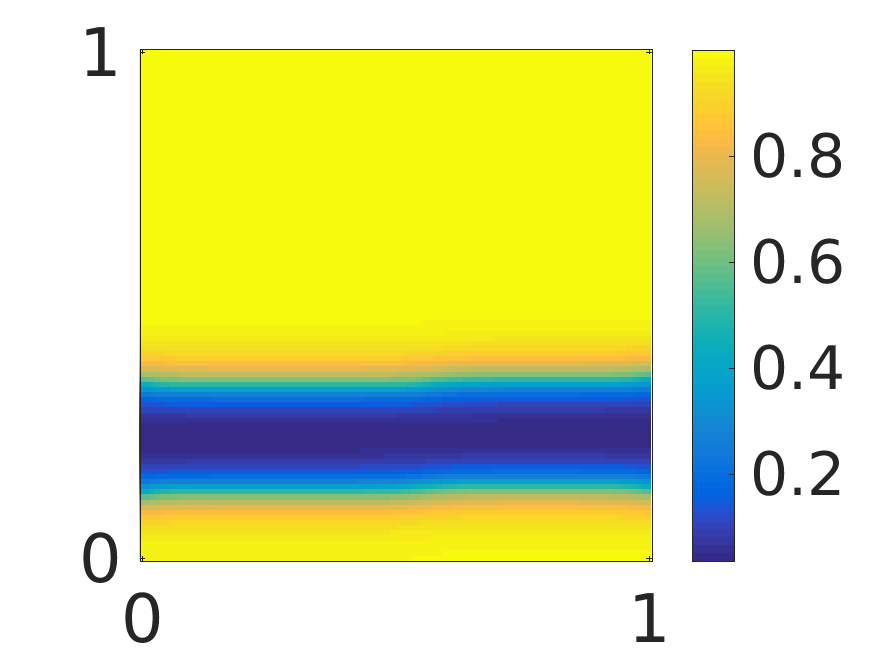} 
\caption{$n=100$, $\delta = 10^{-3}$}
\end{subfigure}\hfill%
\begin{subfigure}{0.24\textwidth}
\includegraphics[width=1\textwidth]{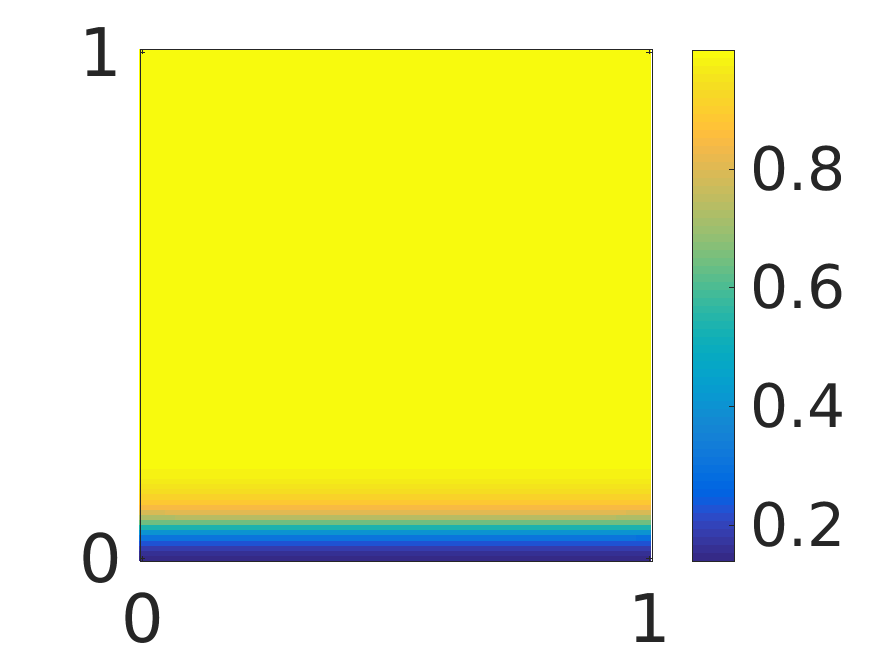}
\caption{PDE, $\delta = 10^{-3}$}
\end{subfigure}
\caption{Plots of the cell density with $t=30$ and $\sigma_c=\sigma_l=10$ demonstrating simple spatial patterning at various values of diffusion, $\delta \geq 10^{-3}$. The size of these channels is consistent for all 300 realizations of different random initial conditions, although the location of the channels does change.}\label{large_diffusion_plots}
\end{figure}

For large values of the parameter ($\delta \geq 10^{-3}$), a single region or channel develops with low cell density, and through which the majority of fluid passes. In the remainder of the domain, there is little fluid flow and the cell density is high; see Figure \ref{large_diffusion_plots}. The location of the large fluid channel, corresponding to regions of low cell density, depends on the initial data and parameter values, but we always observe a single channel for $\delta \geq 10^{-3}$. For brevity we plot only the cell density distributions in the scaffold, as the associated shear stress can be inferred from these plots (the fluid flow is always low in regions of high cell density and vice versa).
\begin{figure}\setlength{\tabcolsep}{0pt}
\centering
\begin{tabular}{rc c c} 
 & $n=25$ & $n=50$ & $n=100$\\

$t=2$ &\raisebox{-.5\height}{\includegraphics[width=.3\textwidth]{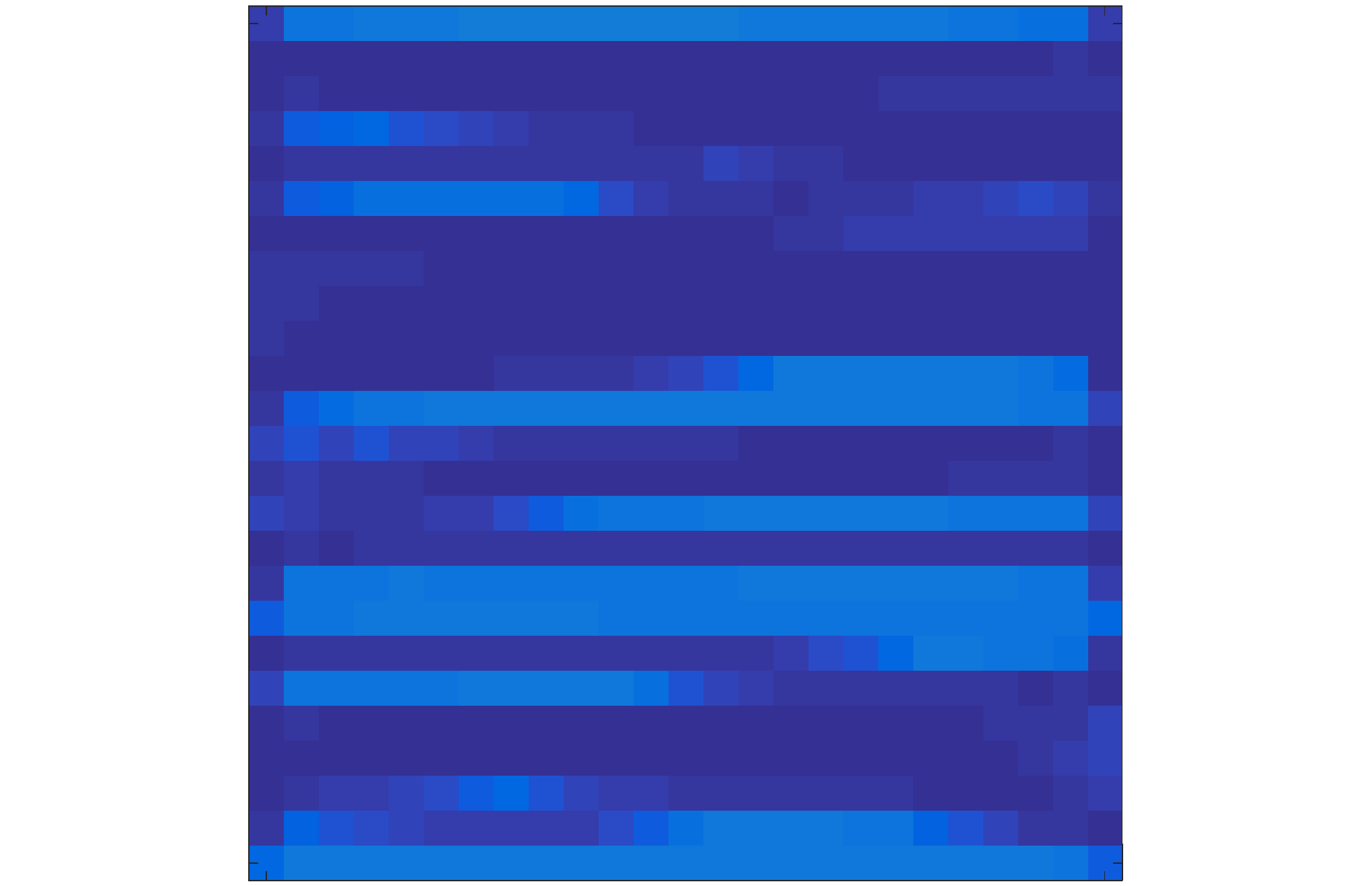}}&
\raisebox{-.5\height}{\includegraphics[width=.3\textwidth]{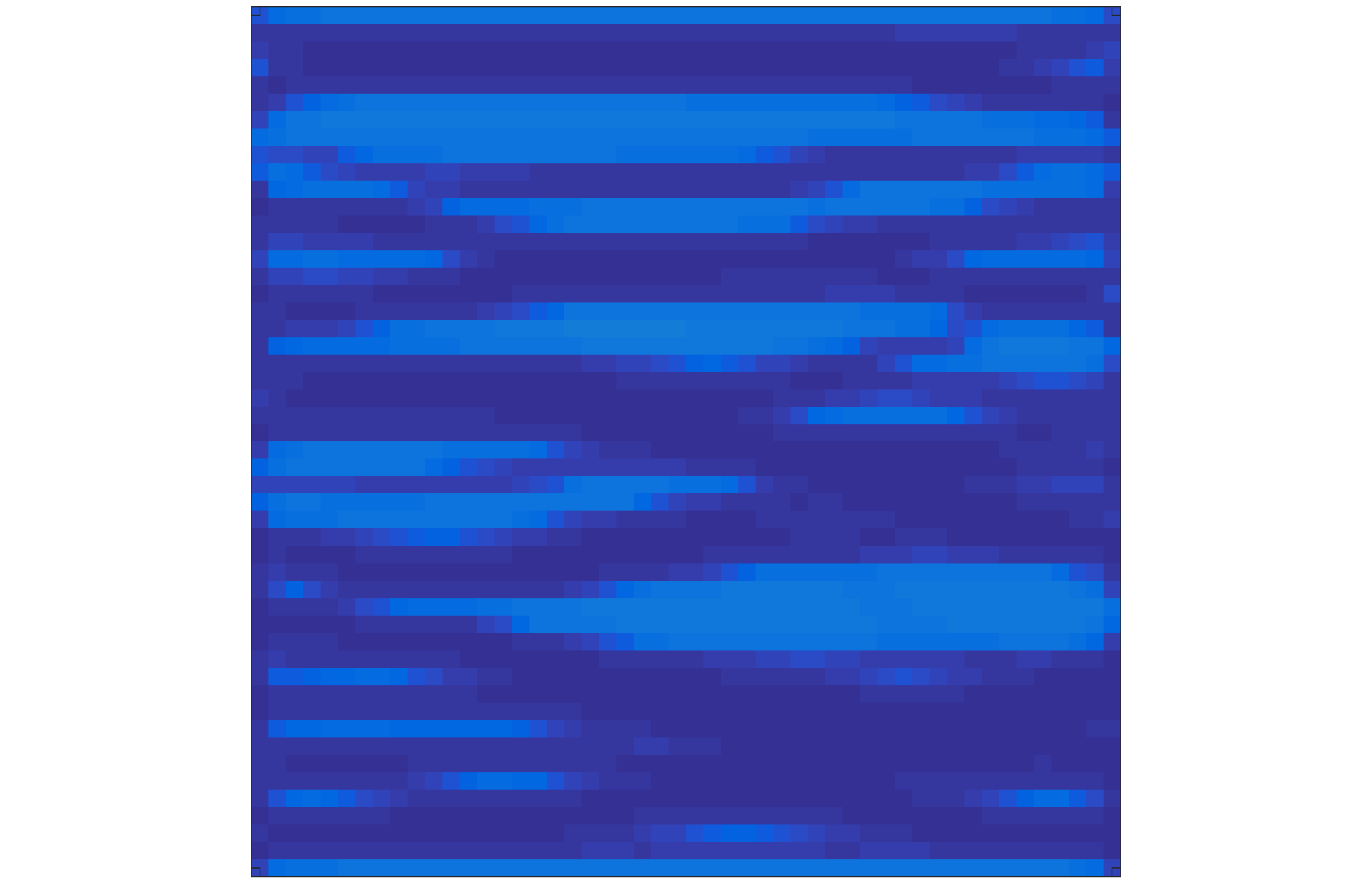}}&
\raisebox{-.5\height}{\includegraphics[width=.3\textwidth]{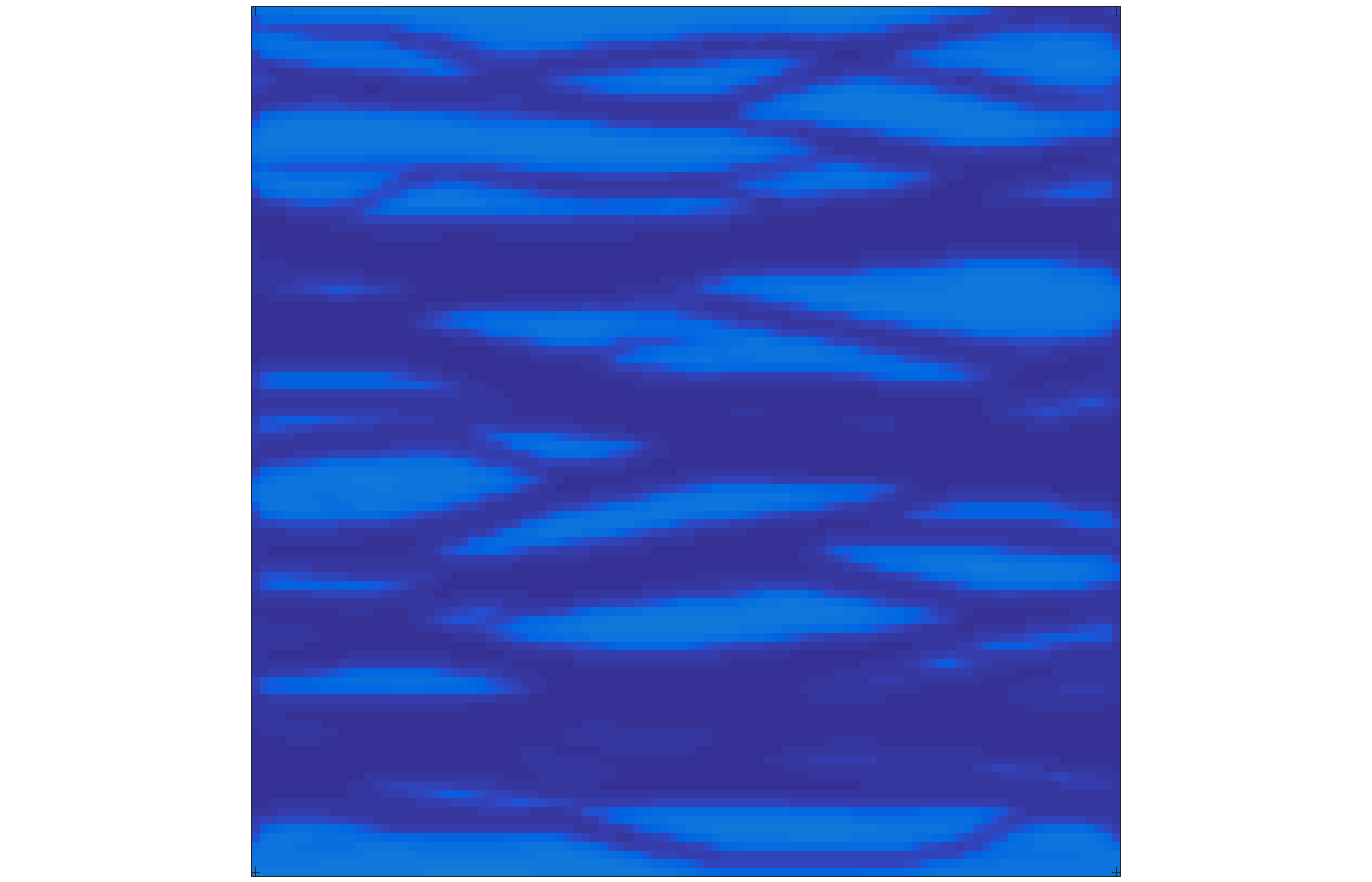}}\vspace{1mm}\\

$t=5$ &\raisebox{-.5\height}{\includegraphics[width=.3\textwidth]{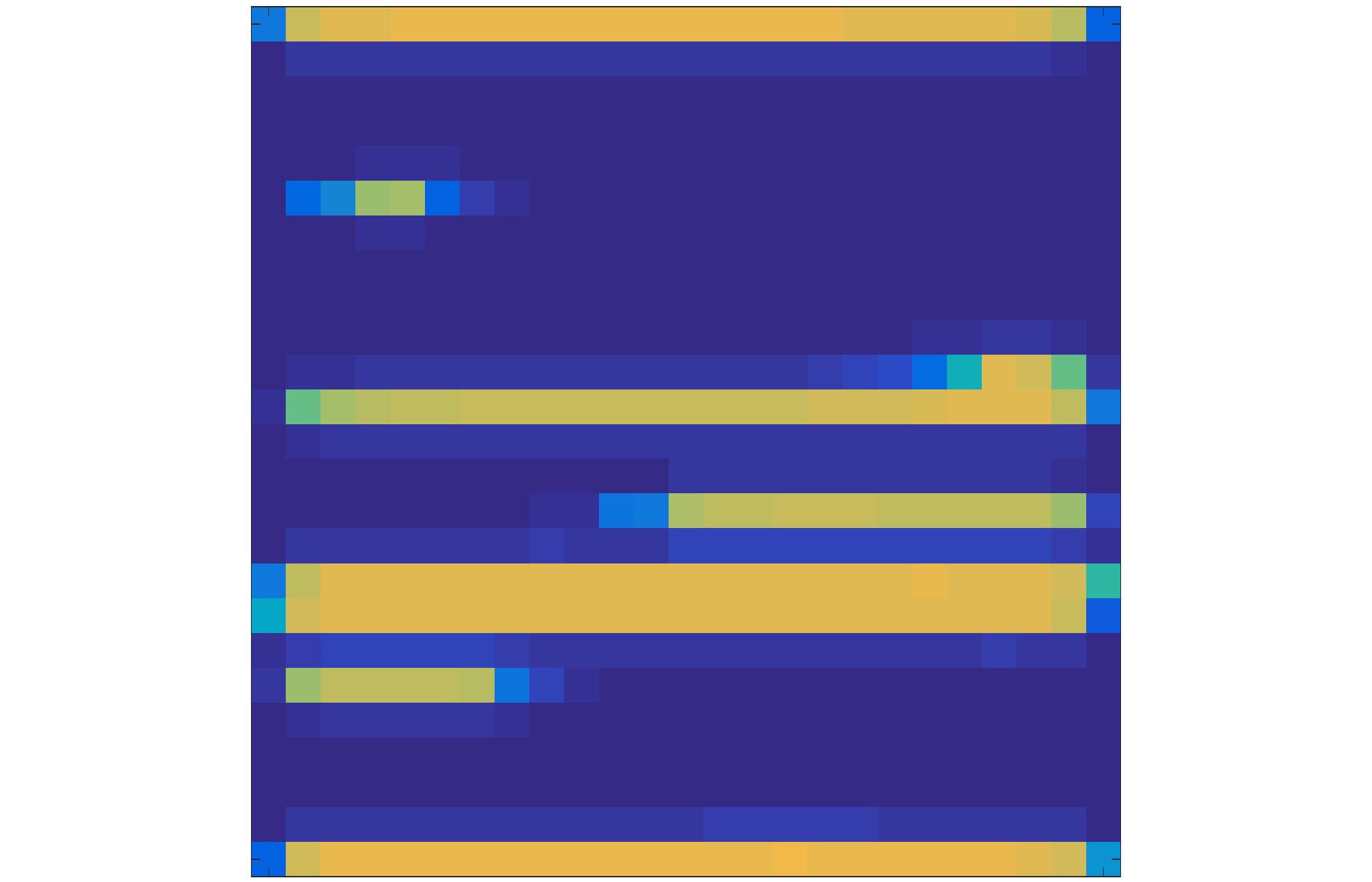}}&
\raisebox{-.5\height}{\includegraphics[width=.3\textwidth]{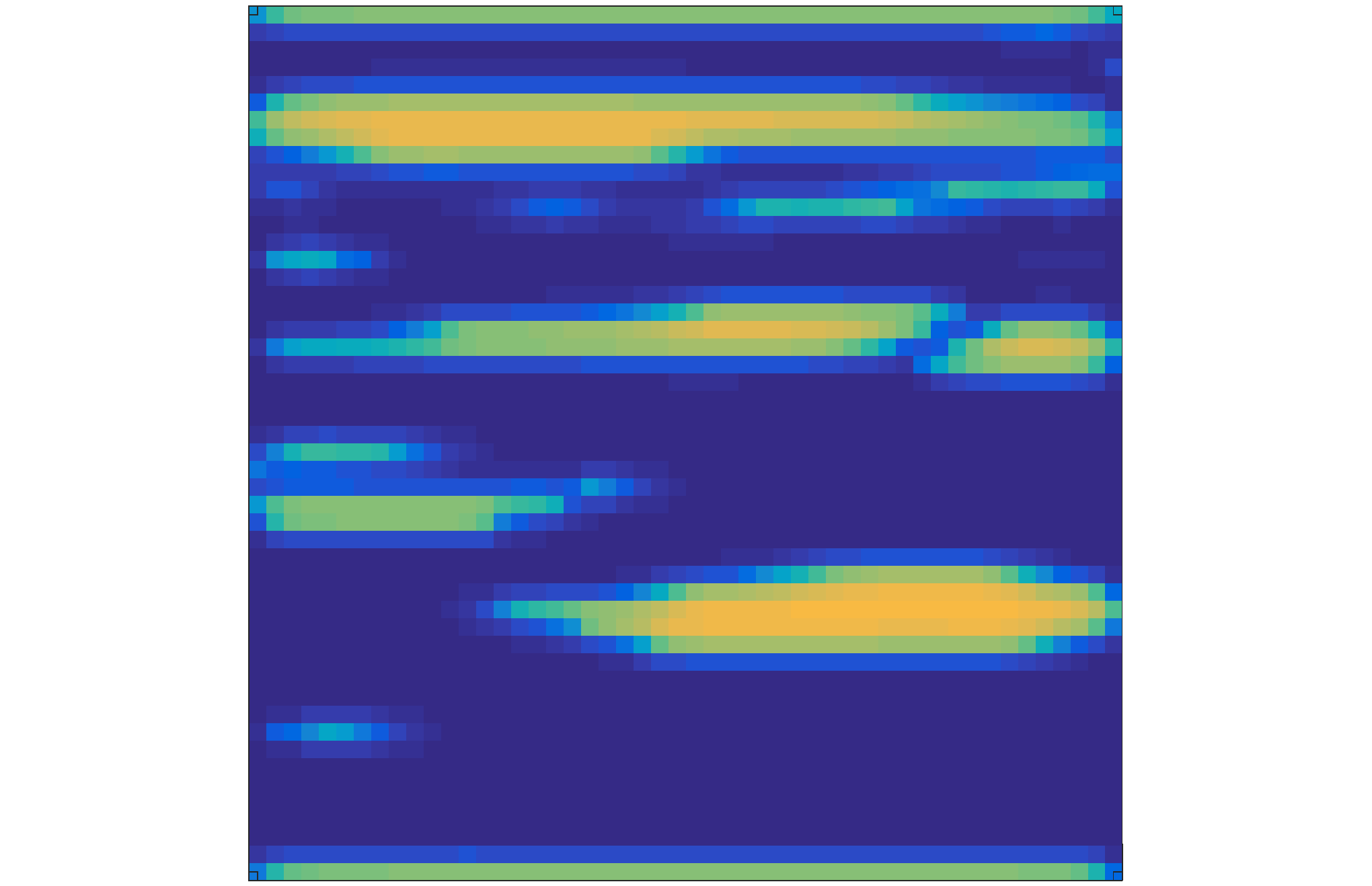}}&
\raisebox{-.5\height}{\includegraphics[width=.3\textwidth]{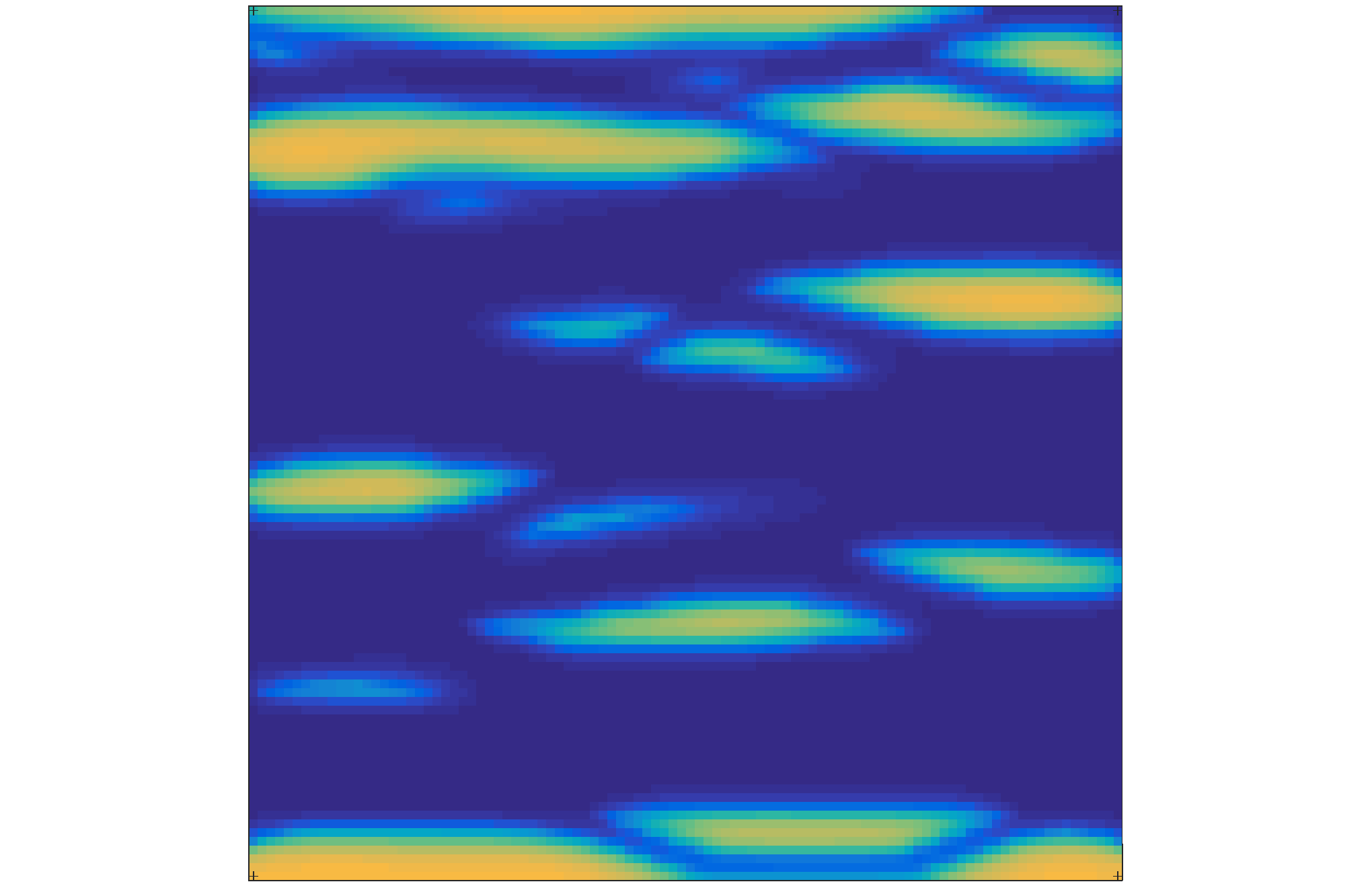}}\vspace{1mm}\\

$t=10$ &\raisebox{-.5\height}{\includegraphics[width=.3\textwidth]{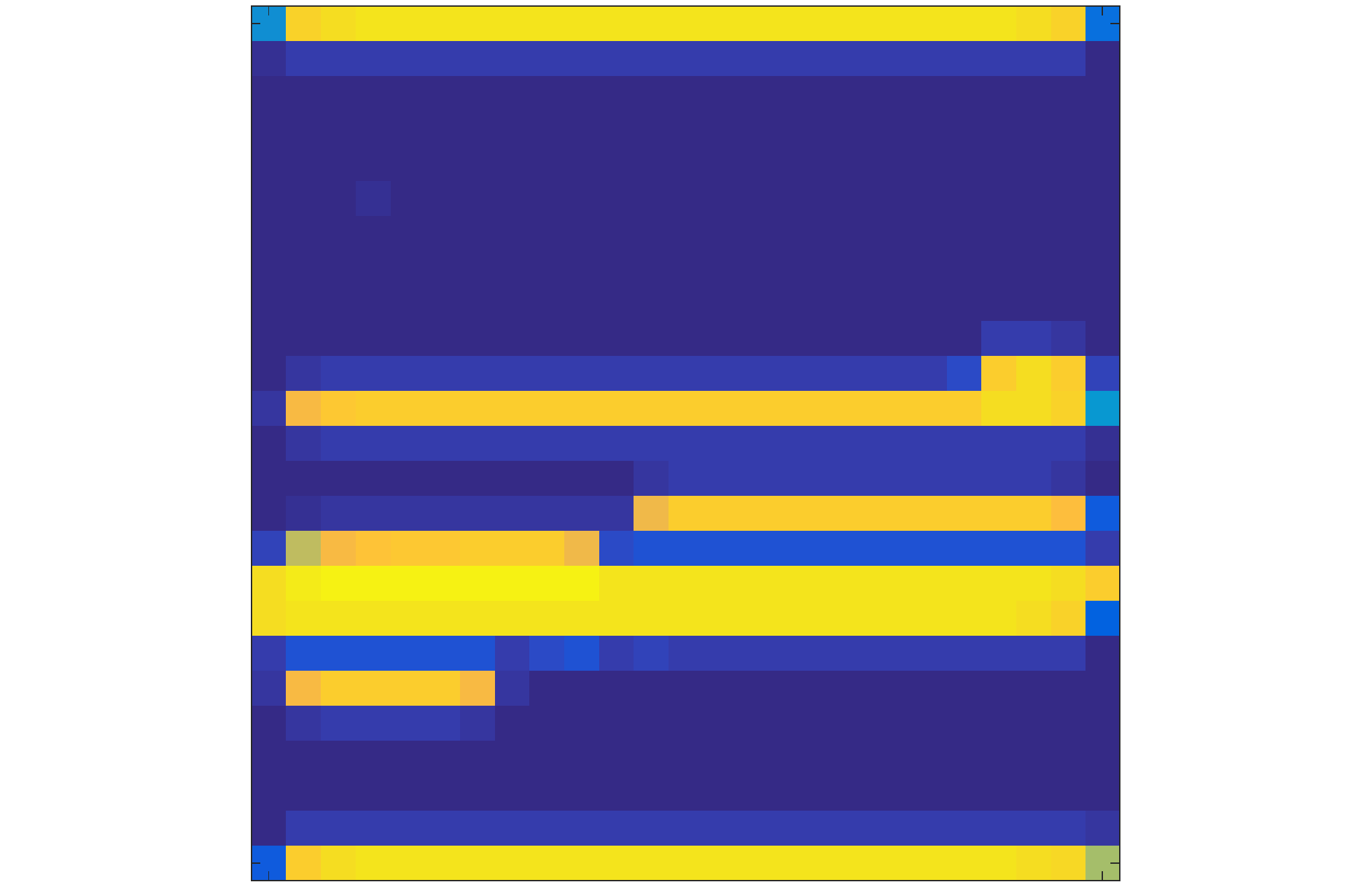}}&
\raisebox{-.5\height}{\includegraphics[width=.3\textwidth]{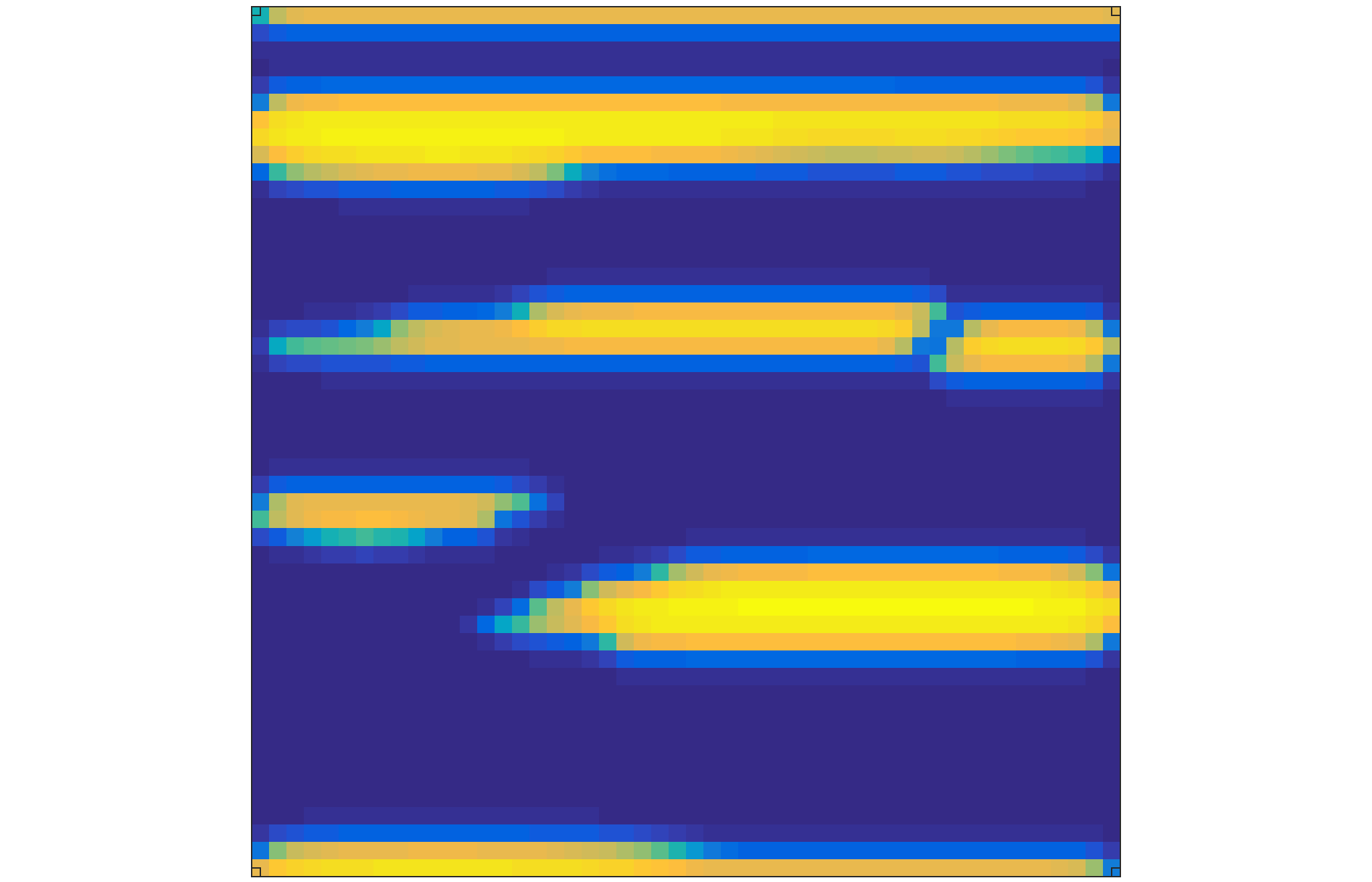}}&
\raisebox{-.5\height}{\includegraphics[width=.3\textwidth]{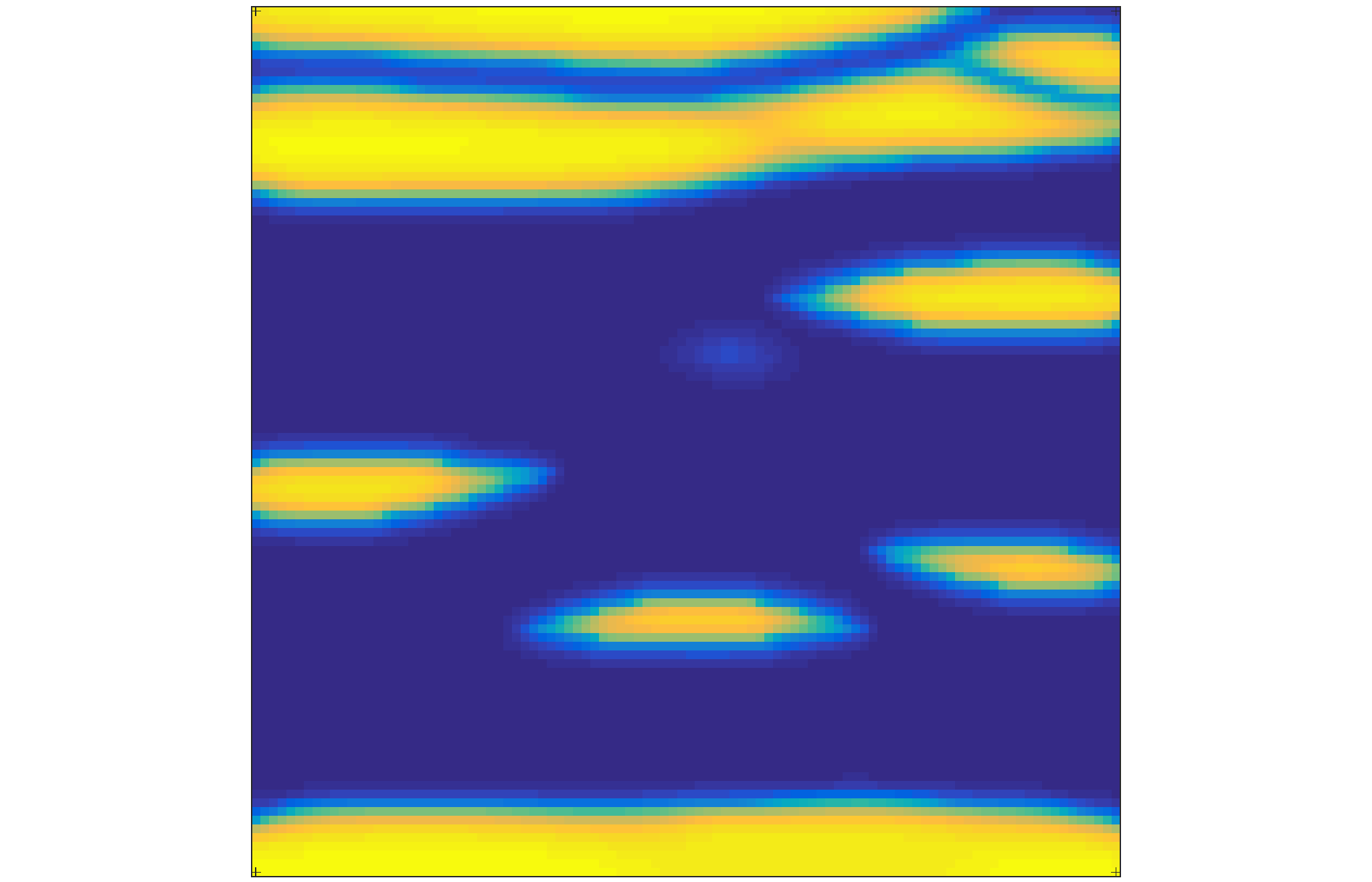}}\vspace{1mm}\\

$t=15$ & \raisebox{-.5\height}{\includegraphics[width=.3\textwidth]{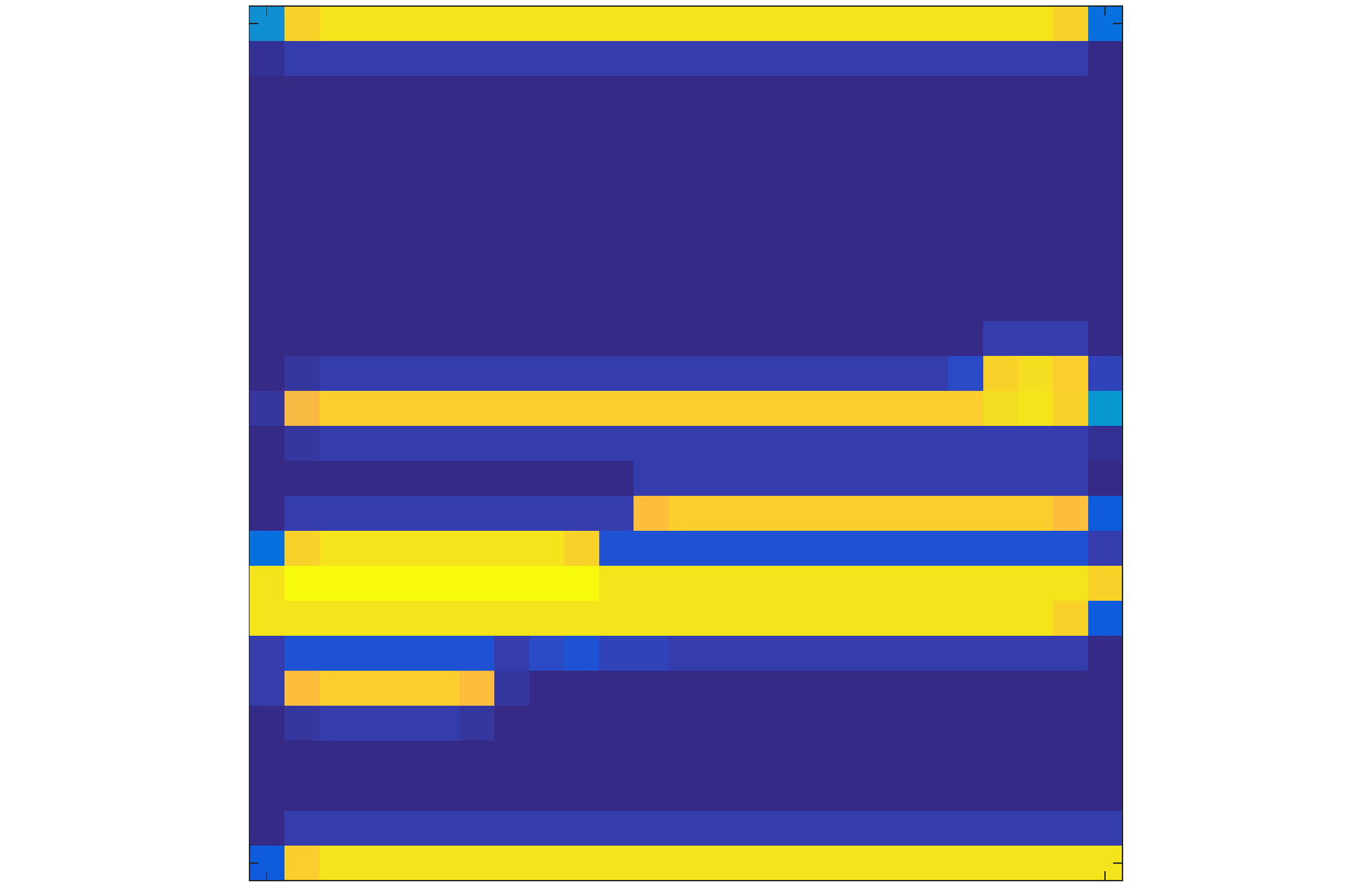}}&
\raisebox{-.5\height}{\includegraphics[width=.3\textwidth]{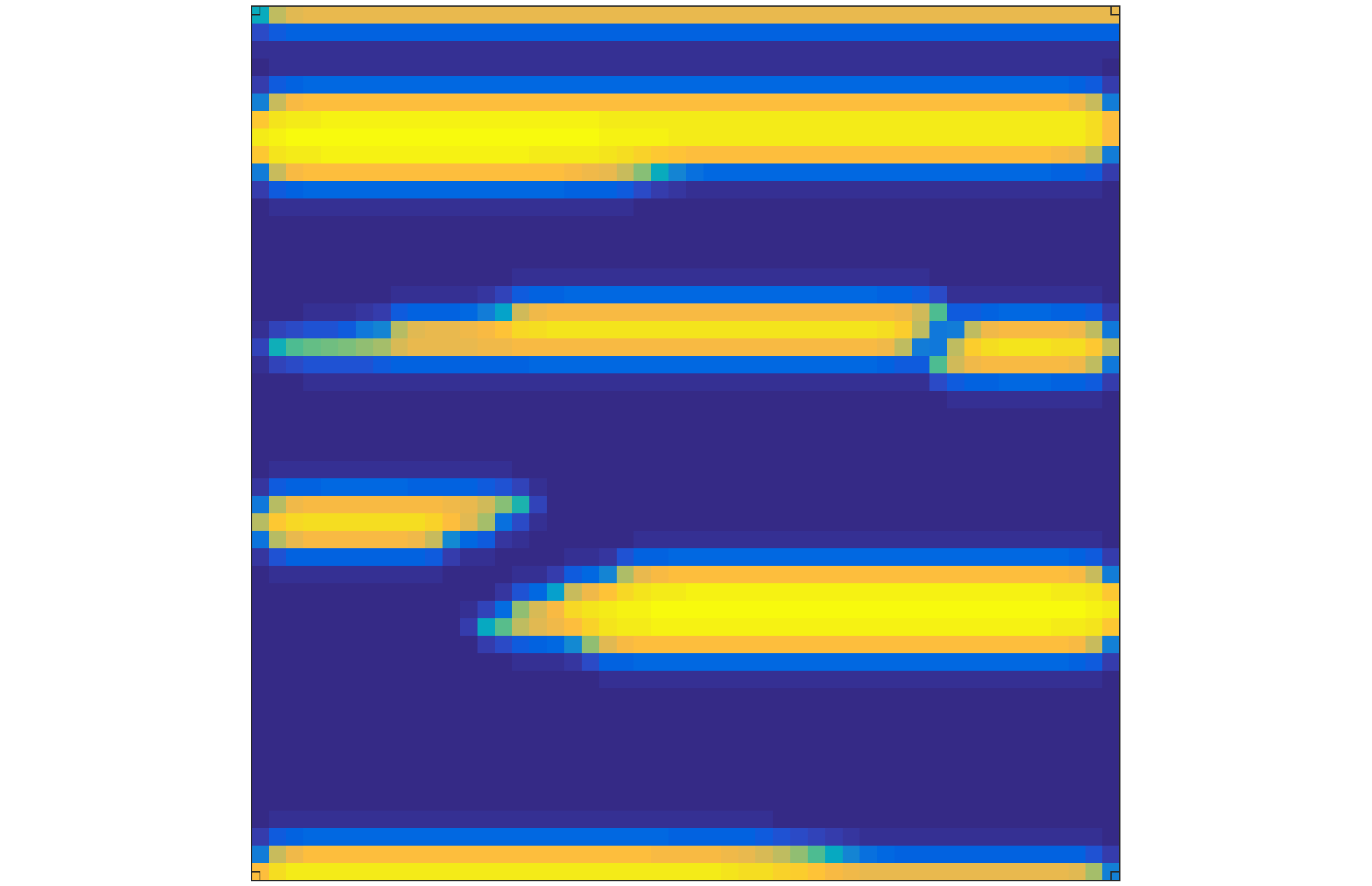}}&
\raisebox{-.5\height}{\includegraphics[width=.3\textwidth]{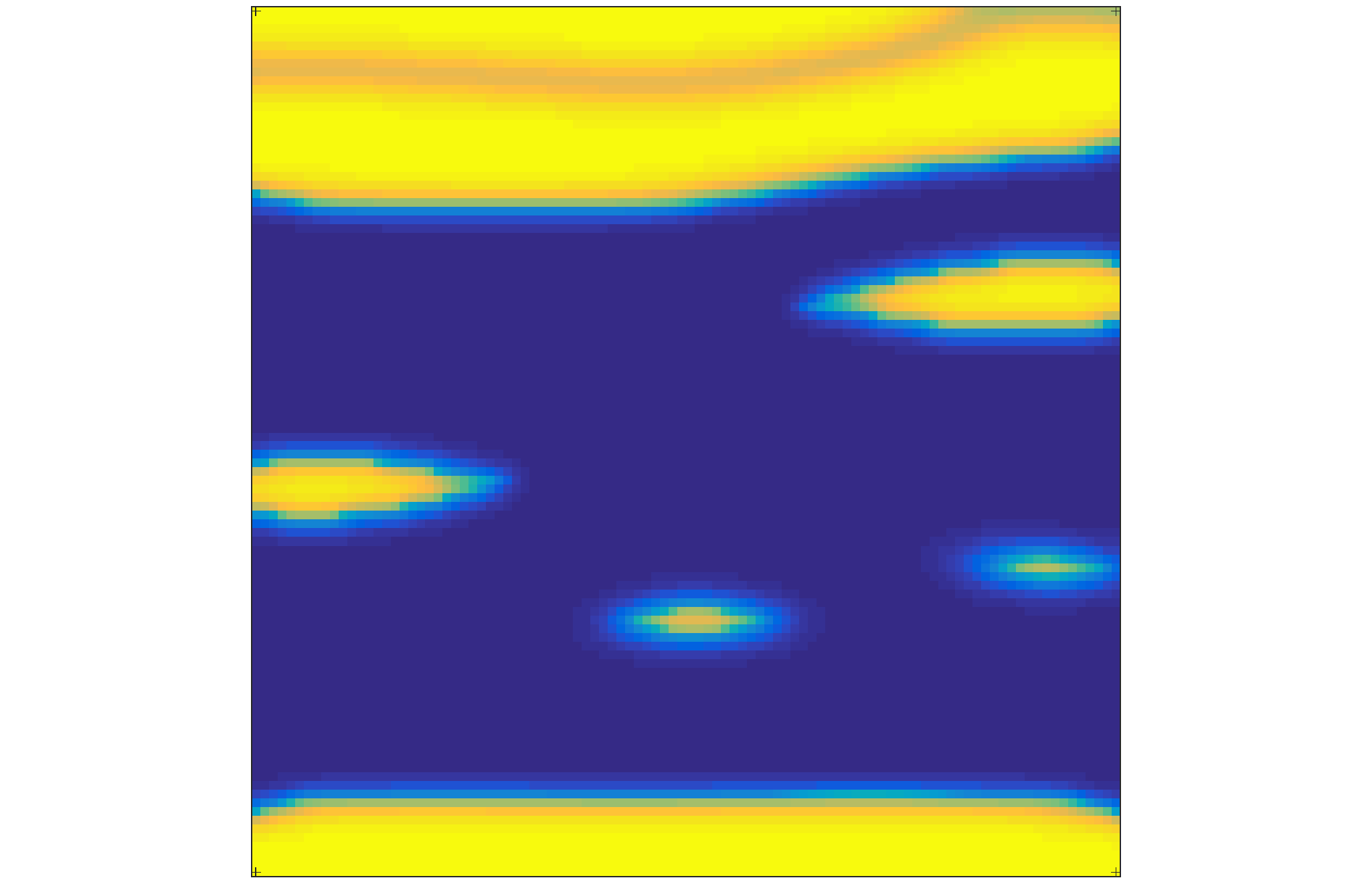}}\vspace{1mm}\\

$t=30$ &\raisebox{-.5\height}{\includegraphics[width=.3\textwidth]{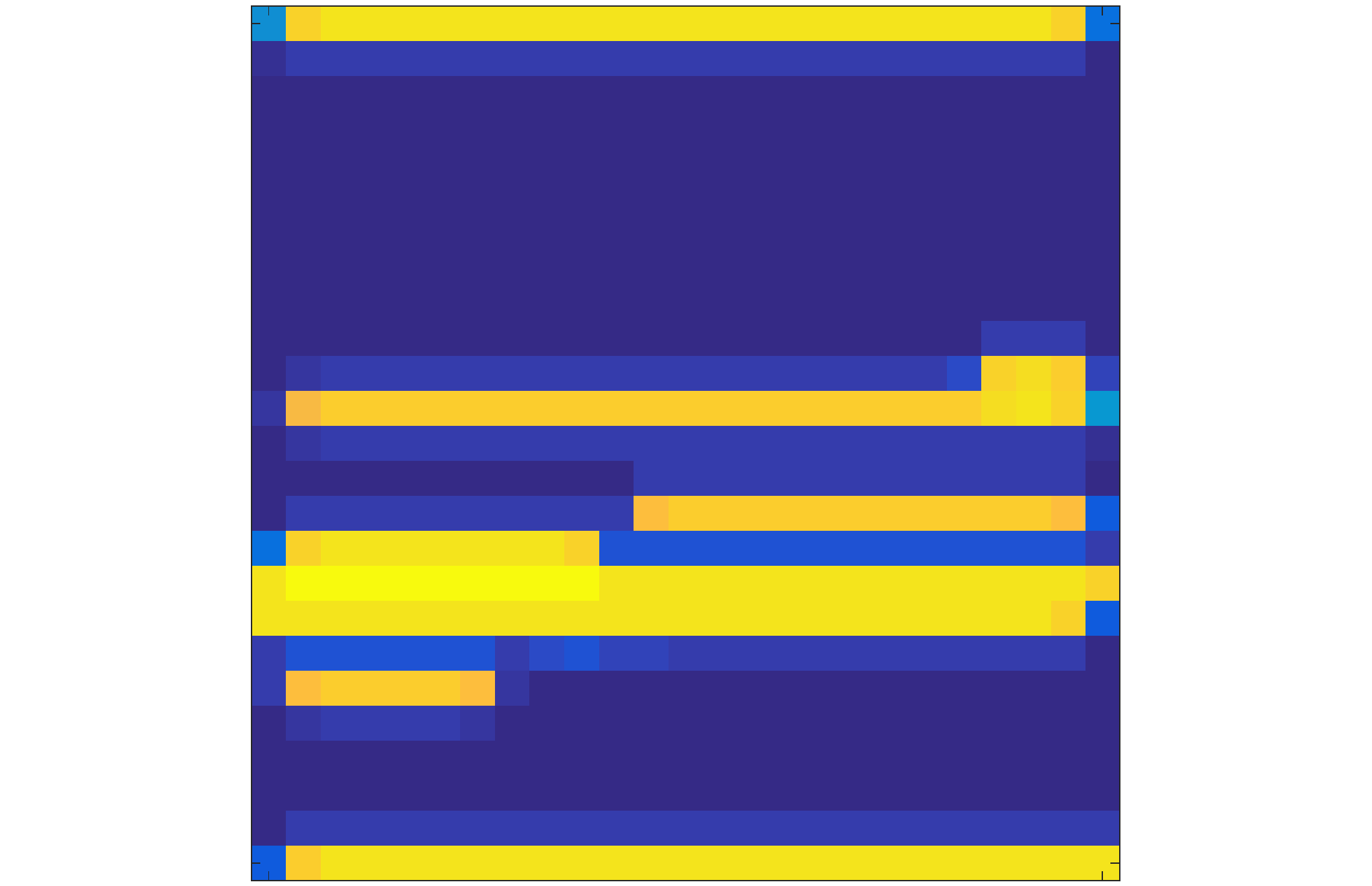}}&
\raisebox{-.5\height}{\includegraphics[width=.3\textwidth]{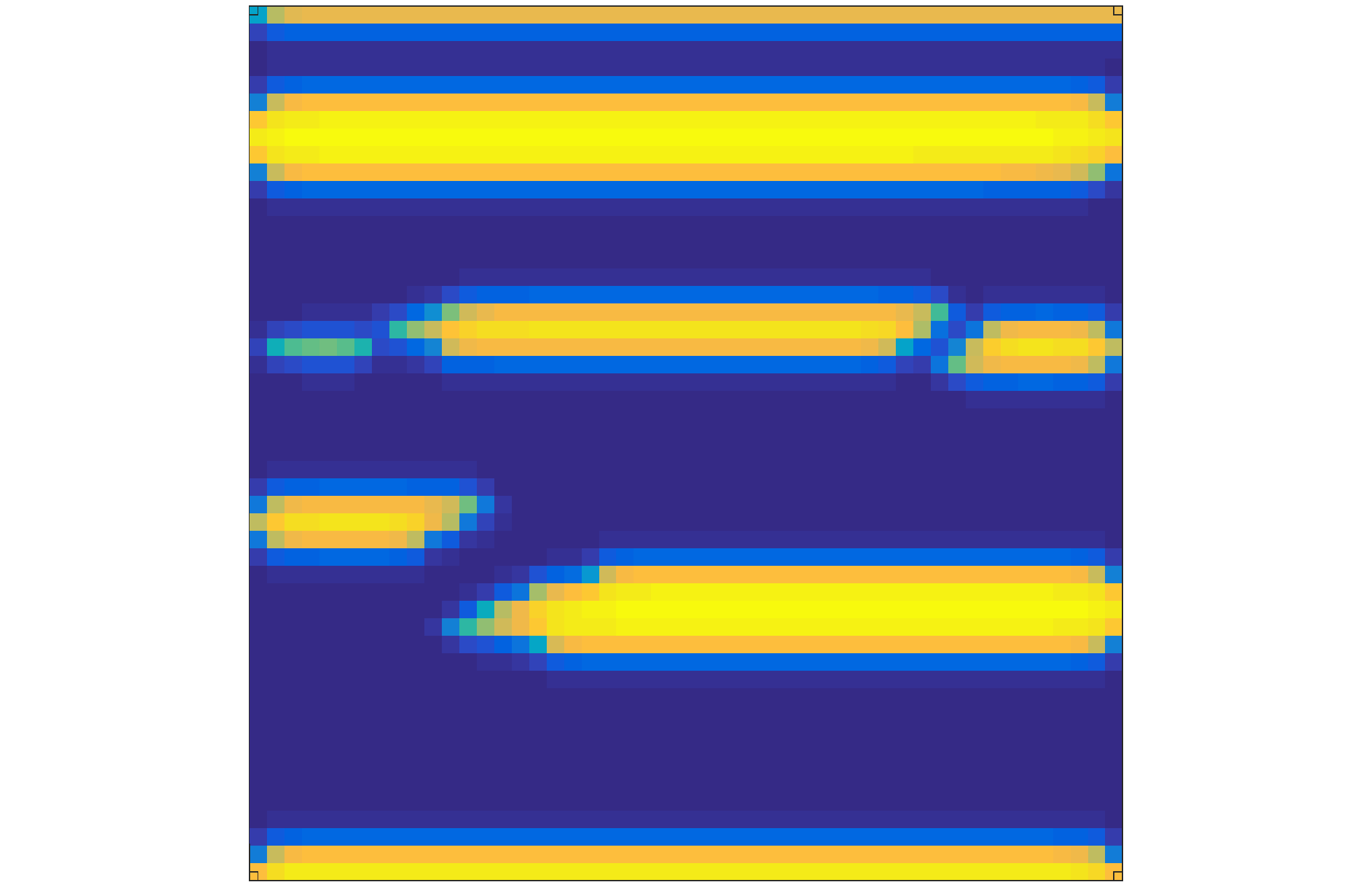}}&
\raisebox{-.5\height}{\includegraphics[width=.3\textwidth]{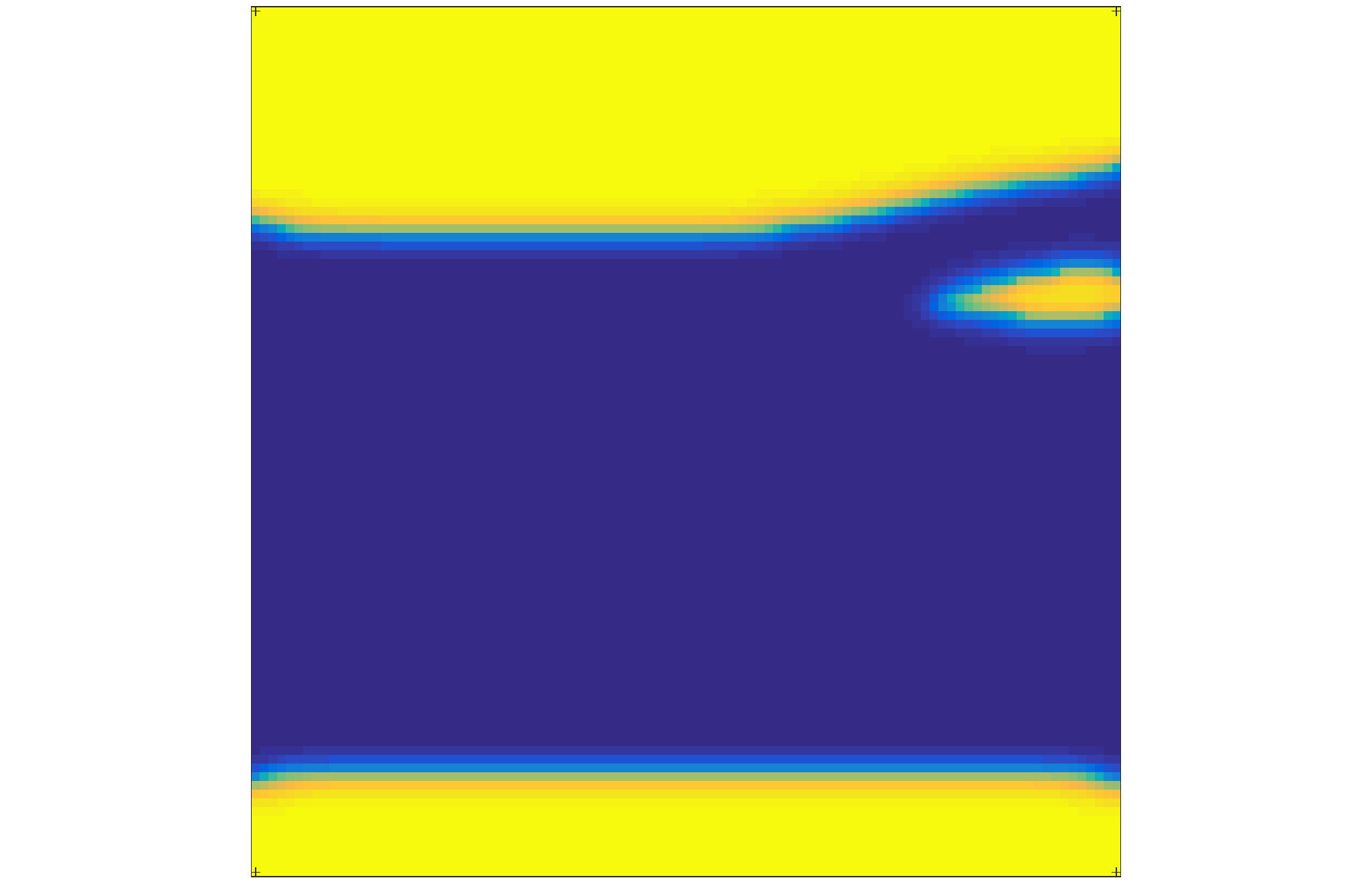}}\vspace{1mm}\\
\end{tabular}
\includegraphics[clip, trim={0 23cm 0 0}, width=.9\textwidth]{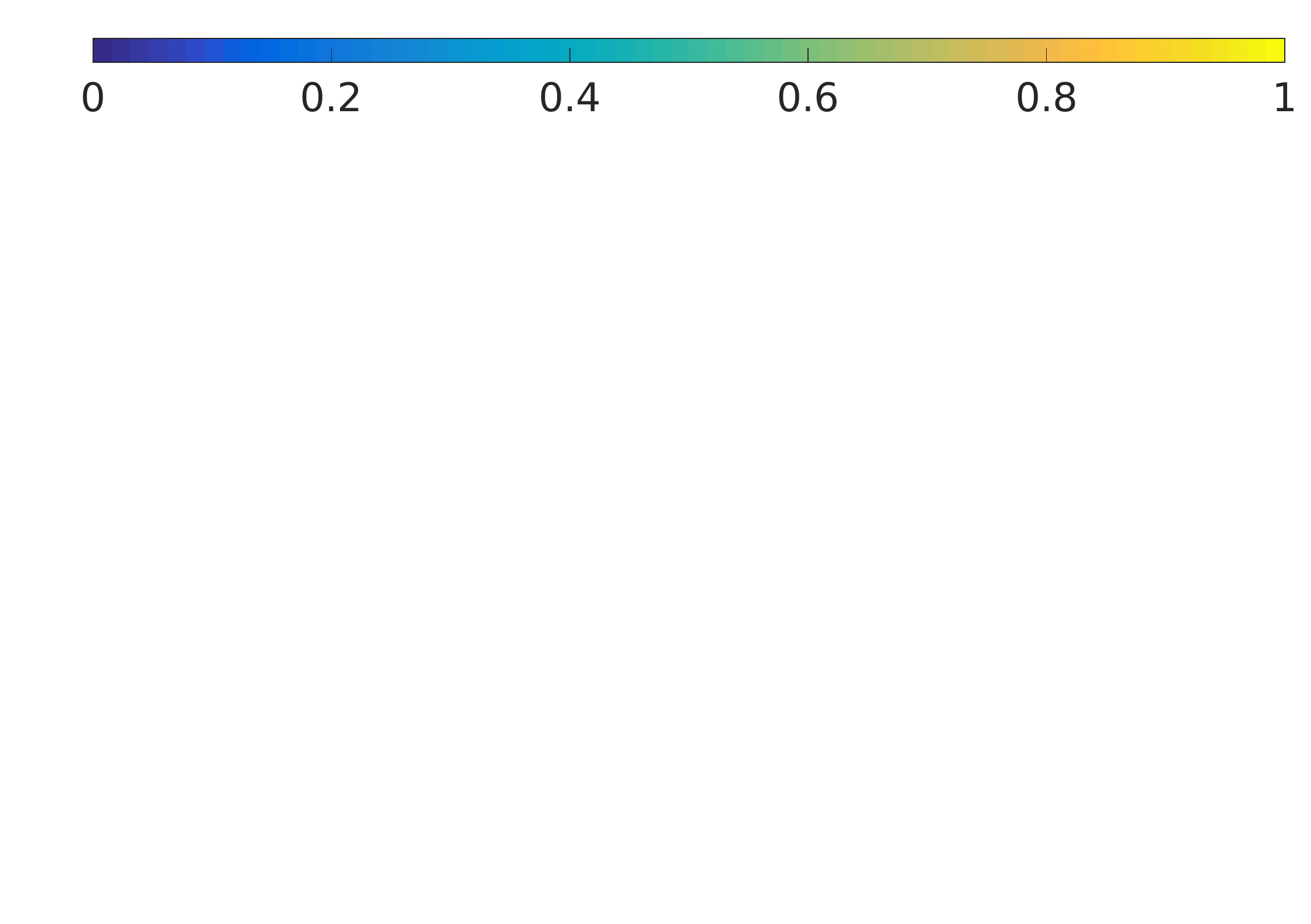}%
\caption{Cell density plots with $\delta=10^{-4}$, $\sigma_l=2.5$ for three lattice sizes at different points in time for one realization of the initial cell density.}
\label{sigma_times2.5}
\end{figure}
\begin{figure}\setlength{\tabcolsep}{0pt}
\centering
\begin{tabular}{rc c c} 
 & $n=25$ & $n=50$ & $n=100$\\

$t=2$ &\raisebox{-.5\height}{\includegraphics[width=.3\textwidth]{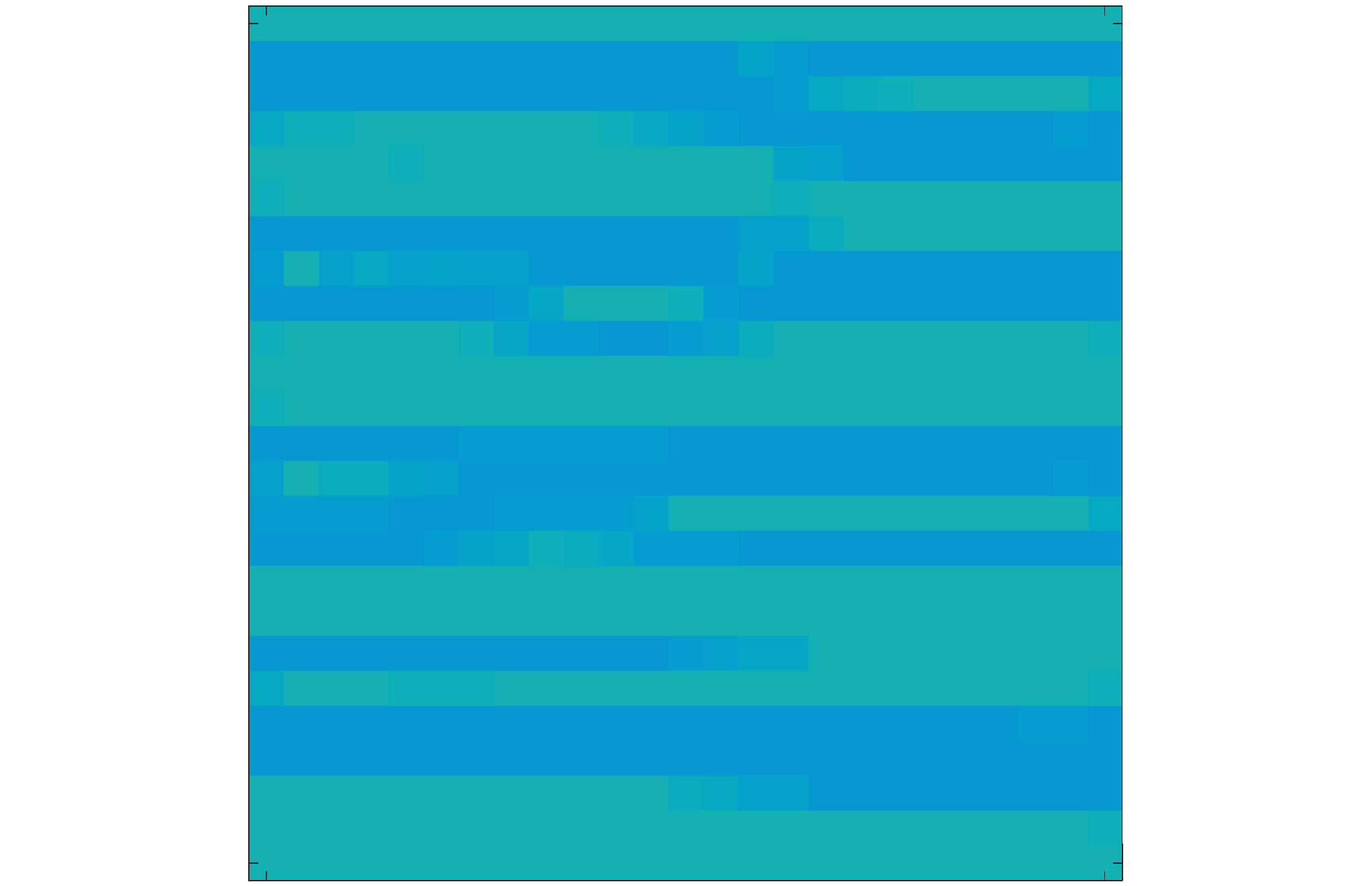}}&
\raisebox{-.5\height}{\includegraphics[width=.3\textwidth]{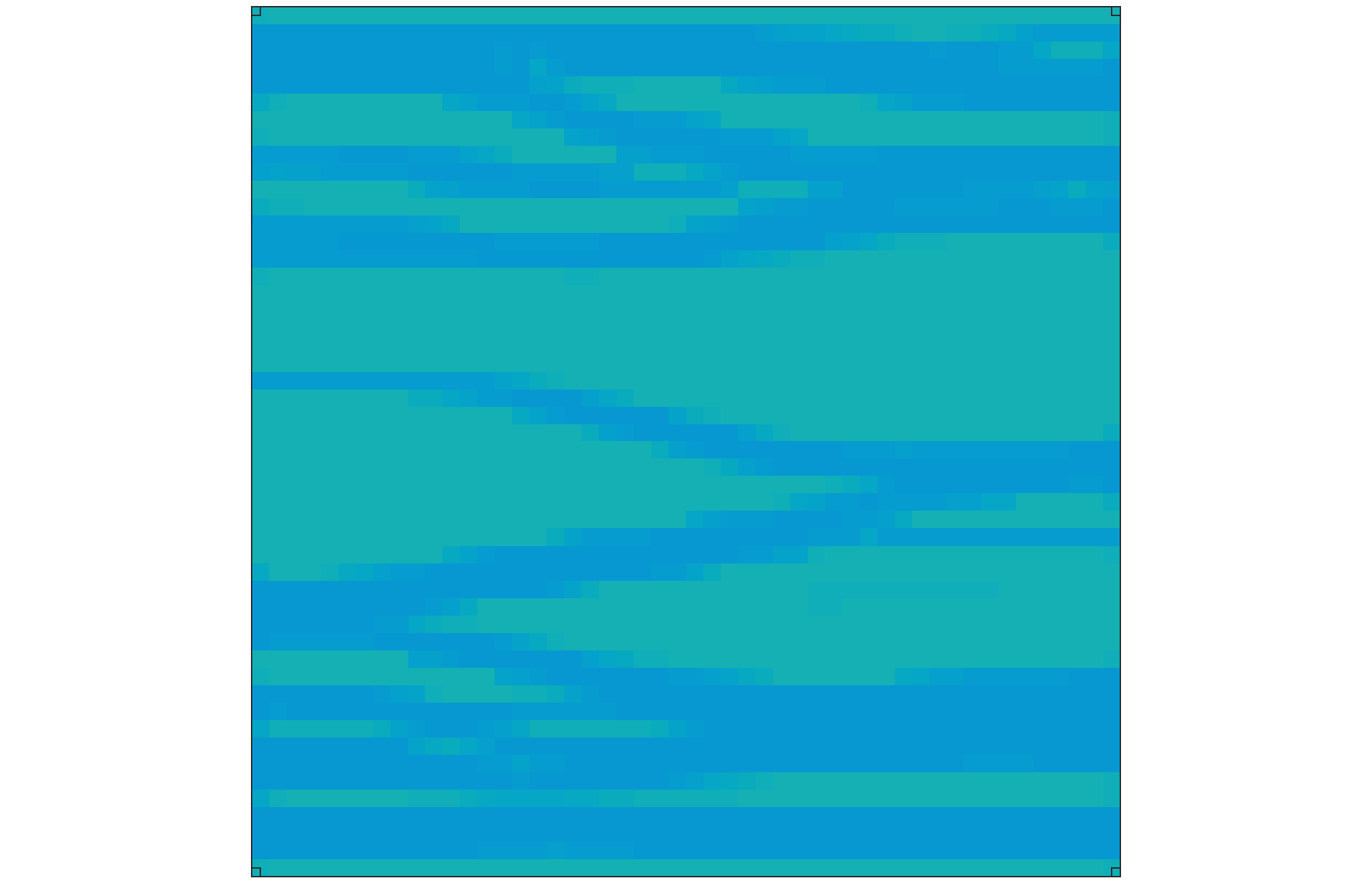}}&
\raisebox{-.5\height}{\includegraphics[width=.3\textwidth]{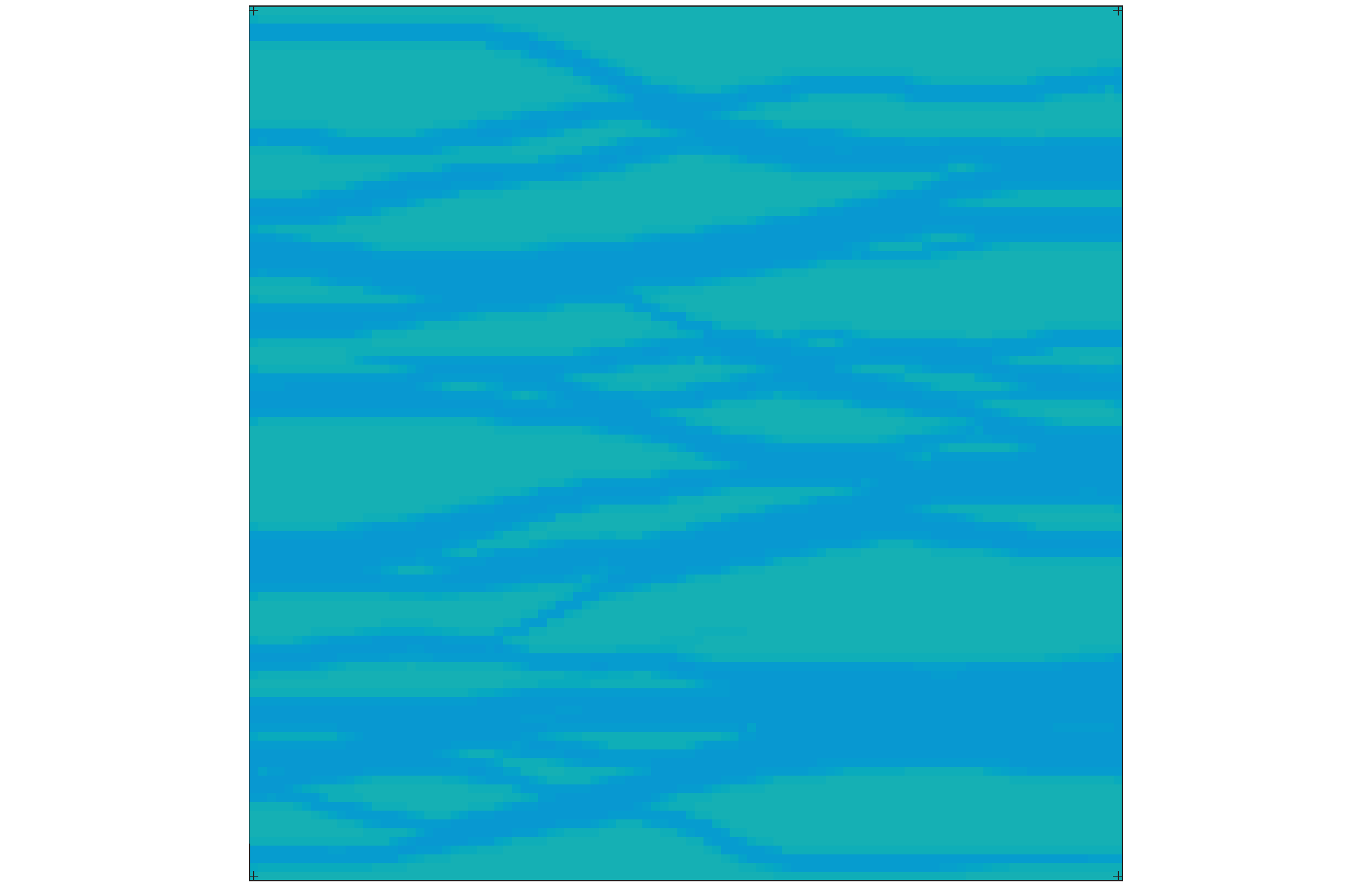}}\vspace{1mm}\\

$t=5$ &\raisebox{-.5\height}{\includegraphics[width=.3\textwidth]{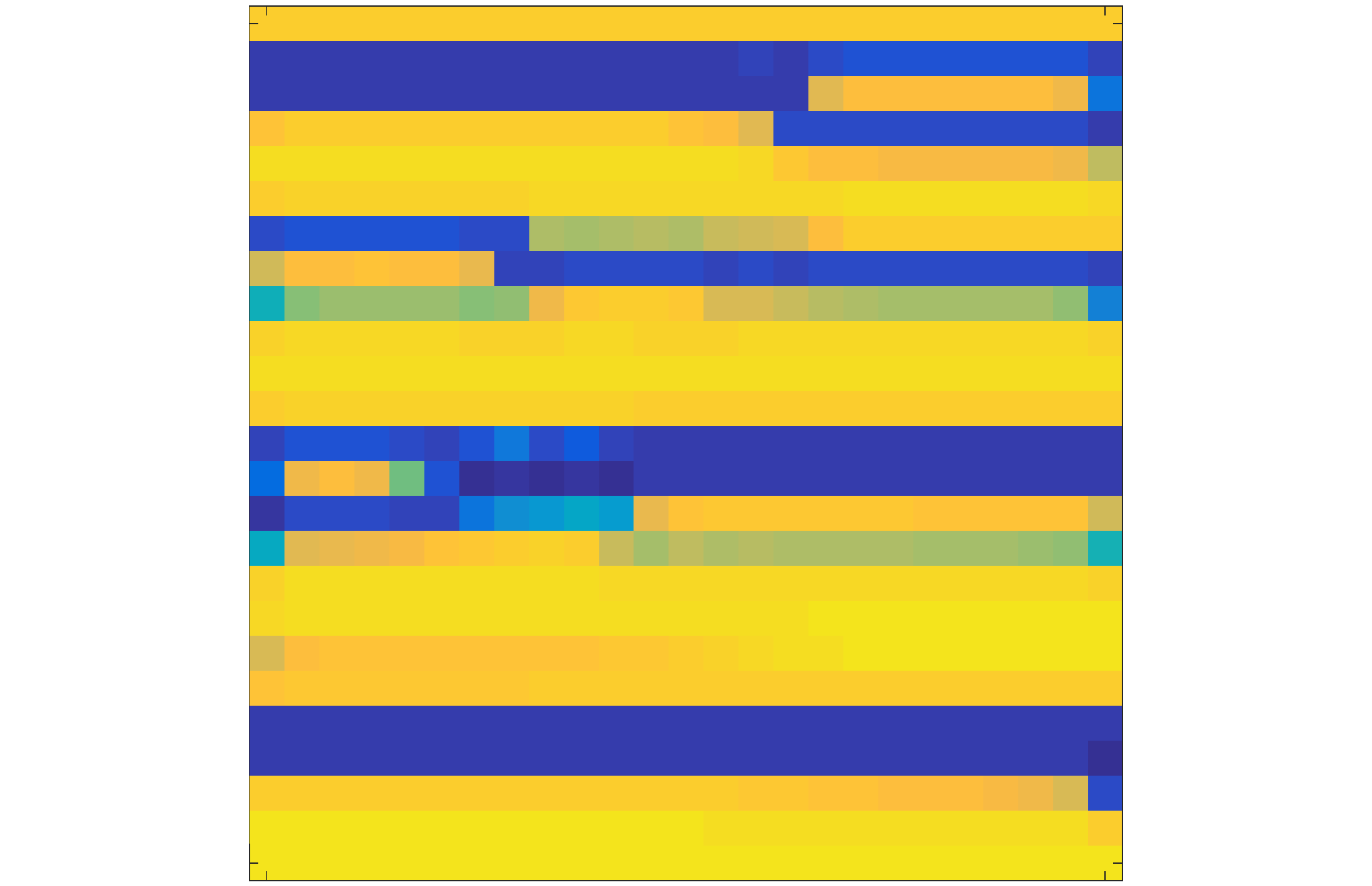}}&
\raisebox{-.5\height}{\includegraphics[width=.3\textwidth]{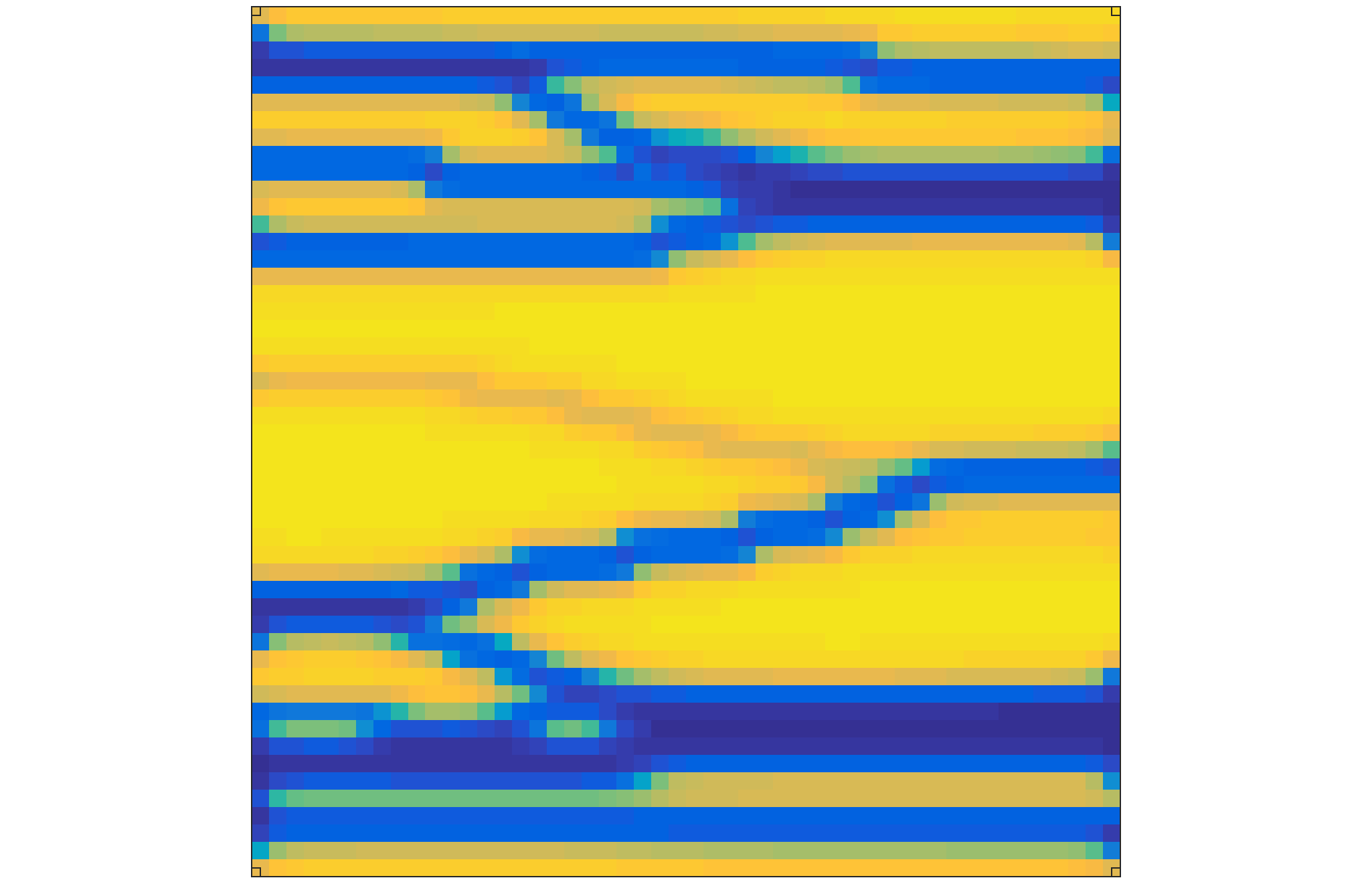}}&
\raisebox{-.5\height}{\includegraphics[width=.3\textwidth]{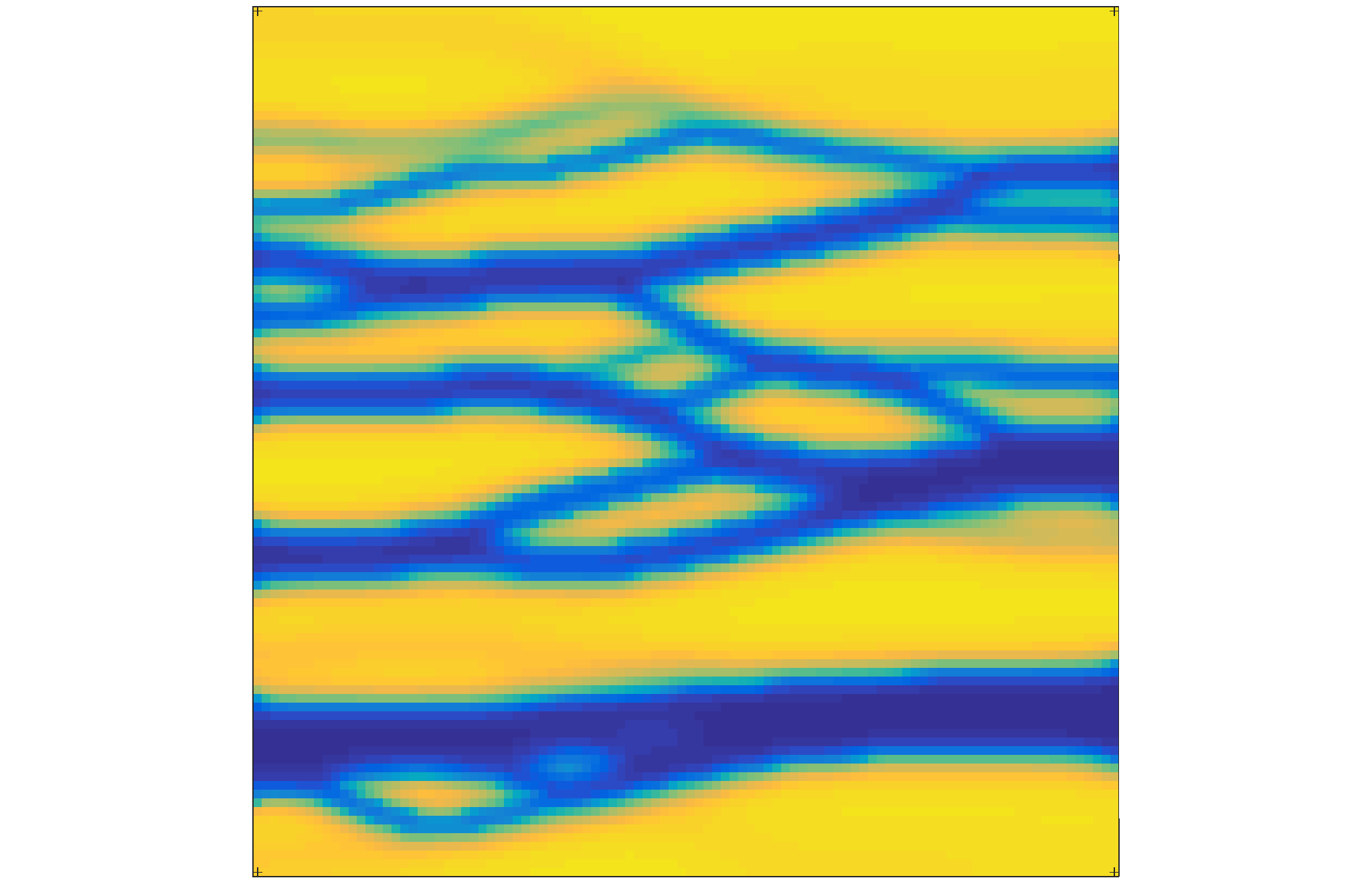}}\vspace{1mm}\\

$t=10$ &\raisebox{-.5\height}{\includegraphics[width=.3\textwidth]{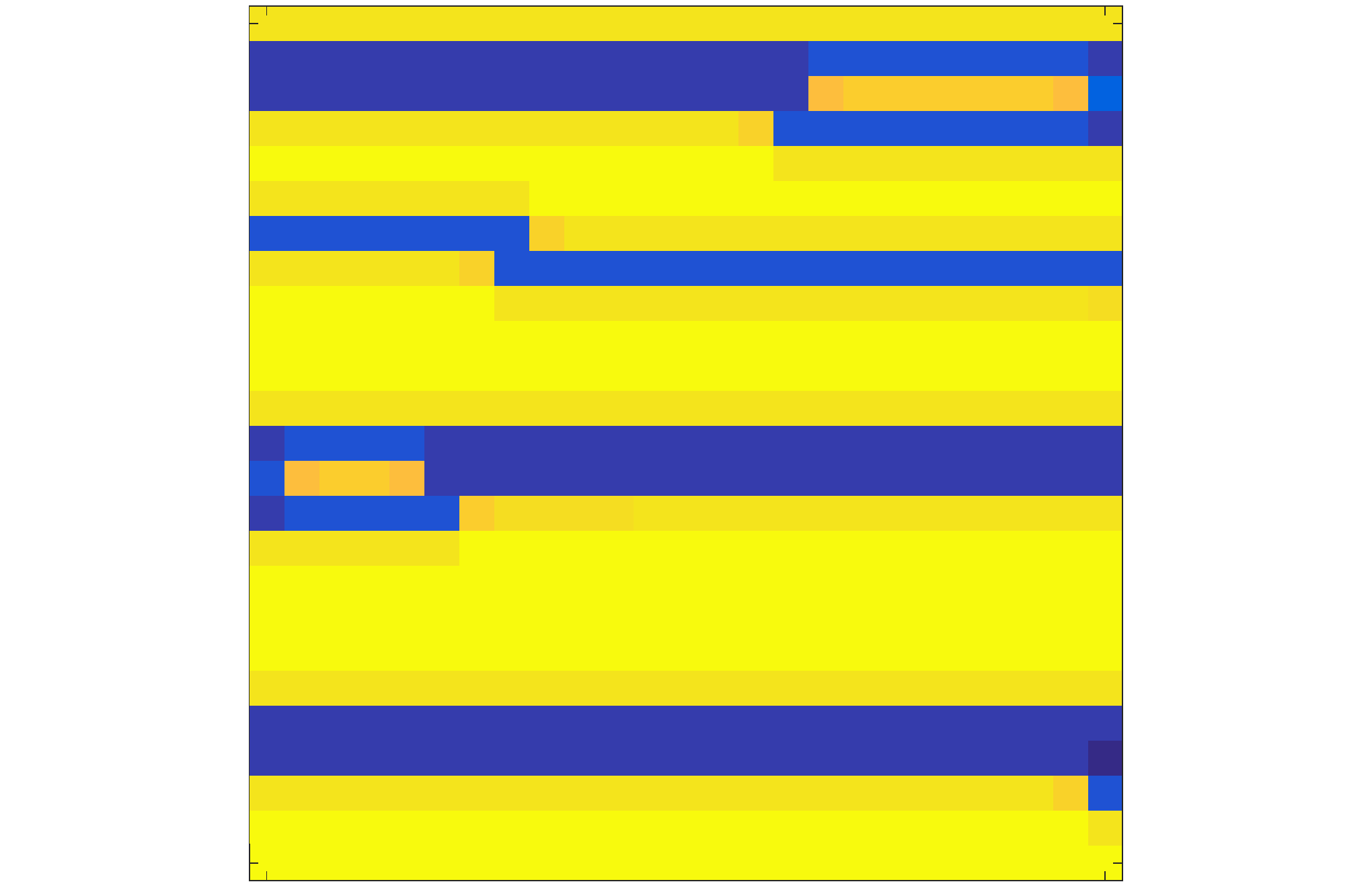}}&
\raisebox{-.5\height}{\includegraphics[width=.3\textwidth]{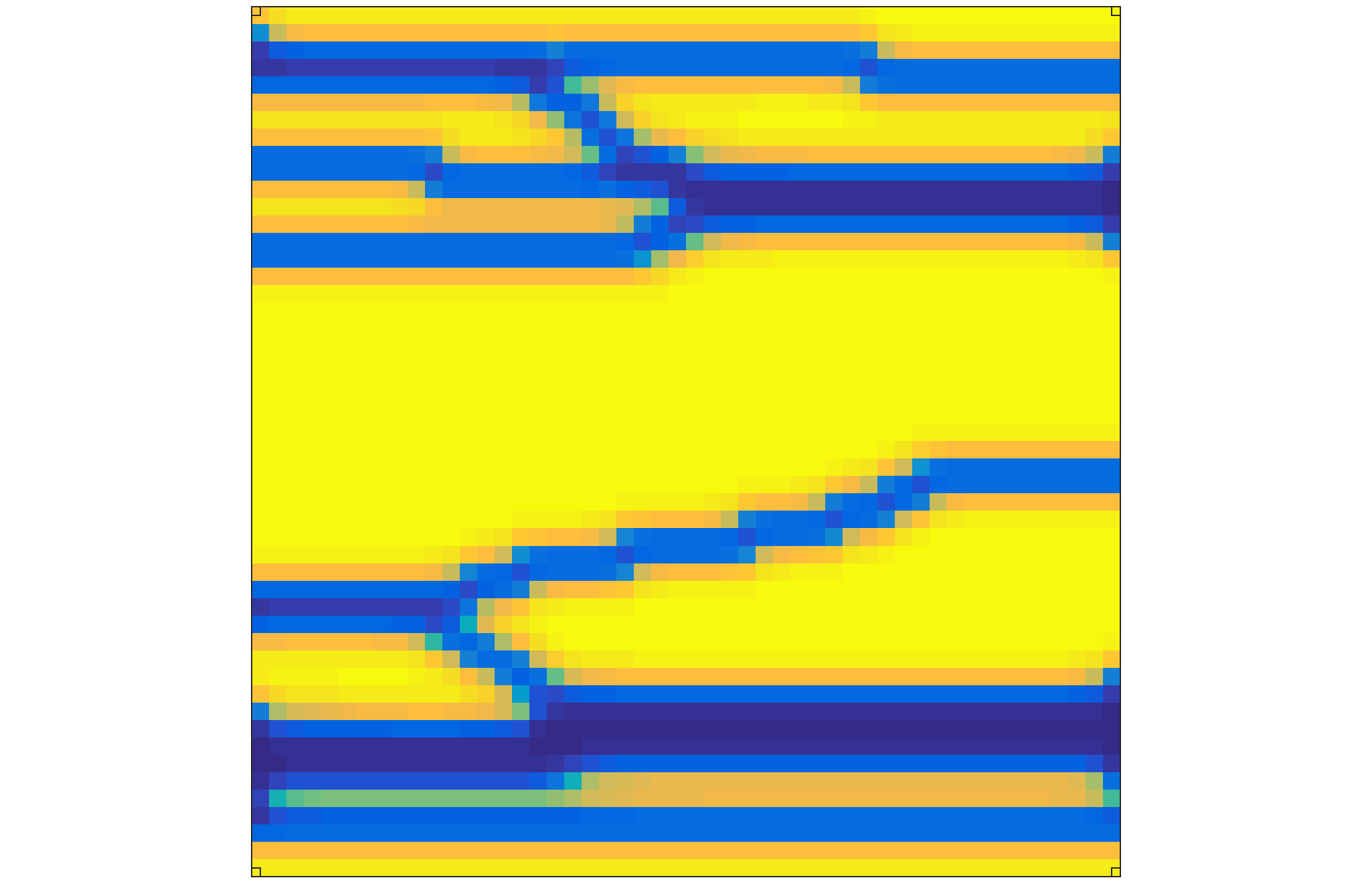}}&
\raisebox{-.5\height}{\includegraphics[width=.3\textwidth]{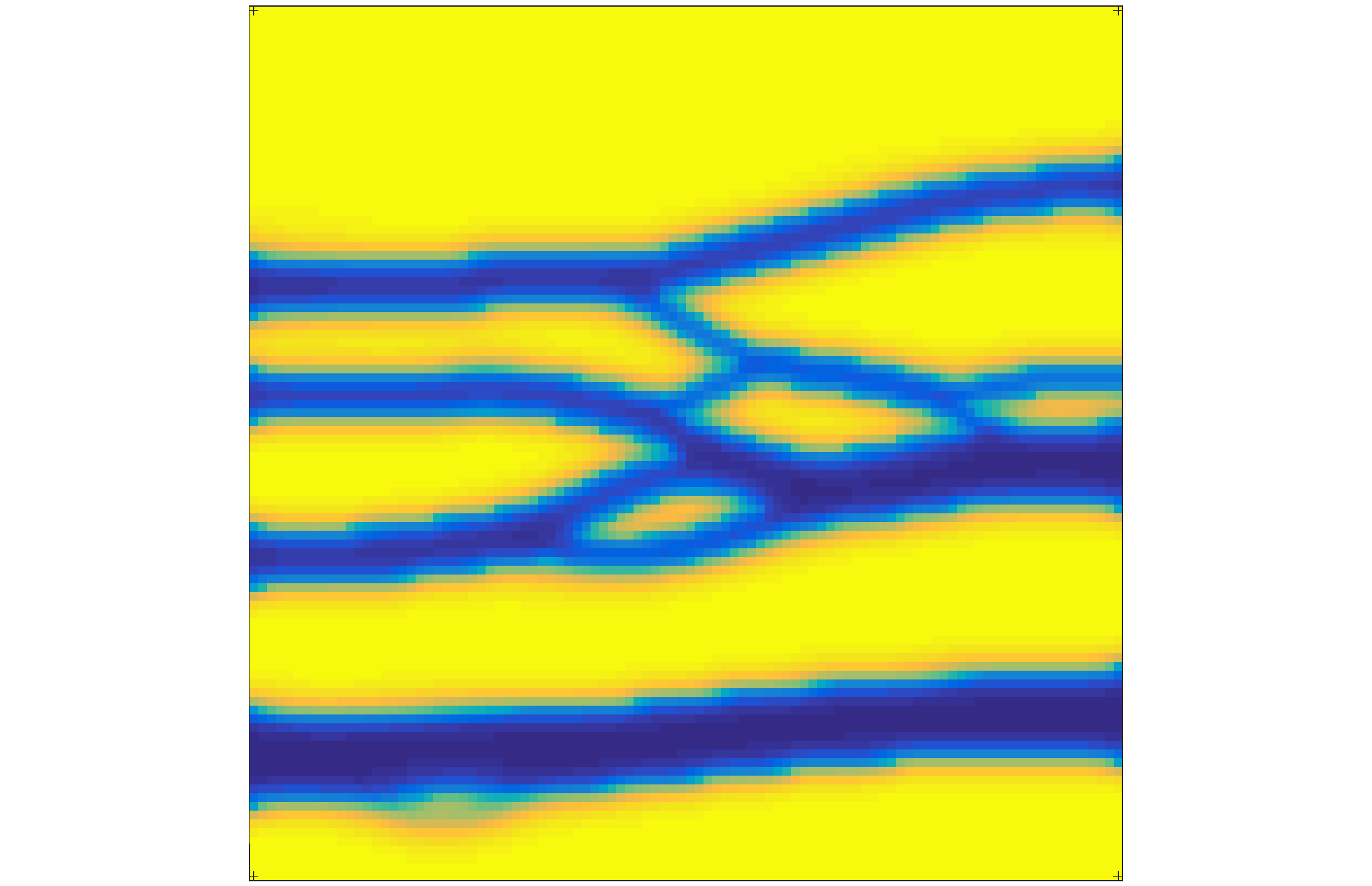}}\vspace{1mm}\\

$t=15$ & \raisebox{-.5\height}{\includegraphics[width=.3\textwidth]{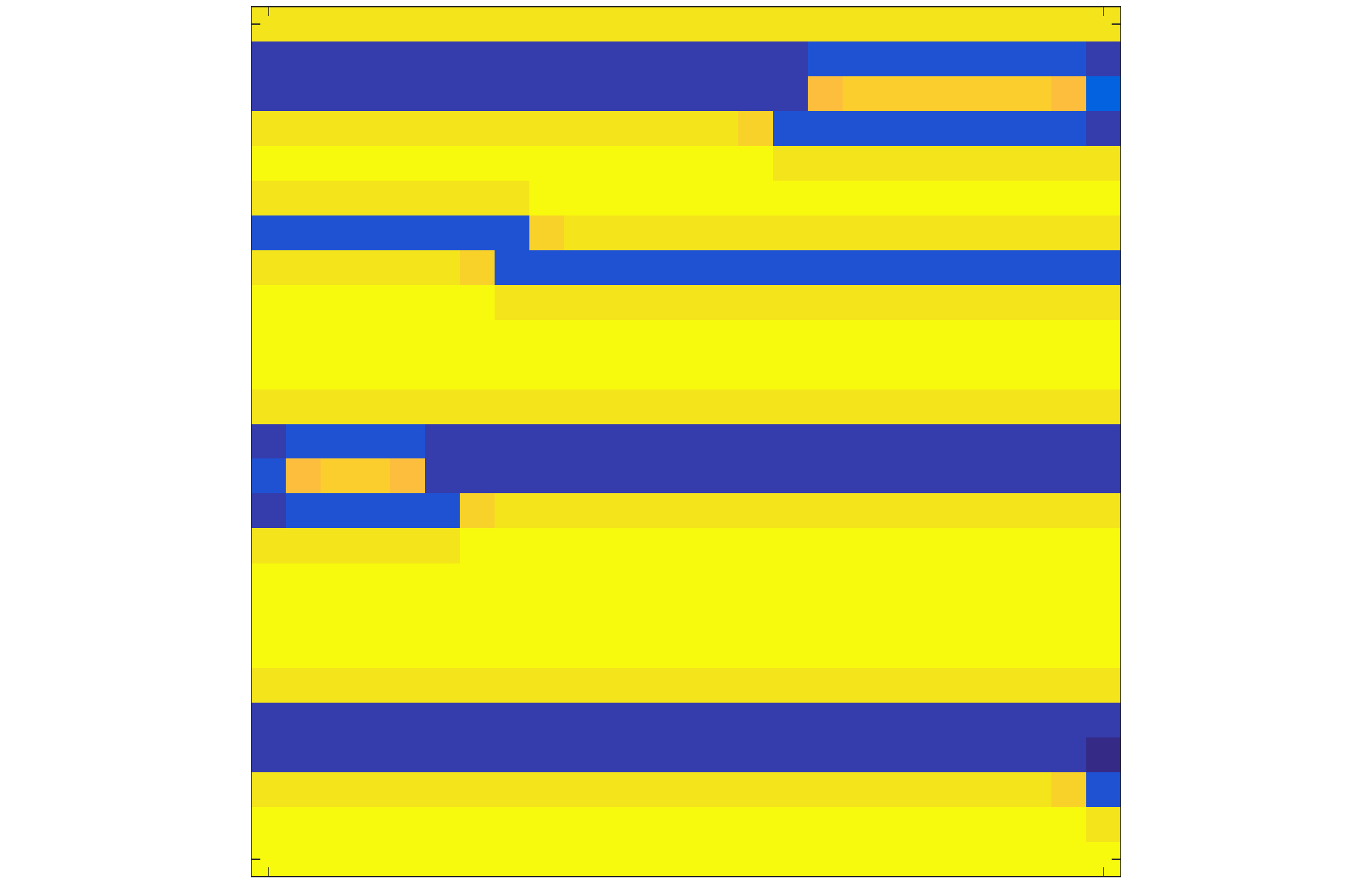}}&
\raisebox{-.5\height}{\includegraphics[width=.3\textwidth]{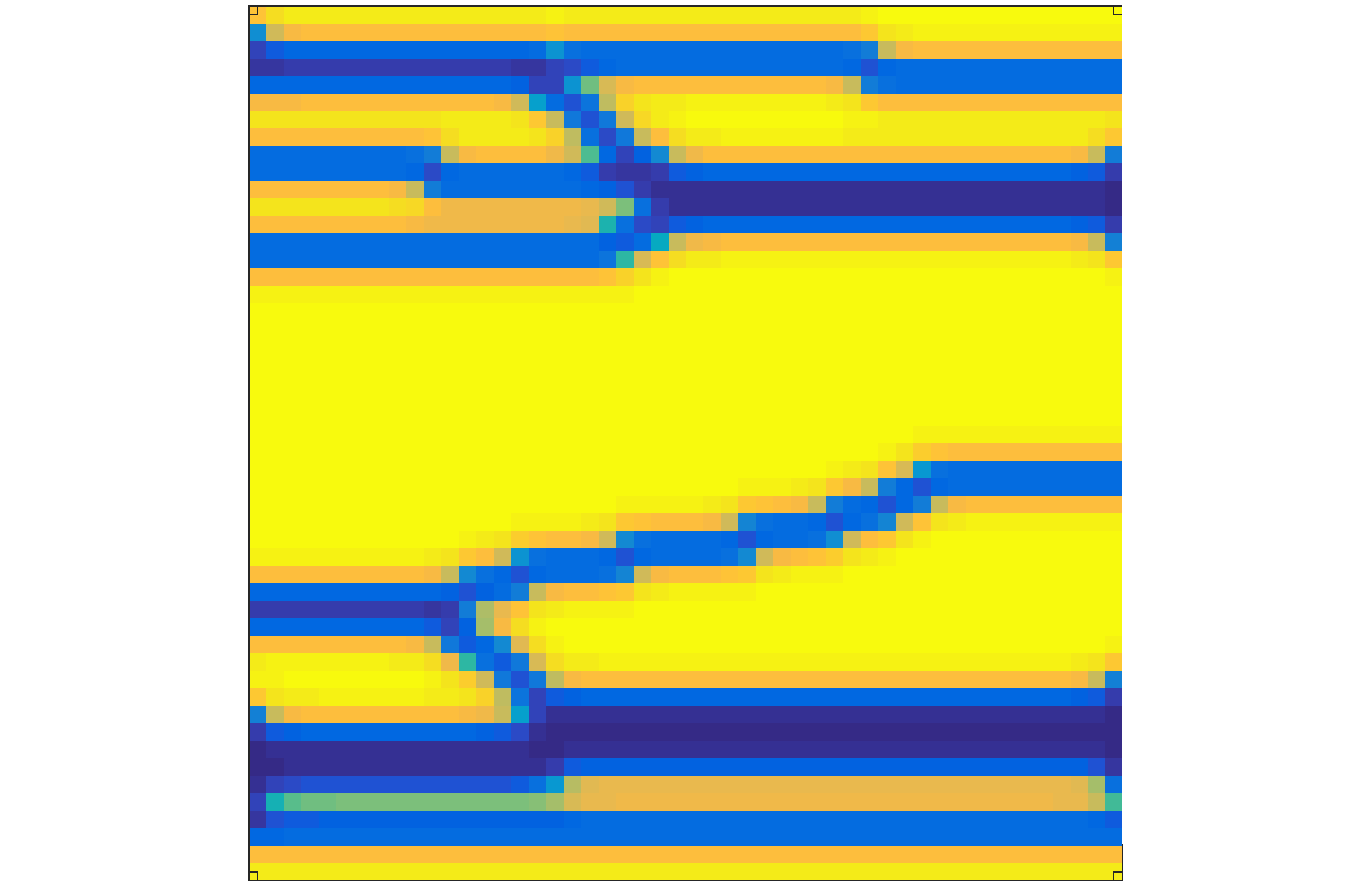}}&
\raisebox{-.5\height}{\includegraphics[width=.3\textwidth]{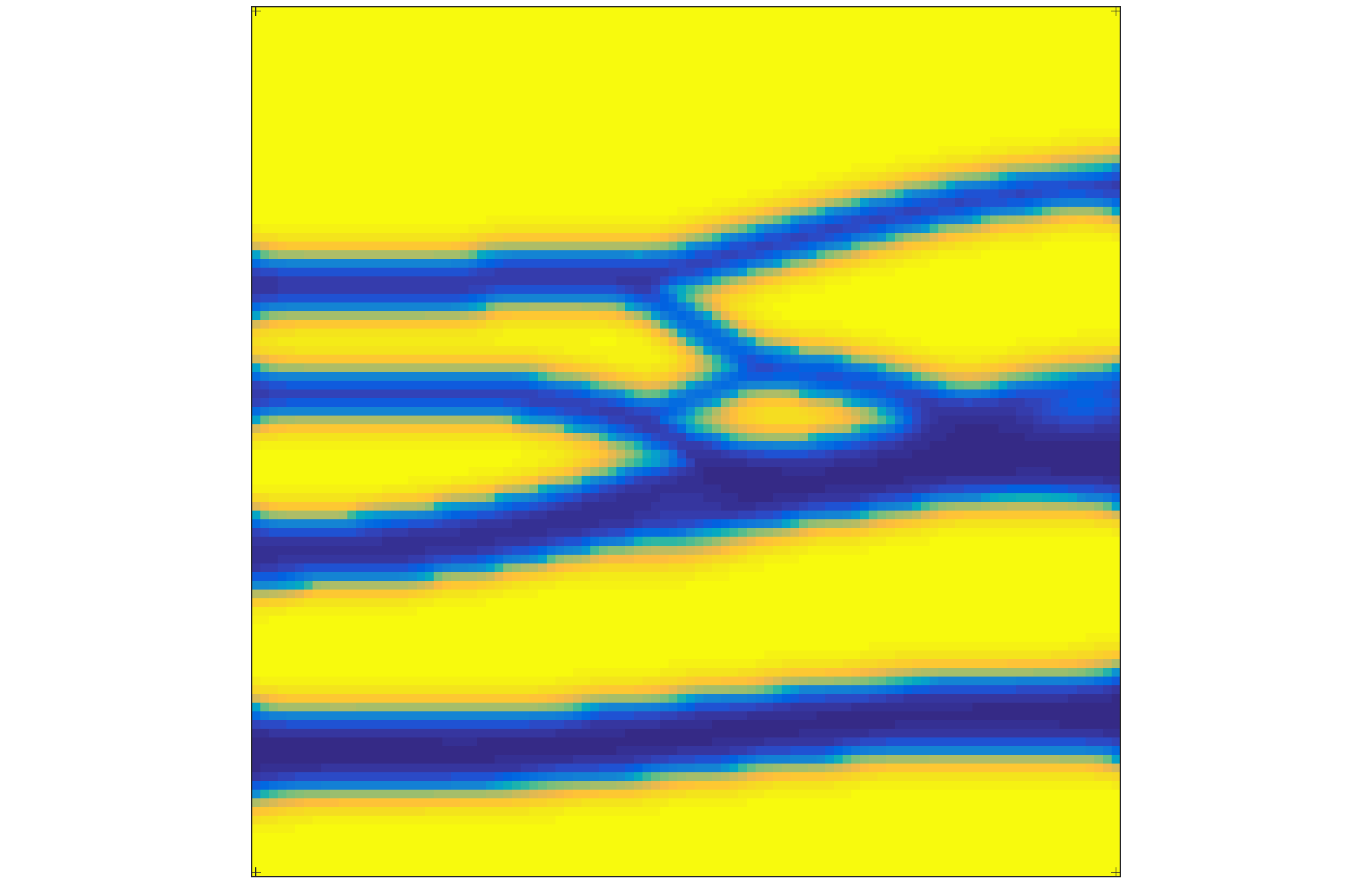}}\vspace{1mm}\\

$t=30$ &\raisebox{-.5\height}{\includegraphics[width=.3\textwidth]{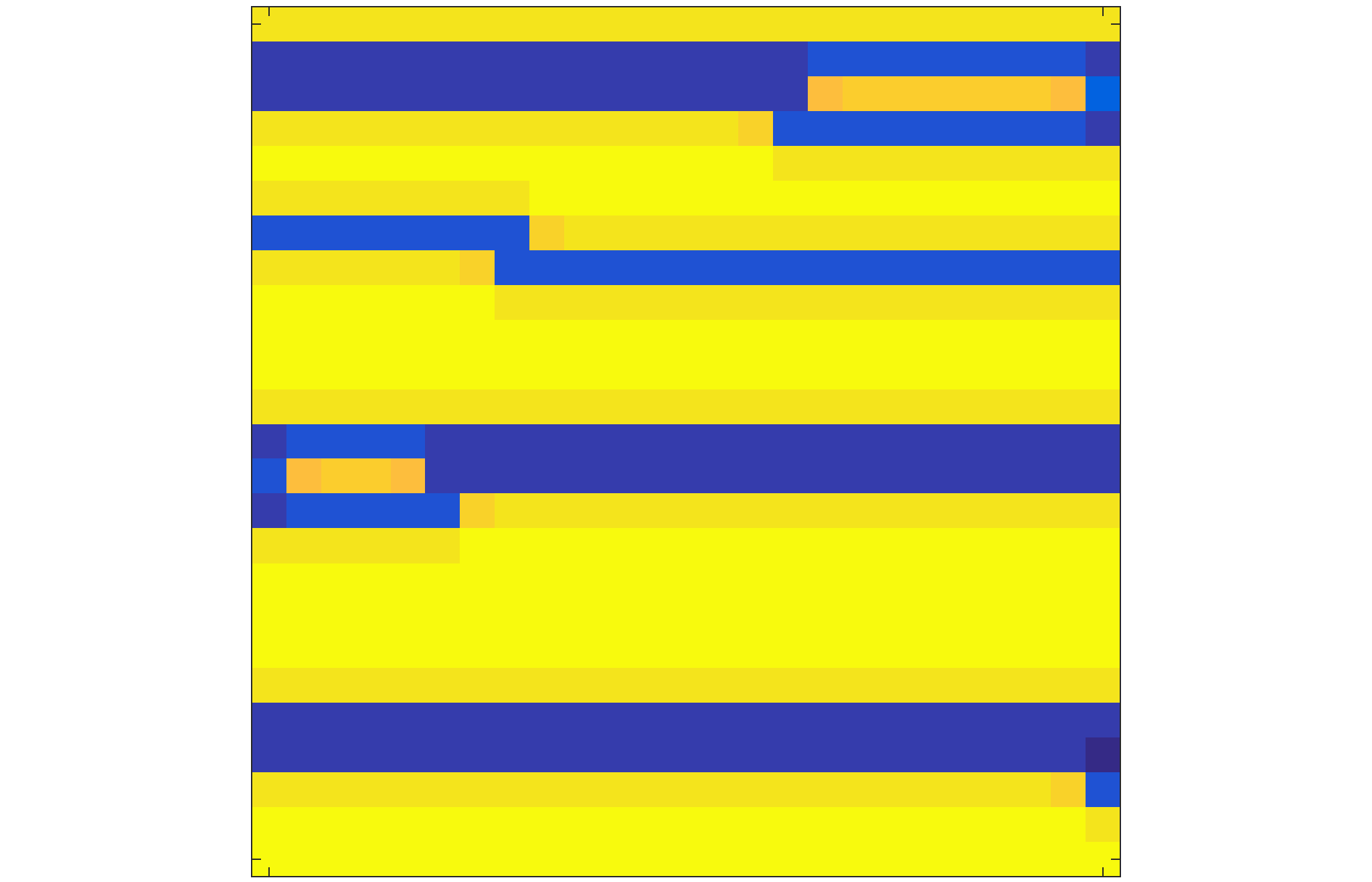}}&
\raisebox{-.5\height}{\includegraphics[width=.3\textwidth]{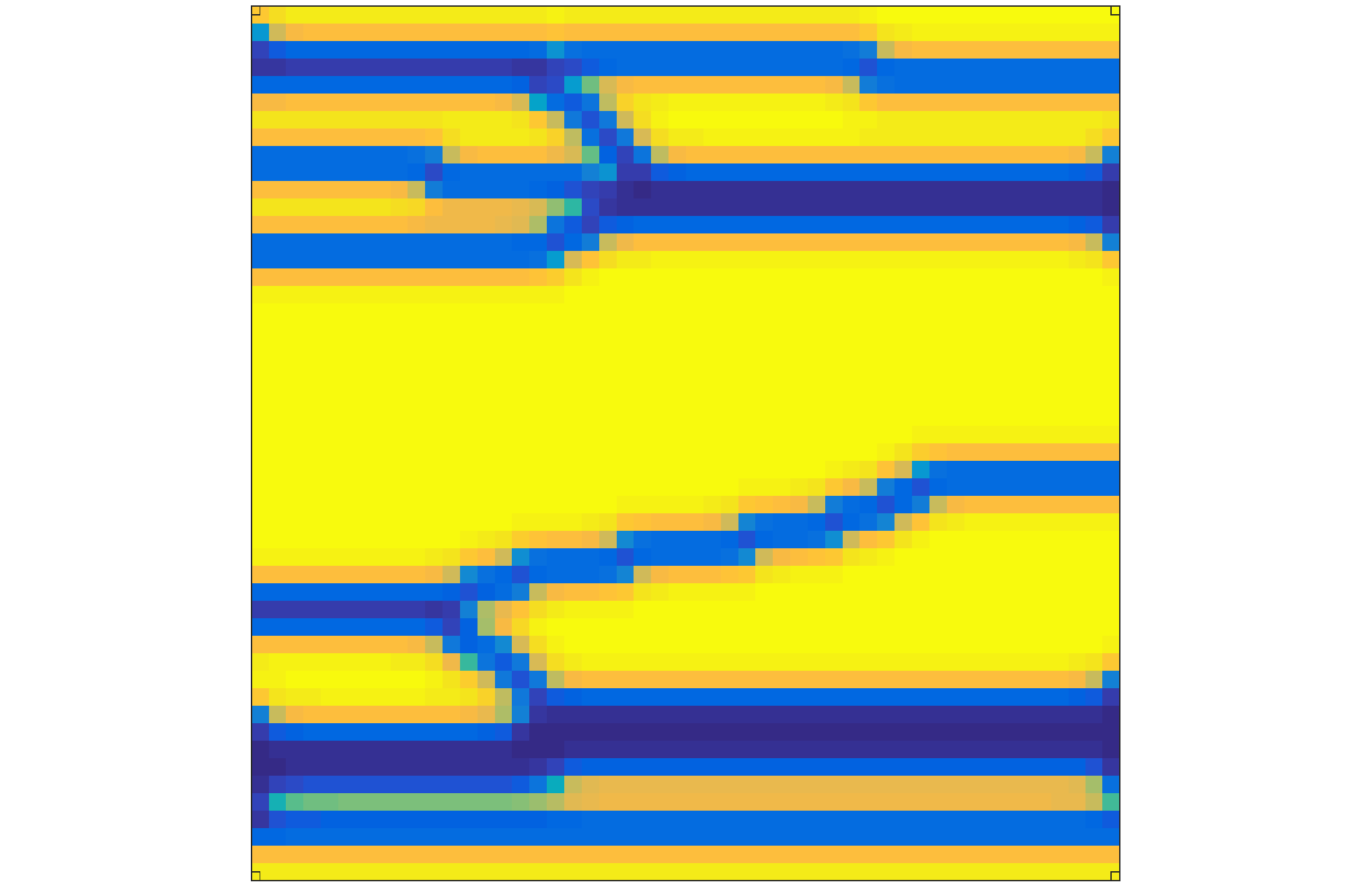}}&
\raisebox{-.5\height}{\includegraphics[width=.3\textwidth]{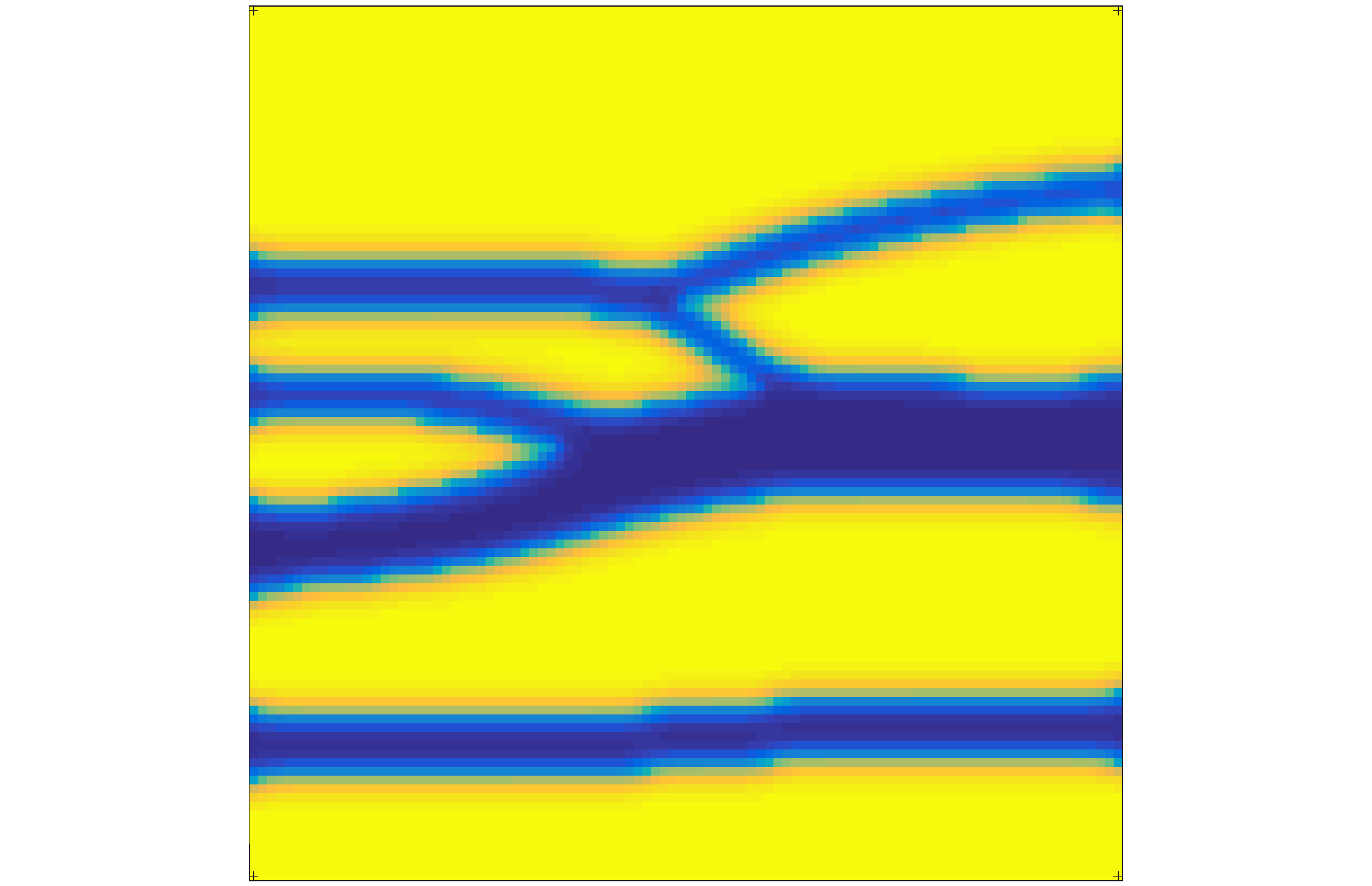}}\vspace{1mm}\\
\end{tabular}
\includegraphics[clip, trim={0 23cm 0 0}, width=.9\textwidth]{latticeshearcolorbar.png}
\caption{Cell density plots with $\delta=10^{-4}$, $\sigma_l=7.5$ for three lattice sizes at different points in time for one realization of the initial cell density.}
\label{sigma_times7.5}
\end{figure}

For $\delta=10^{-4}$, we find significantly more spatial structure in the final cell density distributions. Figure \ref{sigma_times2.5} shows plots of the cell density at various times for lattice sizes $n=25, 50$ and $100$, with threshold parameter $\sigma_l=2.5$ (for one realization of the initial cell density). After a period of growth, the initial uniform state rapidly breaks up into several aggregates of high cell density (e.g. $N_i > 0.8$, typically close to $1$) surrounded by regions of low cell density (e.g. $N_i < 0.2$, typically close to $0$) surrounded by regions of low cell density, which then evolve slowly due to diffusion and shear stress mediated cell growth and death. For the smallest lattice size ($n=25$), this process stabilizes quickly leading to a steady state cell density within the first $t=10$ time units (compare the cell density distributions for $n=25$ and $t=10,15,$ and $30$ in Figure \ref{sigma_times2.5}). For $n=100$, many interior clusters slowly shrink as the cell aggregates along the top and bottom boundaries of the scaffold continue to grow over a longer timescale. Figure \ref{sigma_times7.5} shows cell density distribution plots for $\sigma_l=7.5$. Note that for each $n$ and $t$, the total area of high cell density is larger in Figure \ref{sigma_times7.5} than in \ref{sigma_times2.5}. For both $\sigma_l=2.5$ and $\sigma_l=7.5$, the number of clusters of high cell density is initially greater in the $n=100$ case after the onset of non-uniform growth, but there is a coarsening over time which reduces the number of distinct fluid channels and disconnected regions of high cell density. In comparison, the number of disconnected regions of high cell density is larger in the smaller lattices by the end of the simulation time in both Figures \ref{sigma_times2.5} and \ref{sigma_times7.5} (compare the $t=30$ plots for $n=25$ and $n=100$ in both Figures). This difference in the number of disconnected regions of high cell density between small and large lattices is robust across multiple realizations. 

Note that in Figure \ref{sigma_times2.5} the coarsening for the $n=100$ lattice involves small clusters of high cell density disappearing, while for Figure \ref{sigma_times7.5} it is primarily channels of low cell density that disappear over time. This difference in long-time behaviour can be understood via the threshold parameter $\sigma_l$. When $\sigma_l=2.5$, cells die at relatively low values of shear stress and the overall cell density is low so cells recede away from regions of high fluid flow. For $\sigma_l=7.5$ cells can withstand much more shear and grow over a much larger proportion of the scaffold, filling in small channels of fluid flow over time.

\begin{figure}\setlength{\tabcolsep}{0pt}
\centering
\begin{tabular}{rc c c c} 
 & $\sigma_c=2.5$ & $\sigma_c=5$ & $\sigma_c=7.5$& $\sigma_c=10$\\
 
$t=3$ &\raisebox{-.5\height}{\includegraphics[width=.23\textwidth]{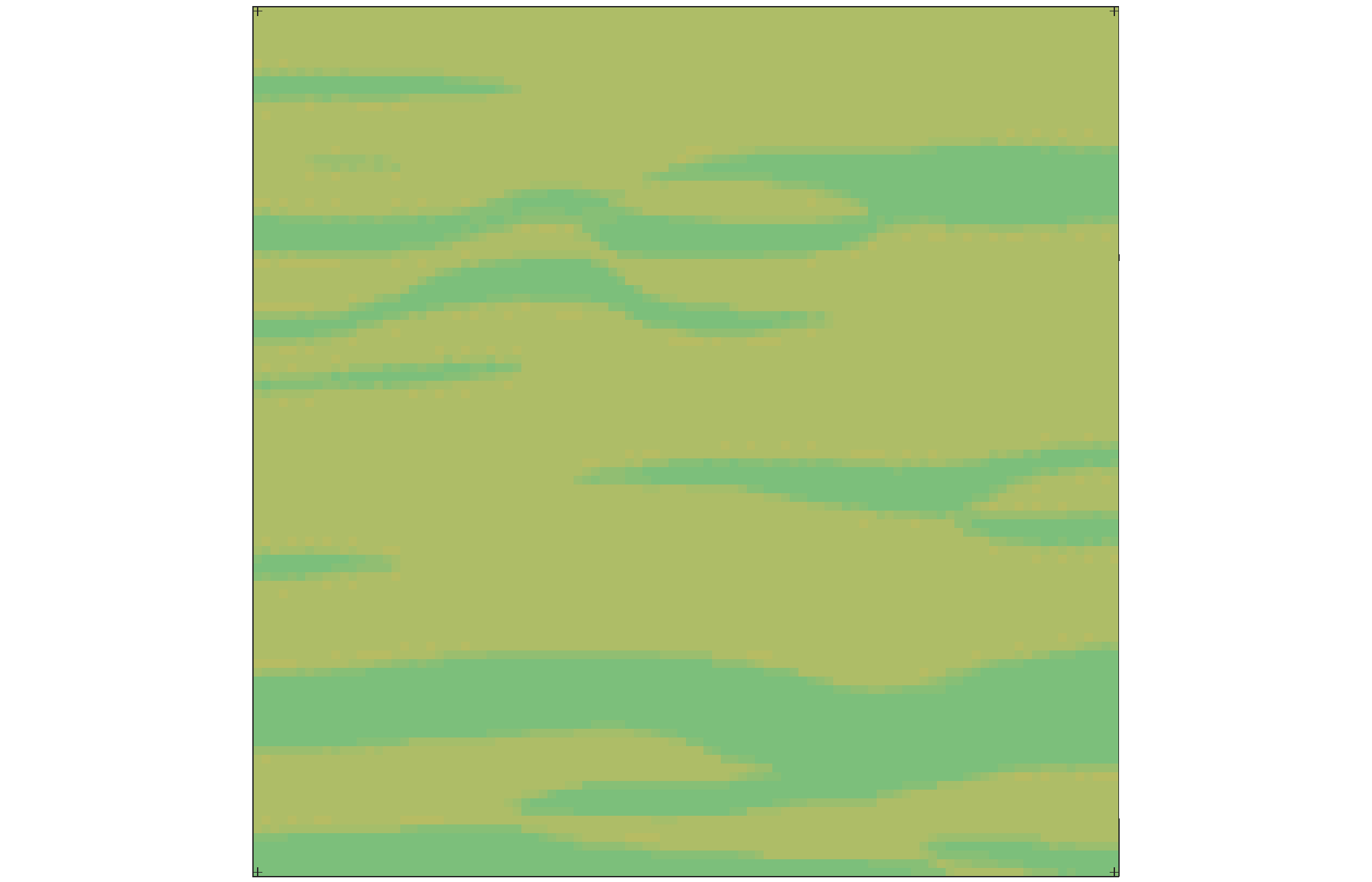}}&
\raisebox{-.5\height}{\includegraphics[width=.23\textwidth]{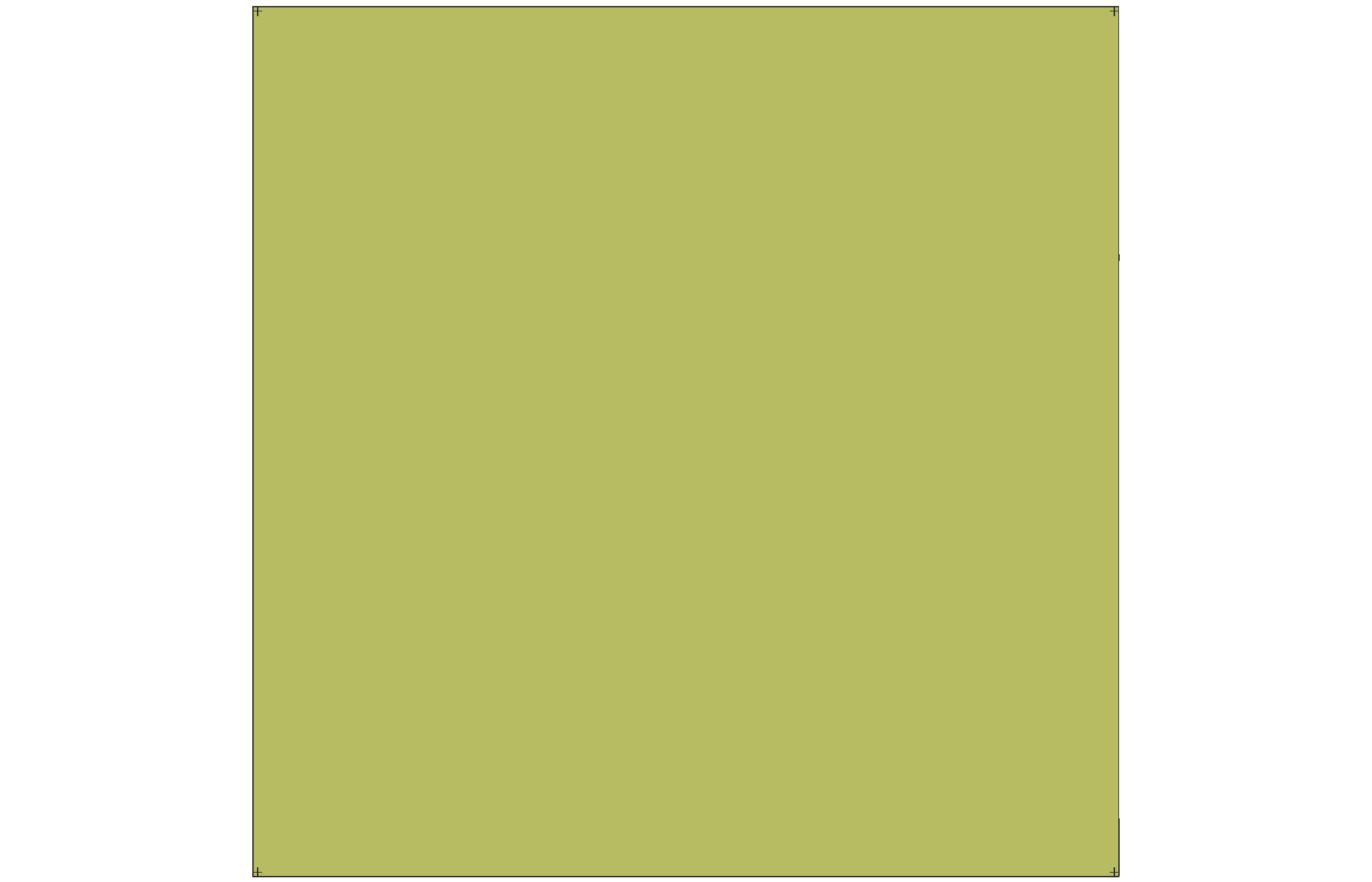}}&
\raisebox{-.5\height}{\includegraphics[width=.23\textwidth]{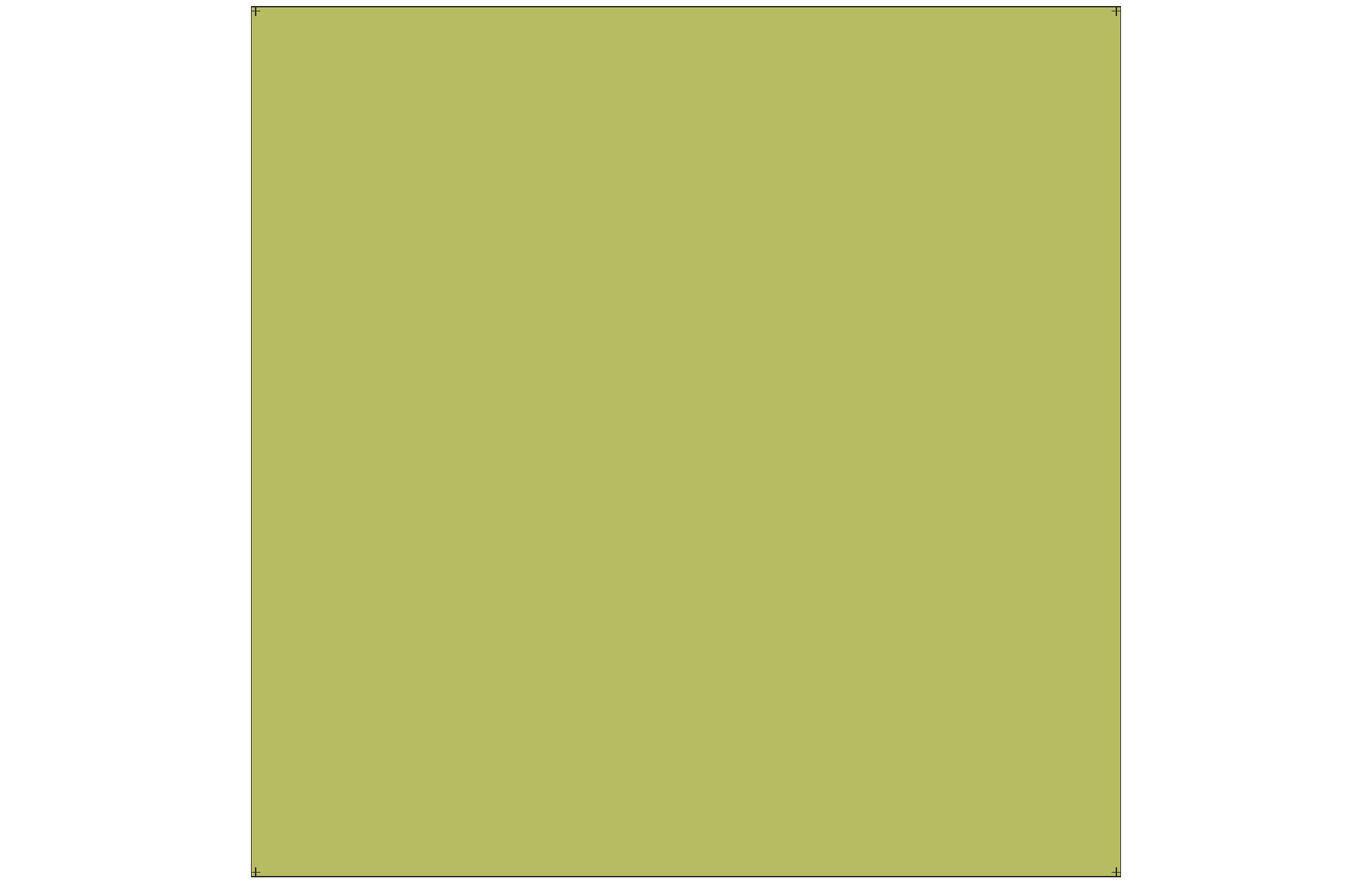}}&
\raisebox{-.5\height}{\includegraphics[width=.23\textwidth]{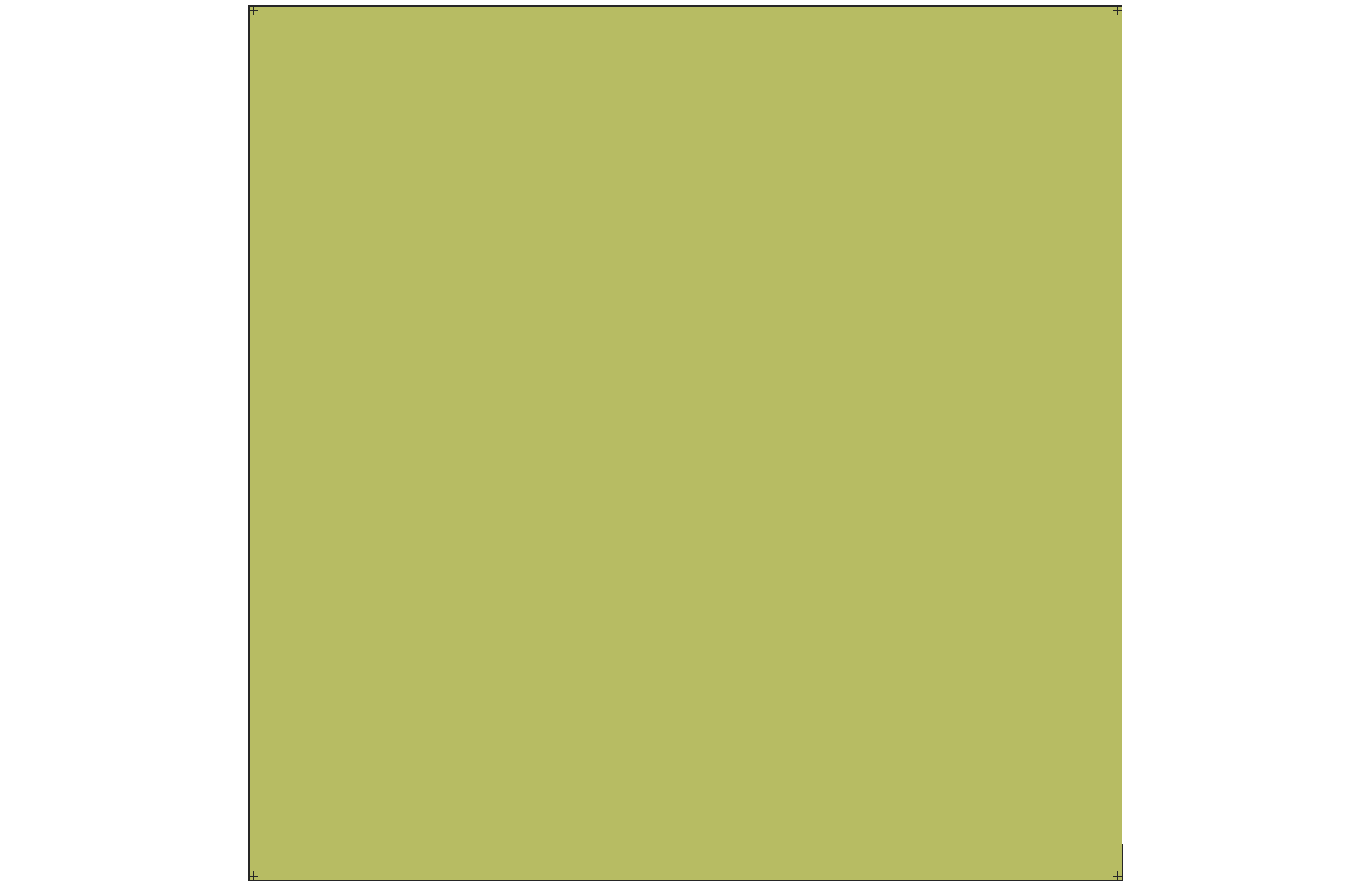}}\vspace{1mm}\\

$t=5$ &\raisebox{-.5\height}{\includegraphics[width=.23\textwidth]{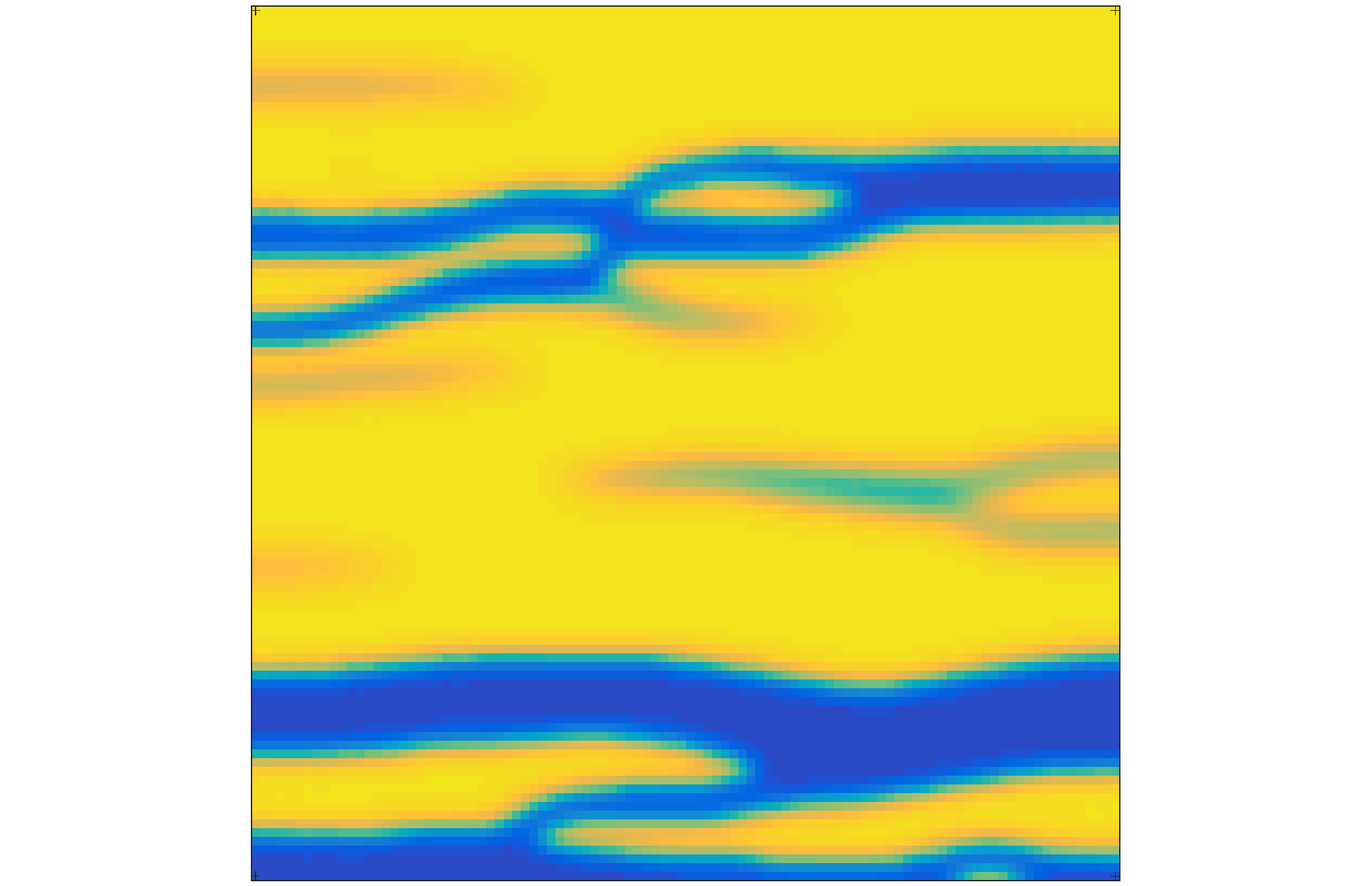}}&
\raisebox{-.5\height}{\includegraphics[width=.23\textwidth]{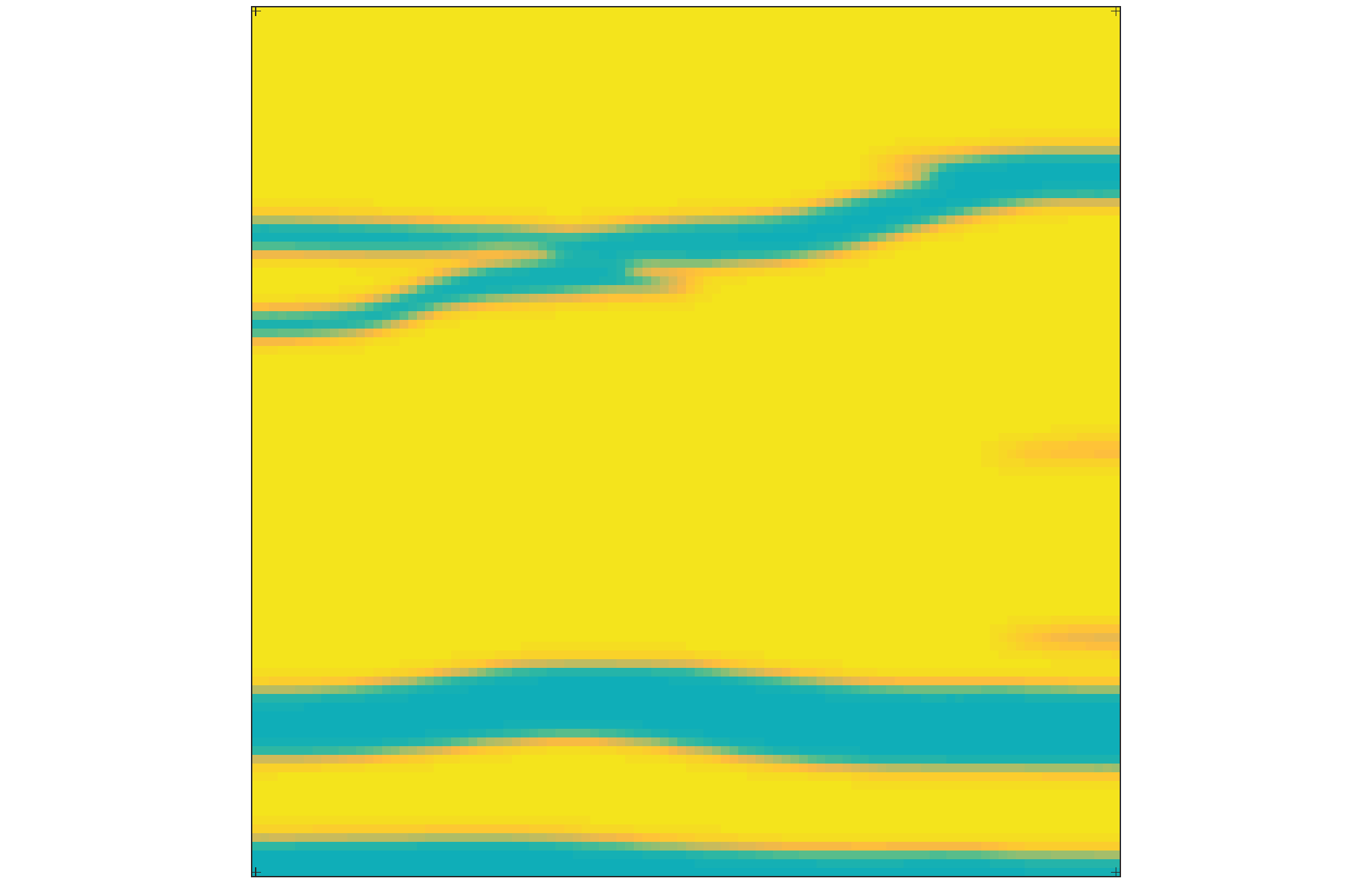}}&
\raisebox{-.5\height}{\includegraphics[width=.23\textwidth]{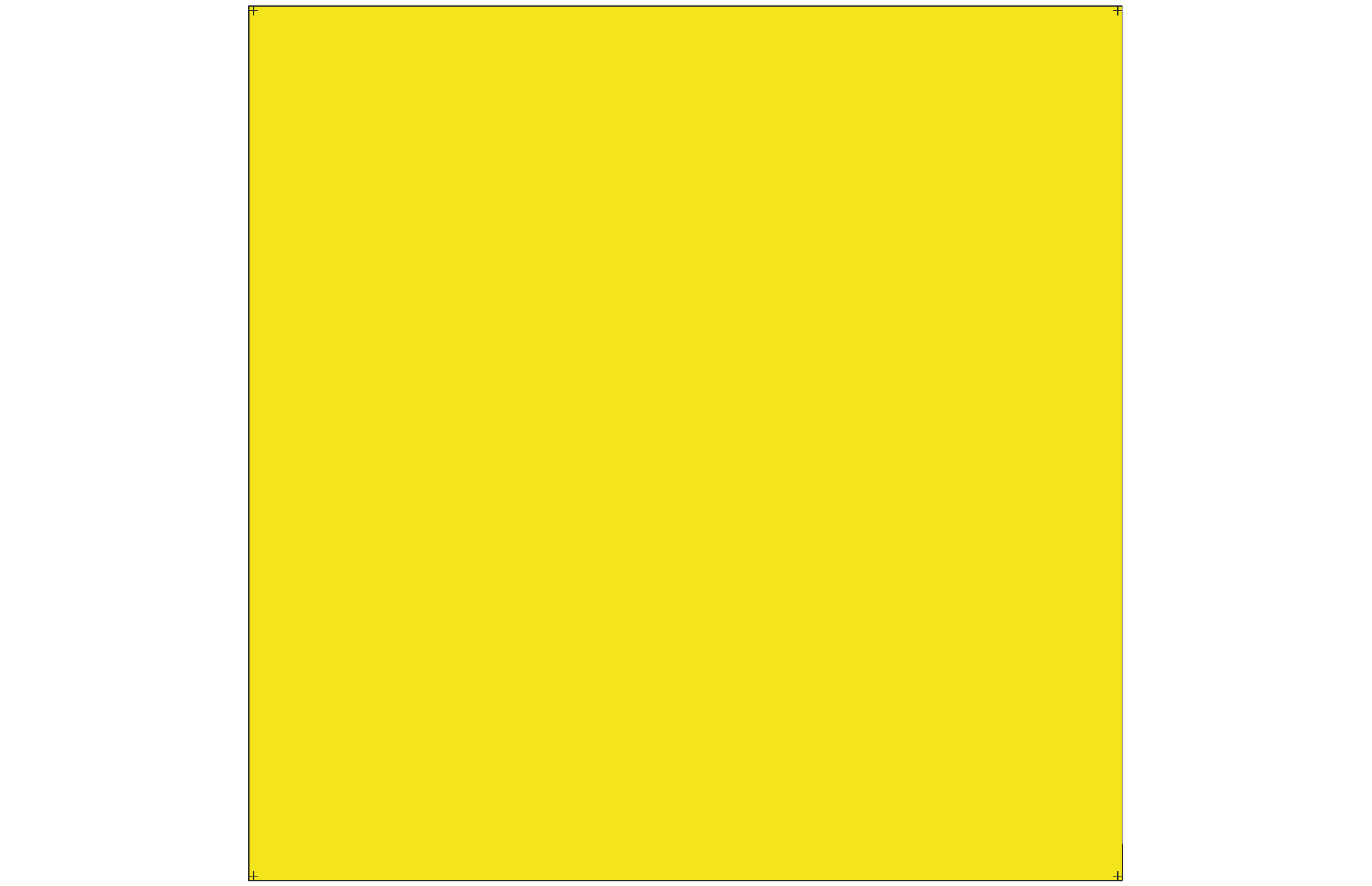}}&
\raisebox{-.5\height}{\includegraphics[width=.23\textwidth]{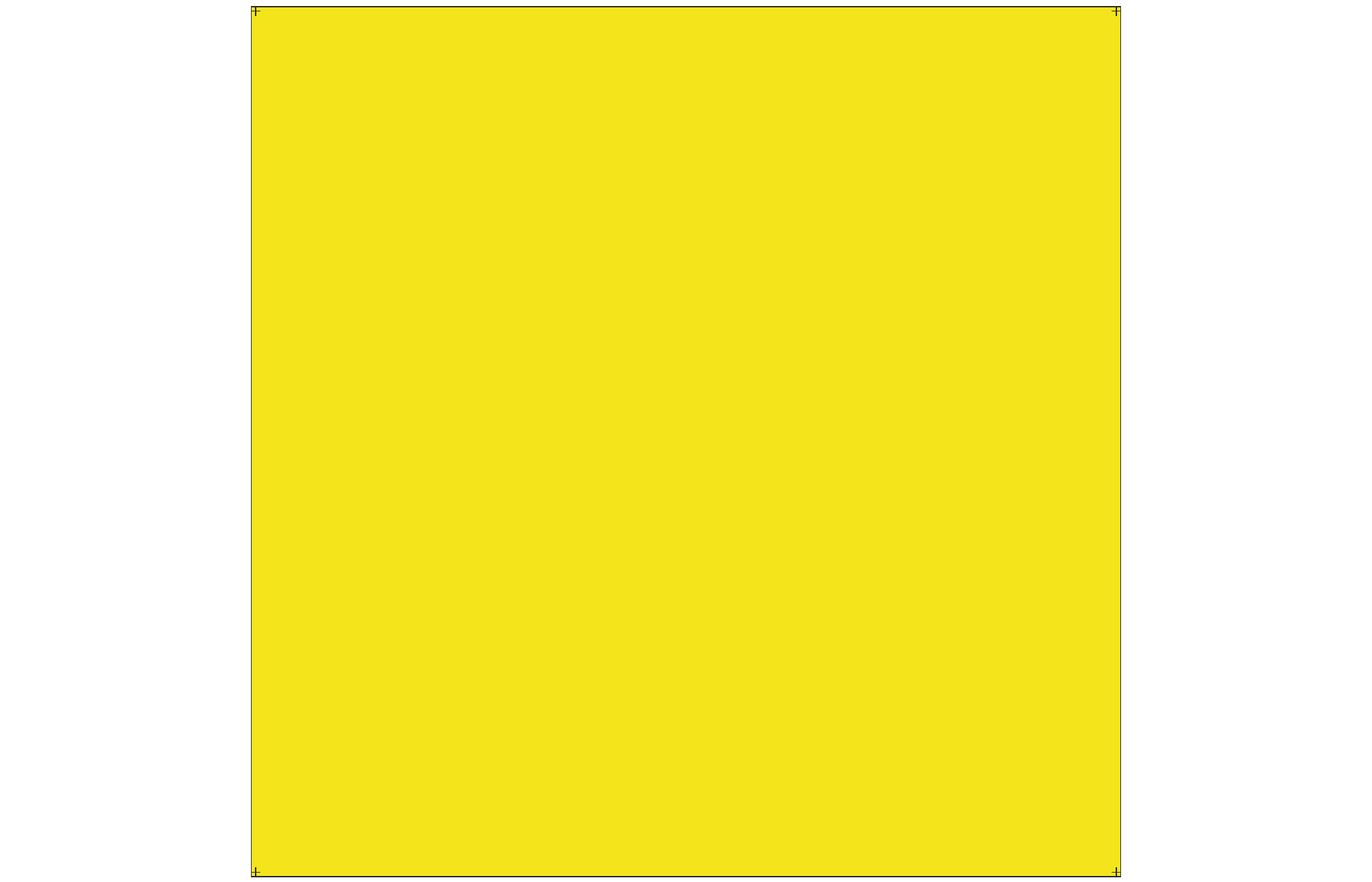}}\vspace{1mm}\\

$t=10$&\raisebox{-.5\height}{\includegraphics[width=.23\textwidth]{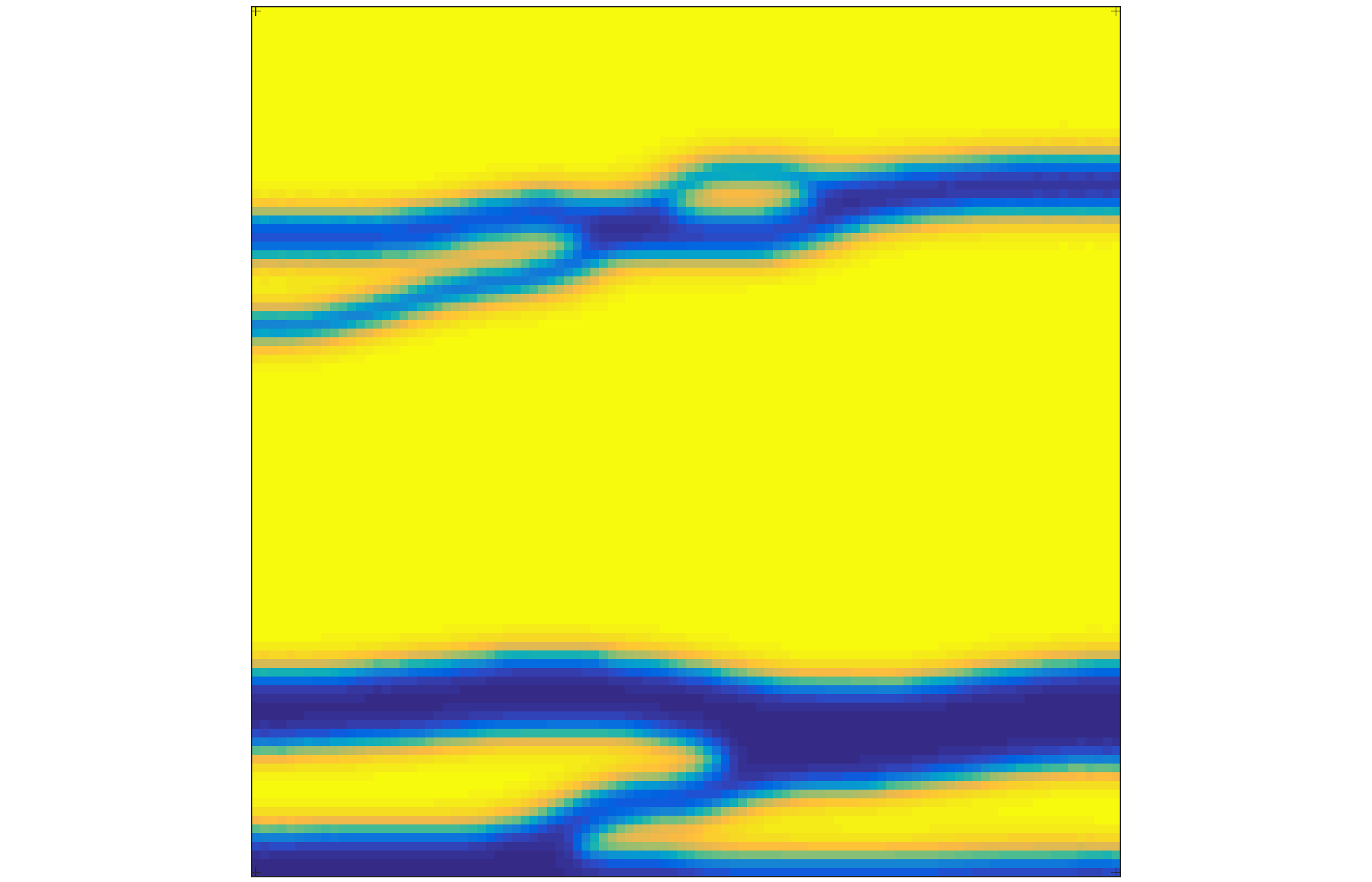}}&
\raisebox{-.5\height}{\includegraphics[width=.23\textwidth]{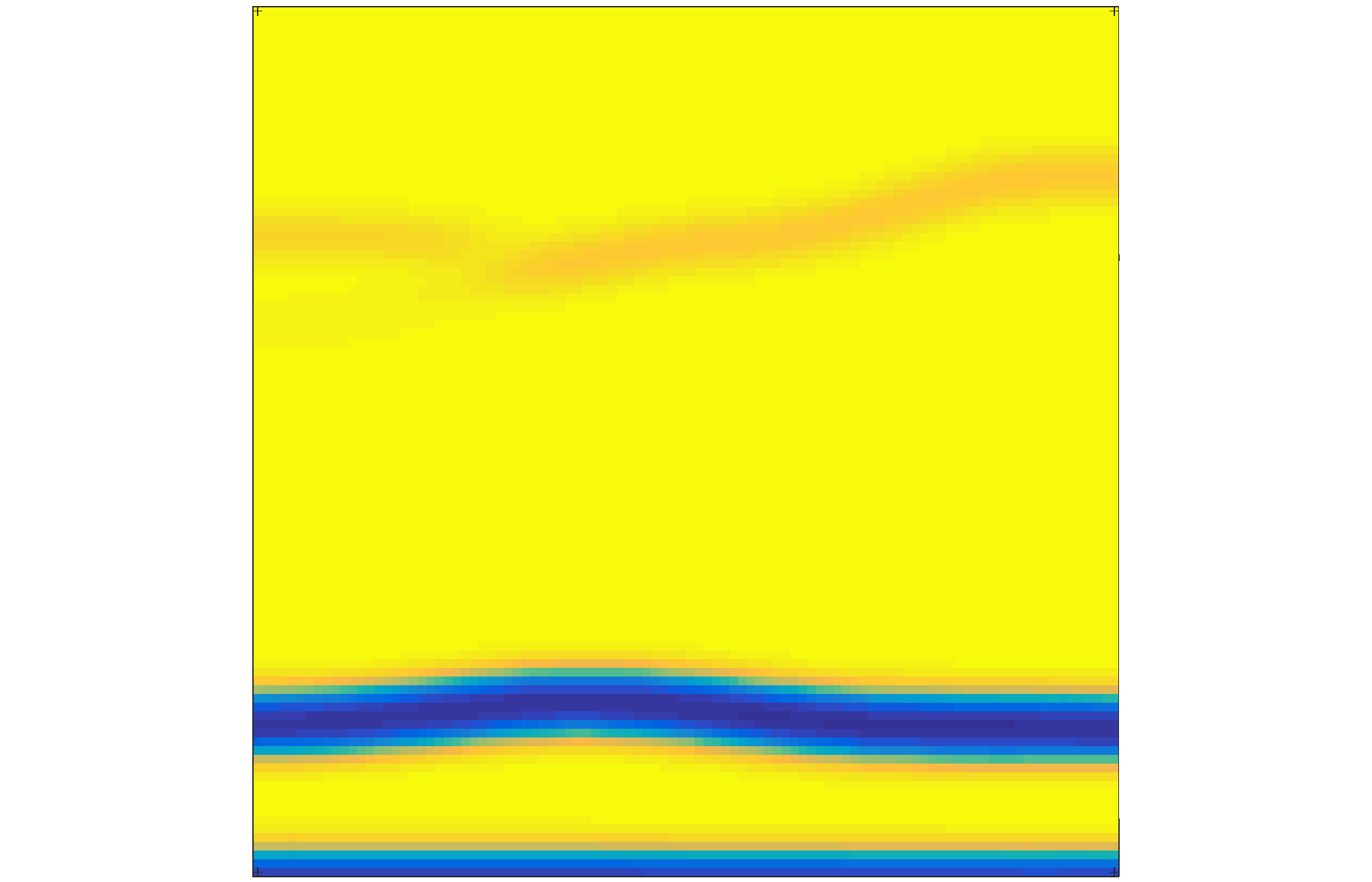}}&
\raisebox{-.5\height}{\includegraphics[width=.23\textwidth]{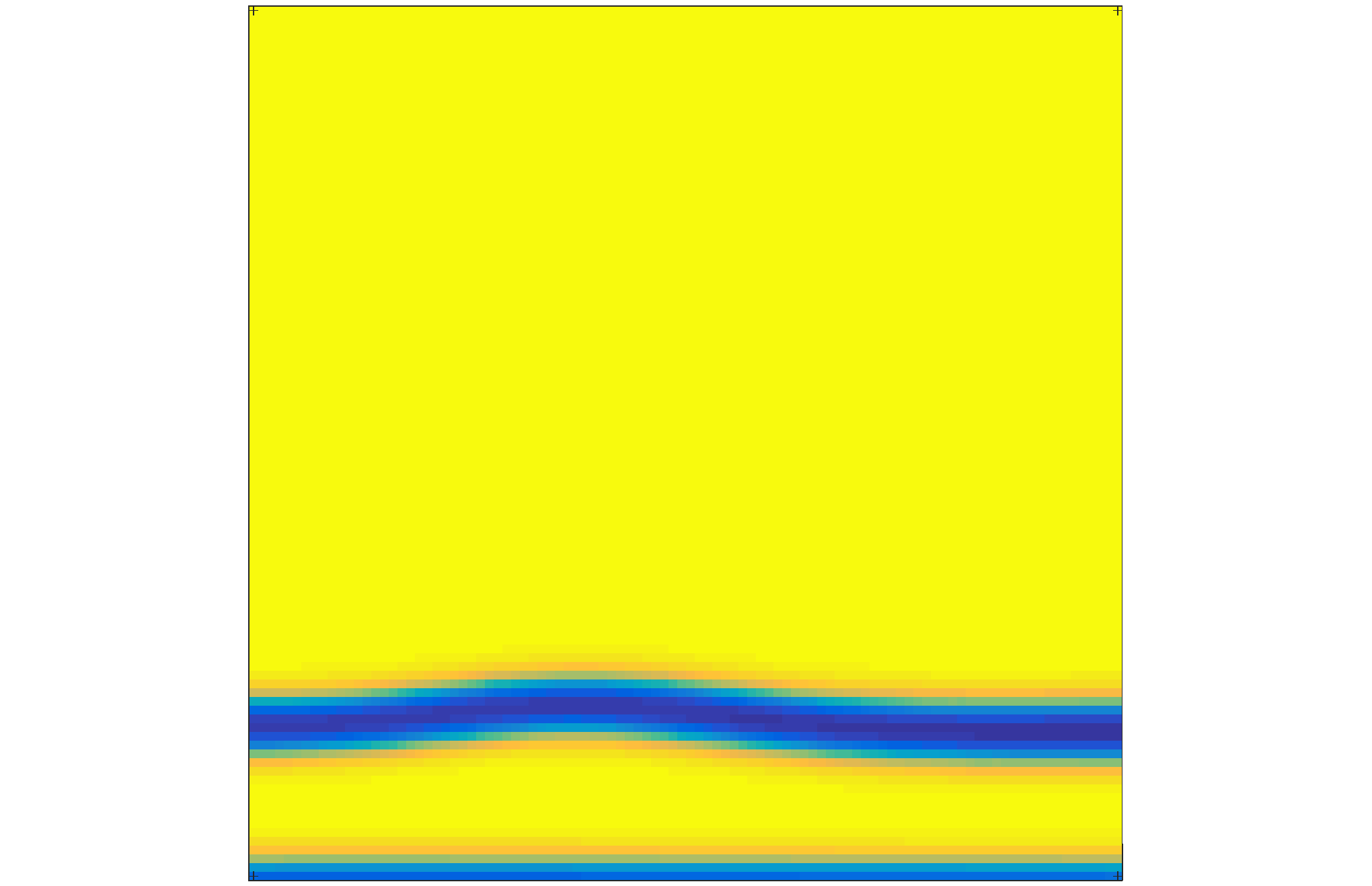}}&
\raisebox{-.5\height}{\includegraphics[width=.23\textwidth]{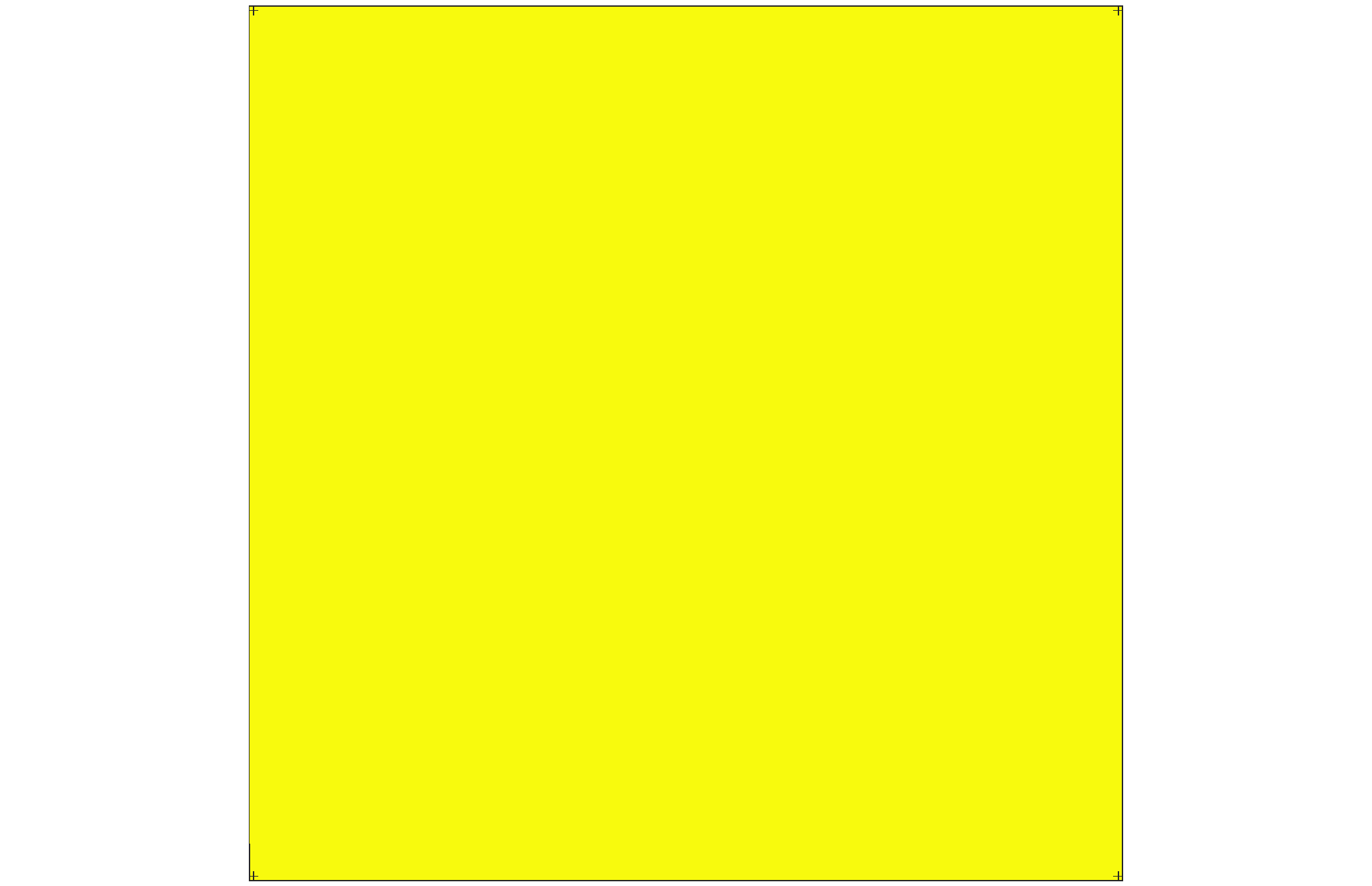}}\vspace{1mm}\\

$t=30$ &\raisebox{-.5\height}{\includegraphics[width=.23\textwidth]{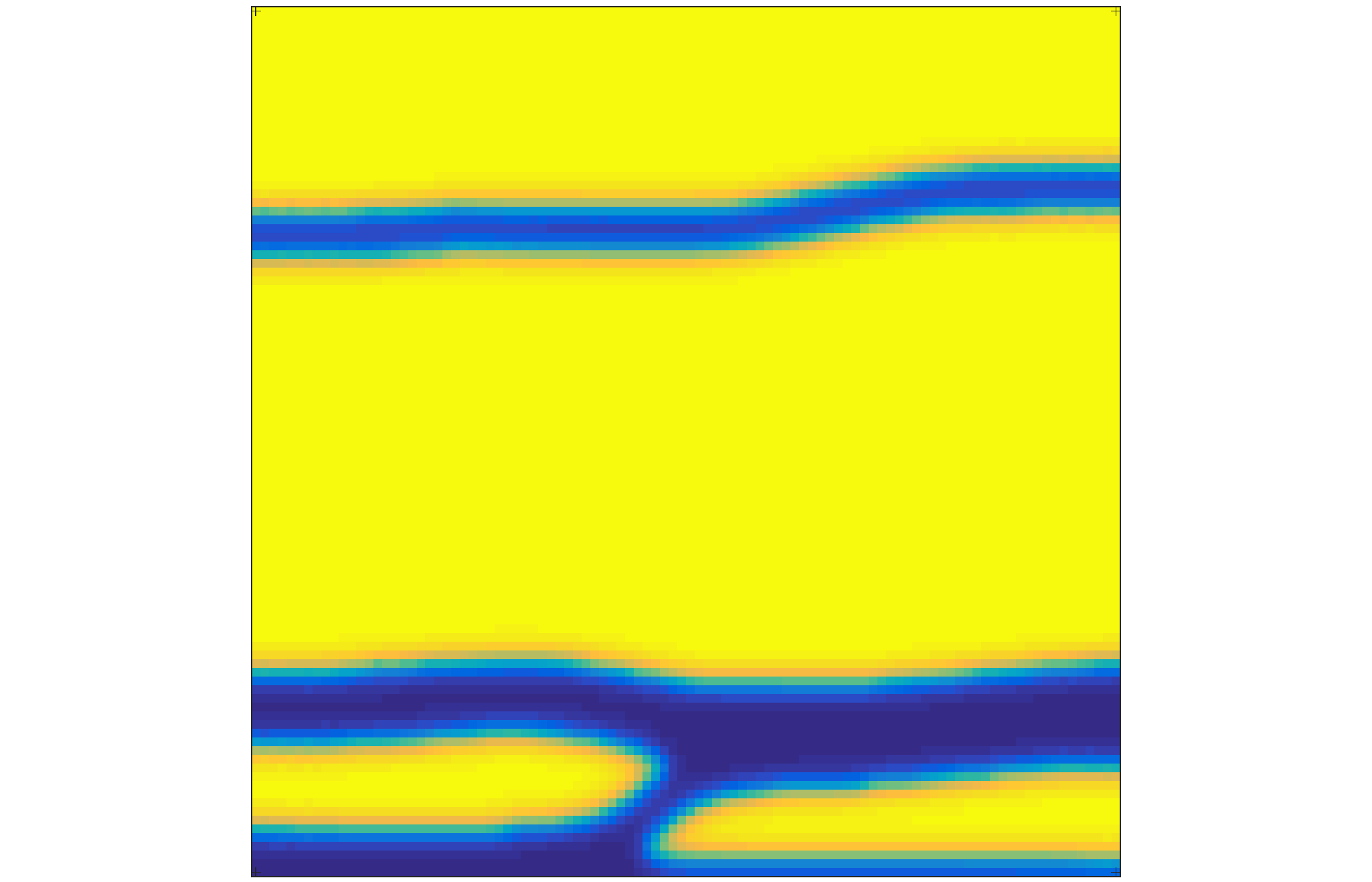}}& 
\raisebox{-.5\height}{\includegraphics[width=.23\textwidth]{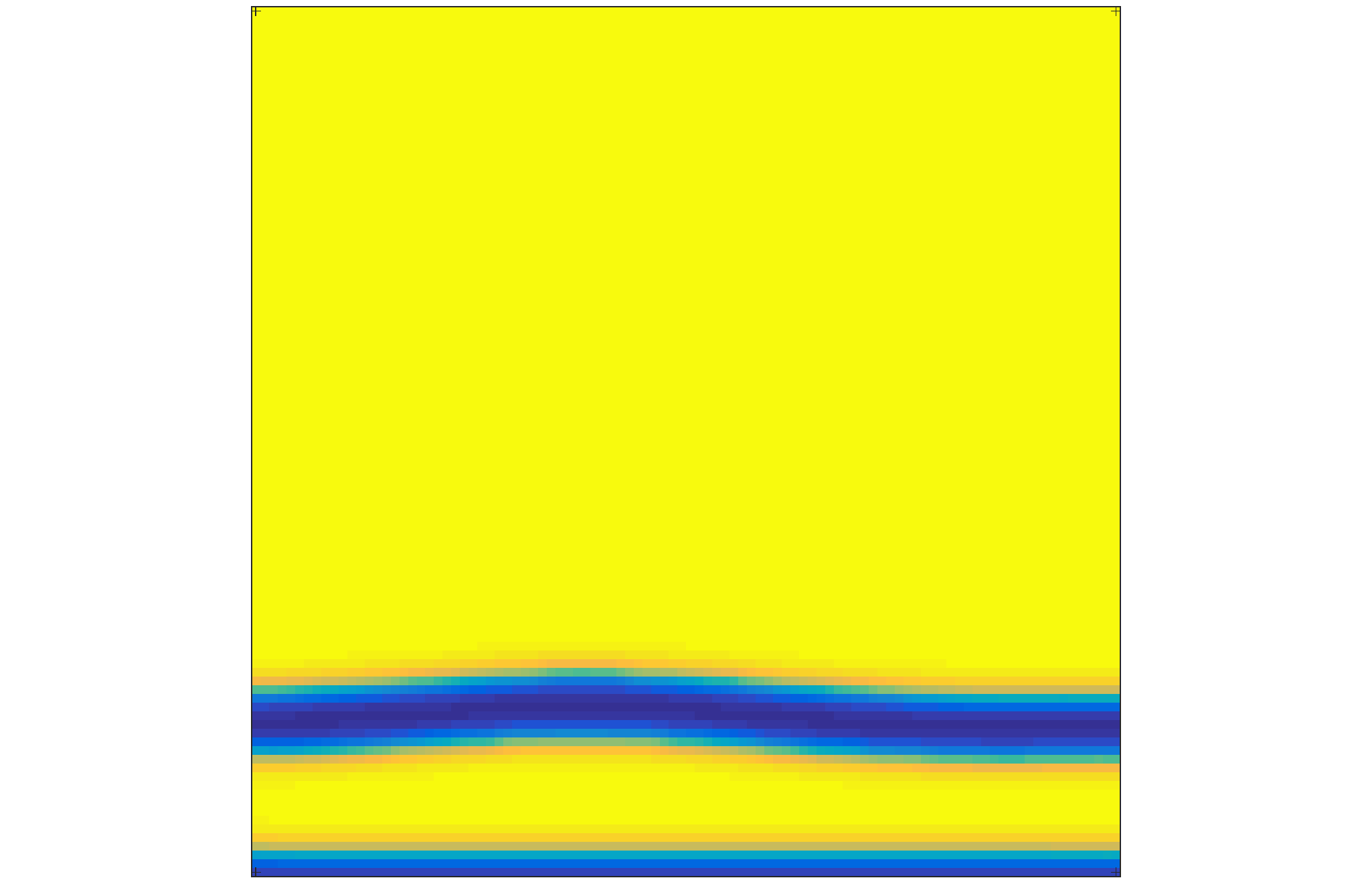}}&
\raisebox{-.5\height}{\includegraphics[width=.23\textwidth]{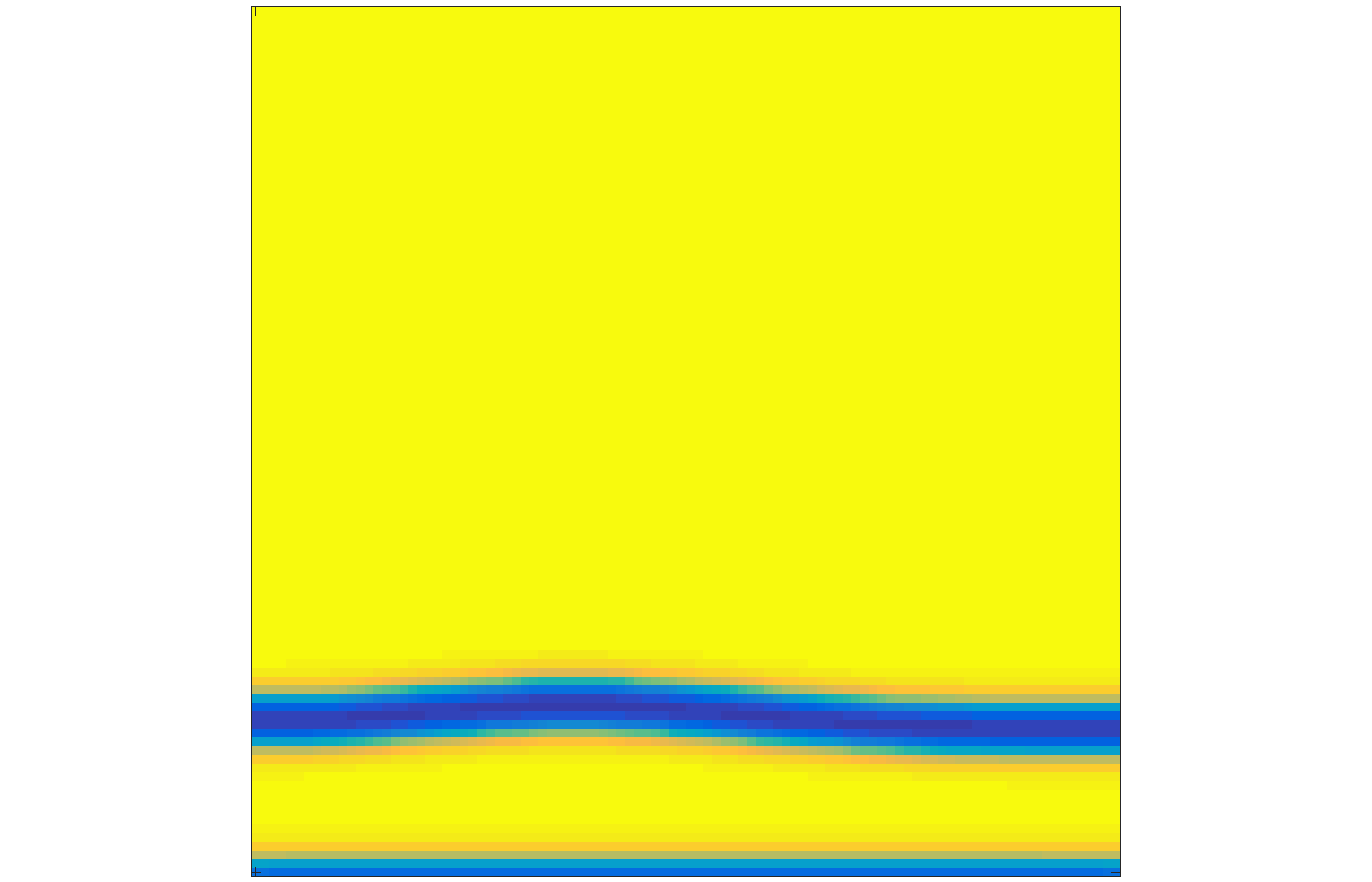}}&
\raisebox{-.5\height}{\includegraphics[width=.23\textwidth]{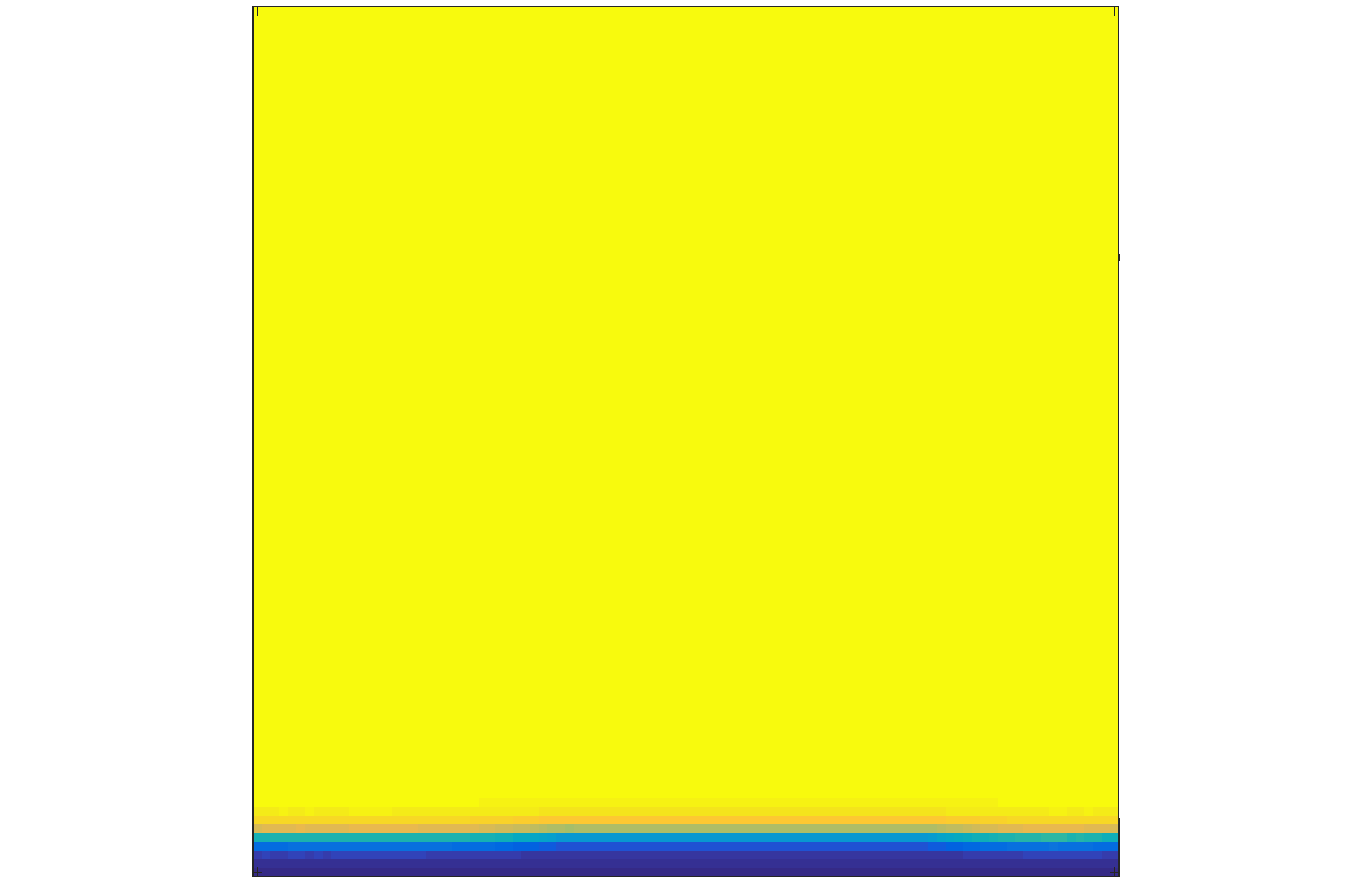}}\\
\end{tabular}
\includegraphics[clip, trim={0 23cm 0 0}, width=.9\textwidth]{latticeshearcolorbar.png}
\caption{Cell density plots with $\delta=10^{-4}$, organized with $\sigma_c=2.5, 5, 7.5$, and $10$ from left to right, and at times $t=2, 5, 10,15$, and $30$ from top to bottom for one realization of the initial cell density.}
\label{sigma_timesPDE}
\end{figure}

These results demonstrate a diverse range of behaviour for the lattice models. Figure \ref{sigma_timesPDE} illustrates predictions from the PDE model for the cell density distribution over the spatial domain at various times and for four values of the threshold parameter $\sigma_c$ for one realization of the initial cell density. This model also exhibits transient dynamics where clusters of high cell density form, and then coarsen into larger regions of high cell density as in the $n=100$ lattice shown in Figures \ref{sigma_times2.5} and \ref{sigma_times7.5}. Overall there are more regions of high cell density in the PDE model compared to the lattice simulations, which is consistent with the shear stress predictions given in Figure \ref{ShearNPlot} (for the same cell density, this plot shows lower predicted shear stress values for the PDE model compared to the lattice model).

\begin{figure}\setlength{\tabcolsep}{1pt}
\centering
\begin{subfigure}{0.49\textwidth}
\includegraphics[width=1\textwidth]{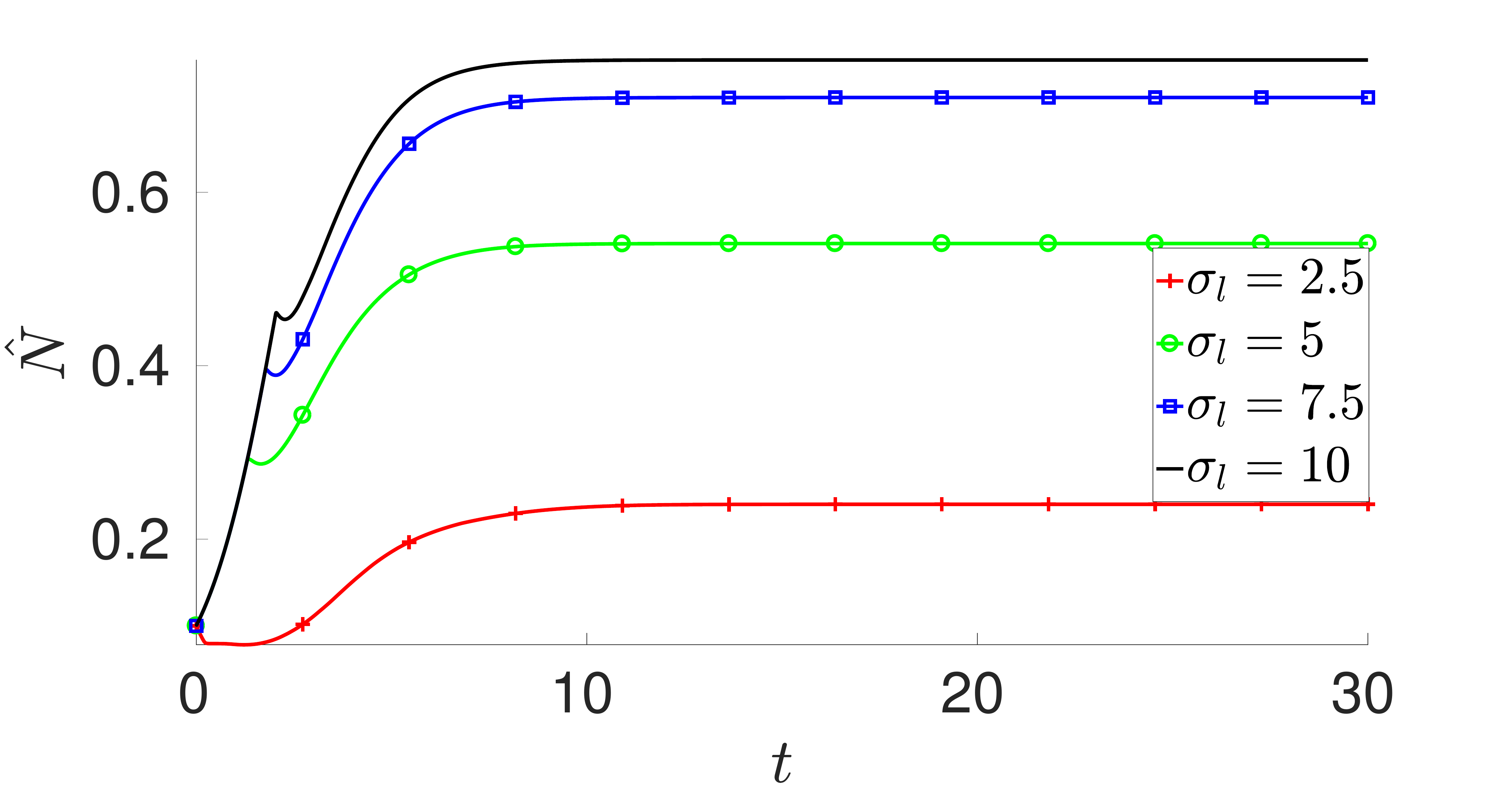}
\caption{$n=25$}\label{time_n25}
\end{subfigure}
\begin{subfigure}{0.49\textwidth}
\includegraphics[width=1\textwidth]{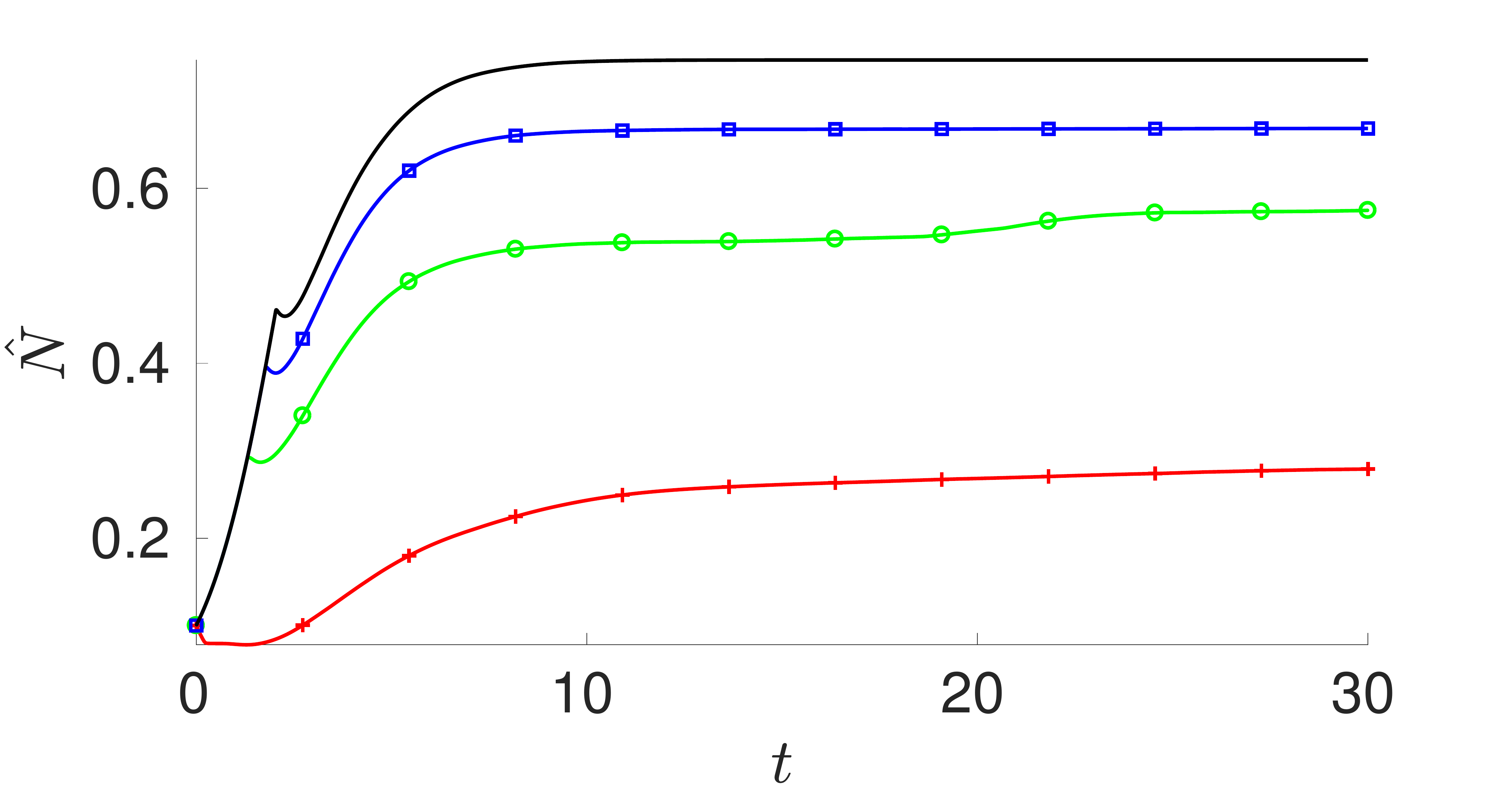}
\caption{$n=50$}
\end{subfigure}

\begin{subfigure}{0.49\textwidth}
\includegraphics[width=1\textwidth]{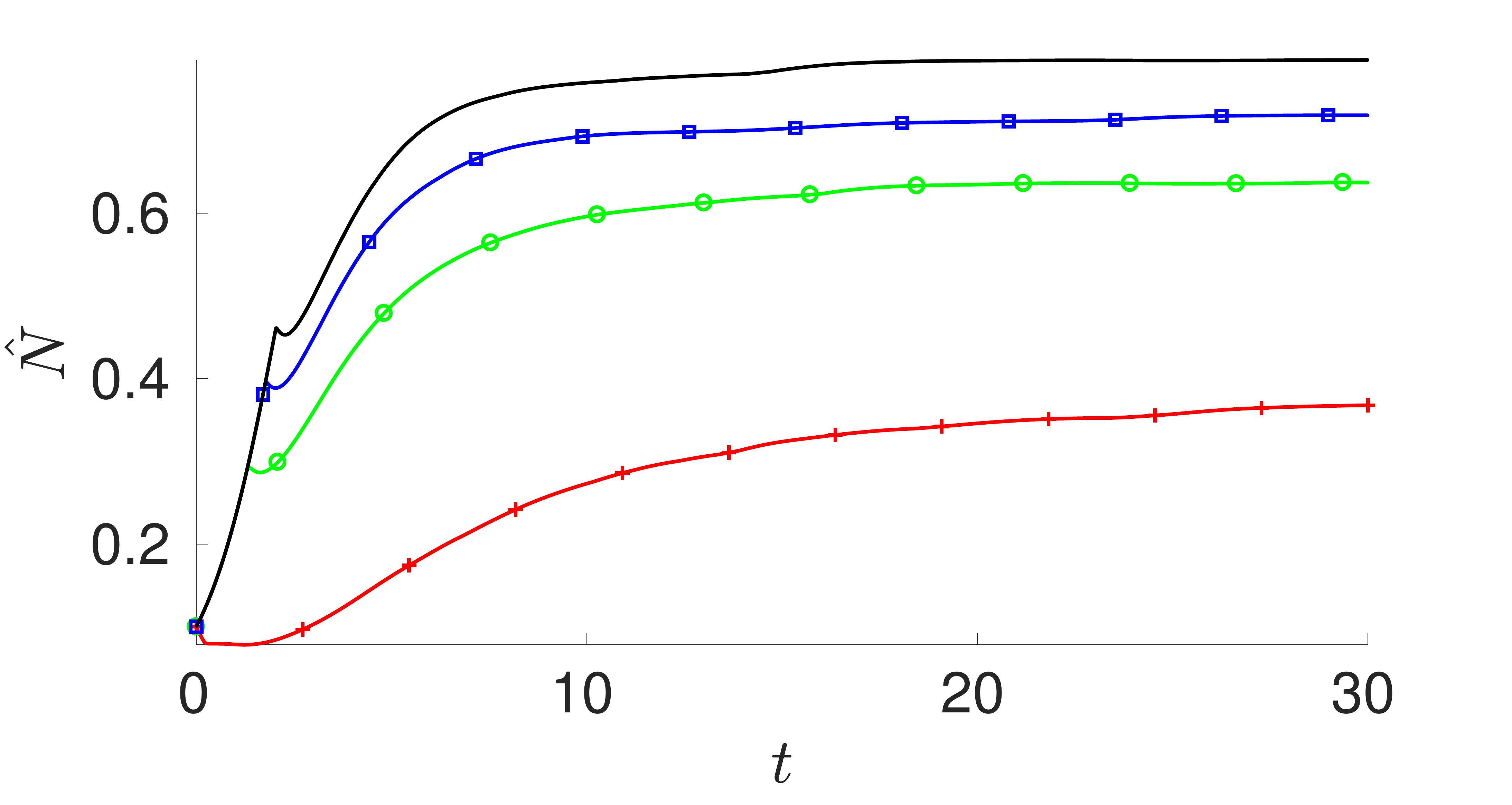}
\caption{$n=100$}\label{time_n100}
\end{subfigure}
\begin{subfigure}{0.49\textwidth}
\includegraphics[width=1\textwidth]{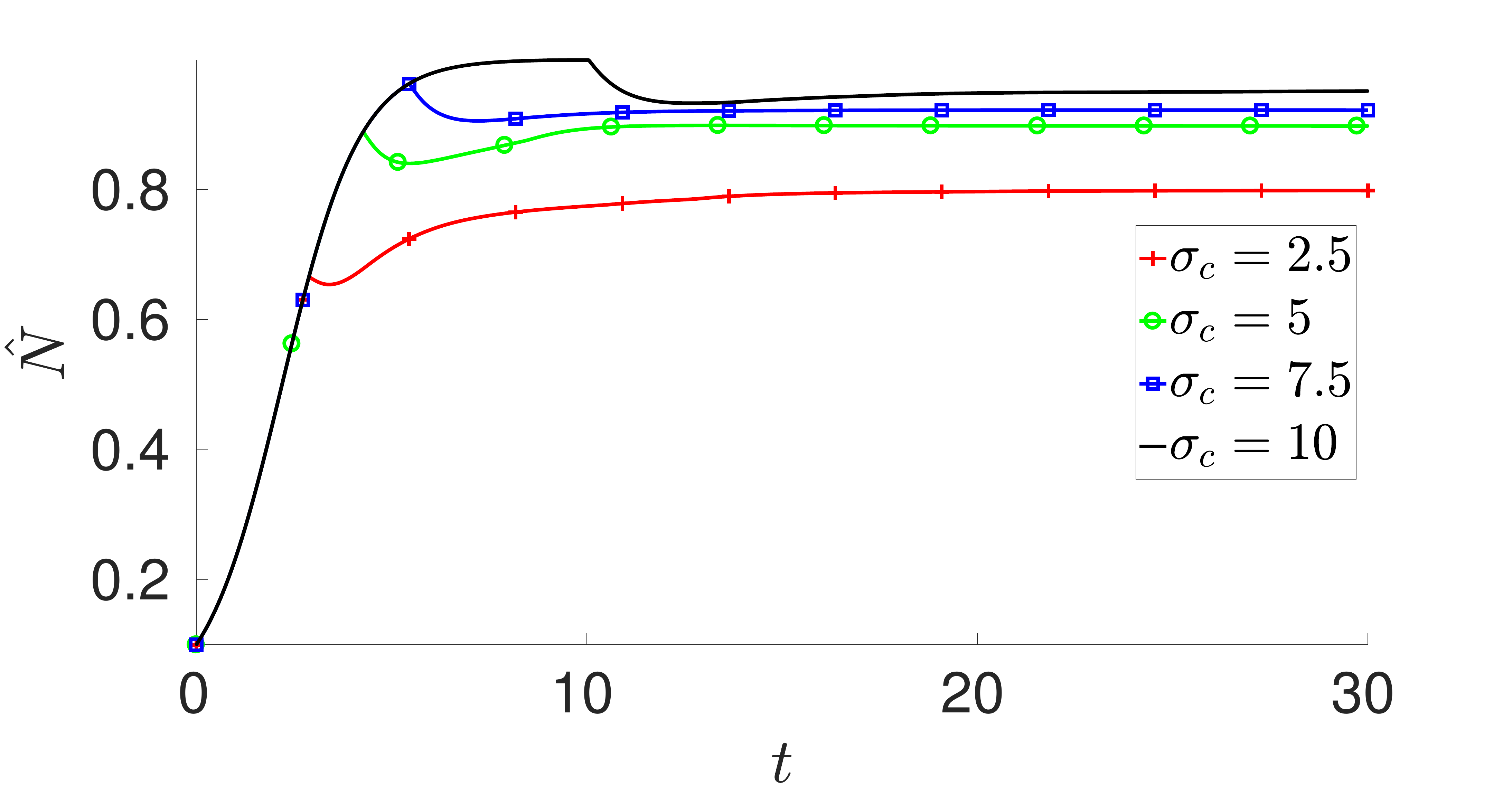}
\caption{PDE}
\end{subfigure}
\caption{Mean cell density plots over time for $\delta=10^{-4}$ for lattice and for the same realization of the initial cell densities used in Figures \ref{sigma_times2.5}-\ref{sigma_timesPDE}.}\label{sigma_time_series}
\end{figure}

In Figure \ref{sigma_time_series}, we plot time series of mean cell densities, $\hat{N}(t)$, for lattice simulations with $n=25, 50,$ and $100$, as well as for the PDE for the same realization of the initial cell densities used in Figures \ref{sigma_times2.5}-\ref{sigma_timesPDE}. The $n=25$ lattice quickly reaches equilibrium for each $\sigma_l$, whereas the larger $n=100$ simulations show growth and death processes over a longer timescale, and in some cases these have not reached an equilibrium value at the end of the simulations at $t=30$ (compare the long time behaviour in Figures \ref{time_n25} and \ref{time_n100}). We truncate the numerical experiments here partly because temporal variations in mean cell density beyond this point for all simulations were small, and partly because this period of time would exceed most \emph{in vitro} tissue engineering experiments as $t=30$ would typically be on the order of months. All simulations shown in Figure \ref{sigma_time_series} grow logistically at the same rate until some region of the scaffold reaches the shear stress threshold $\sigma_l$ or $\sigma_c$ leading to non-uniform growth, and eventually to differences in final mean cell densities.

We observe less spatial structure in the PDE simulations than in the lattice simulations (compare Figures \ref{sigma_times2.5} and \ref{sigma_times7.5} with Figure \ref{sigma_timesPDE}). We quantify spatial variation in cell density with a heterogeneity function. For simplicity we choose
\be\label{hetero}
E(N_1, \dots, N_{n^2}) = \frac{1}{n}\sum_{j=1}^{n^2} \sum_ {i=1}^{n^2} A_{ij}\abs{N_i - N_j},
\ee
which is the total absolute difference in cell density between adjacent nodes throughout the lattice. The function satisfies $E \geq 0$ with $E=0$ only for a homogeneous cell density distribution. The PDE solutions were interpolated onto a square grid with a size of $100$ by $100$, so the function defined by \eqref{hetero} can be seen as a discretization of the functional $\int_0^1\int_0^1 \abs{\partial_x N(x,y)} + \abs{\partial_y N(x,y)}dxdx$ which is a measure of anisotropic total variation \citep{van_gennip_$gamma$-convergence_2012}. This heterogeneity function allows for a quantitative visualization of the coarsening behaviour described earlier that was qualitatively observed in Figures \ref{sigma_times2.5}-\ref{sigma_times7.5}. 

\begin{figure}
\centering
\includegraphics[width=.8\textwidth]{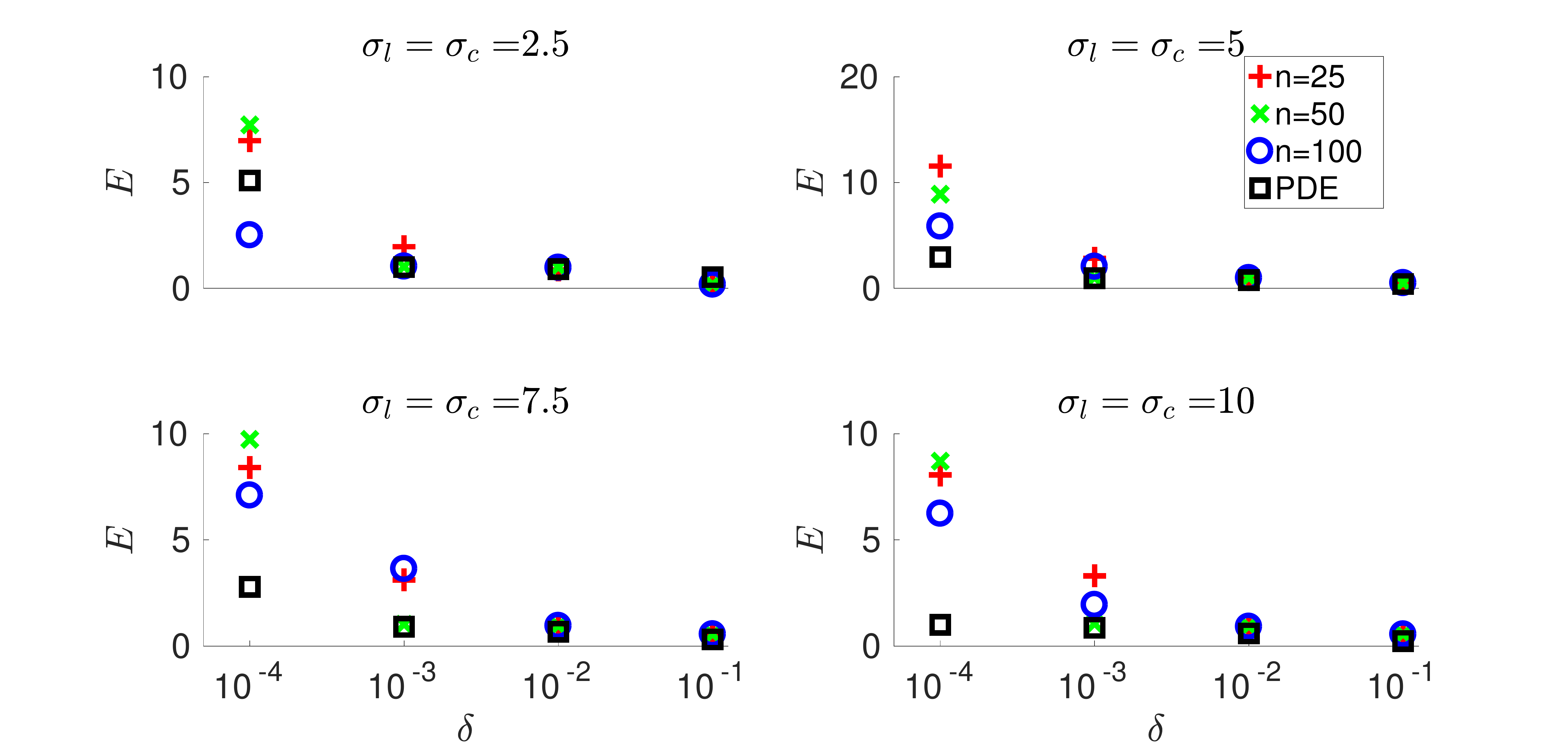}
\caption{Plots of the heterogeneity of cell density defined by \eqref{hetero} over the four values of $\delta$ for several parameter combinations for the PDE and lattice models at $t=30$. Note the variation in the ranges of the heterogeneity function $E$. The standard deviation of $E$ between 300 realizations is, in each case, less than 1\% of the reported mean value across all realizations.}
\label{cell_hetero}
\end{figure}

\begin{figure}
\centering
\includegraphics[width=.8\textwidth]{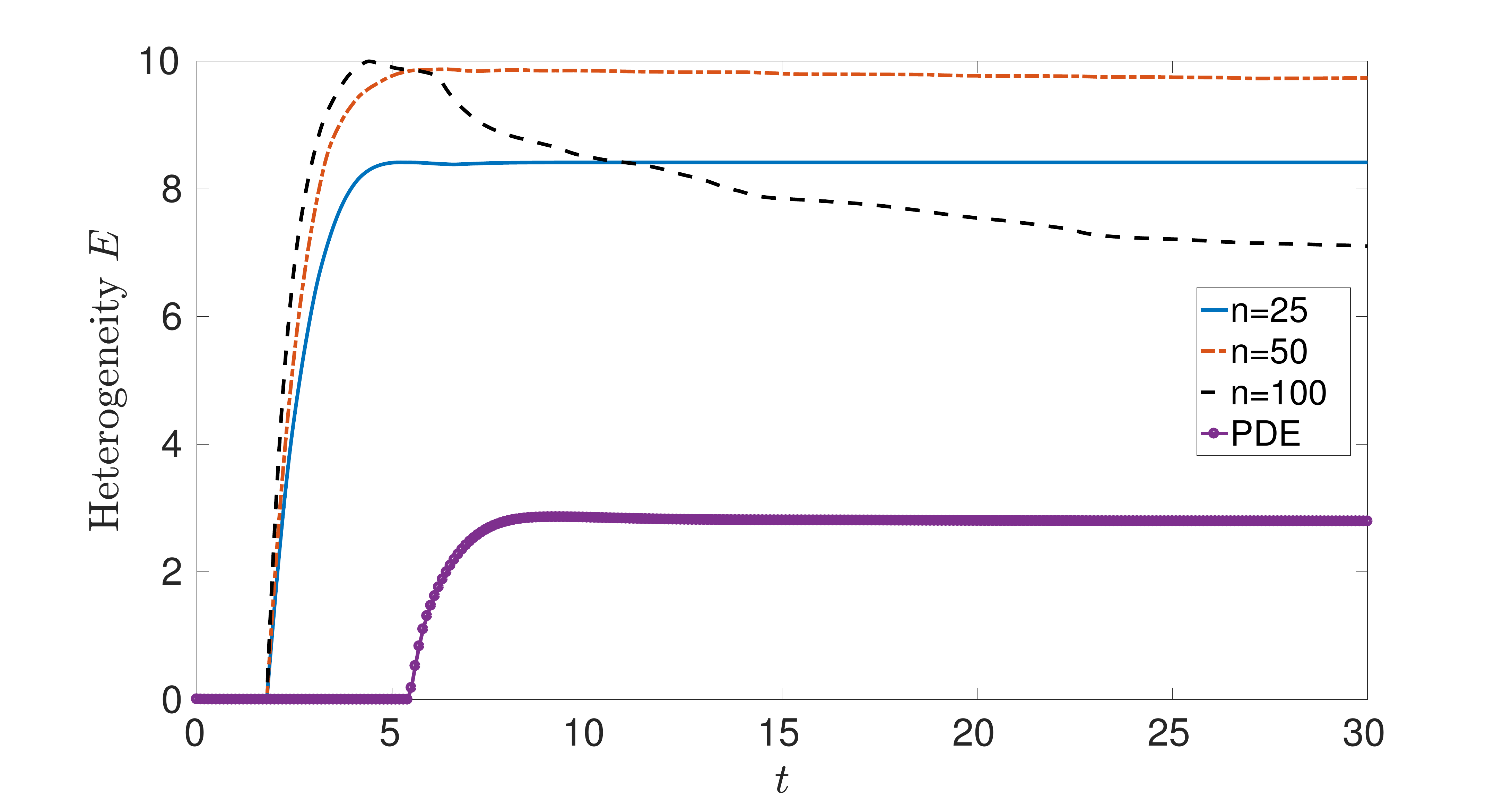}
\caption{Plots of the heterogeneity of cell density defined by \eqref{hetero} over the simulation time for $\delta=10^{-4}$ and $\sigma_l=\sigma_c=7.5$ for one realization of the initial condition. Qualitatively similar behaviour occurred in other realizations.}
\label{cell_hetero_time_series}
\end{figure}

In Figure \ref{cell_hetero} we plot values of this function for the simulated cell densities at $t=30$ averaged over $300$ realizations of the initial cell densities. For the majority of parameter combinations simulated, the value of the heterogeneity given by \eqref{hetero} was lower for the PDE simulations than for any of the lattice simulations with the same parameters. Larger values of diffusion show significantly smaller values of spatial heterogeneity, and for most combinations of the parameters the larger lattice of size $n=100$ has a lower value of $E$ than the other two lattice sizes. Figure \ref{cell_hetero_time_series} shows a time series of this function for one choice of parameters for the lattice simulations and for a corresponding PDE simulation. For $n=25,50$, and for the PDE, the difference between the final heterogeneity and its maximal value is small, while for $n=100$ the maximum value of $E$ occurs around $t \approx 4$, and then slowly falls over time to a value that is significantly smaller (compare this to Figure \ref{sigma_times7.5} for $n=100$ and $t \geq 5$). 

There is less spatial heterogeneity in the final cell densities of the PDE simulations than in the corresponding lattice model for all values of $n$ and all values of $\sigma_l$ and $\sigma_c$ except for $\sigma_l=\sigma_c=2.5$, where we note that for the cell density distributions in Figure \ref{sigma_times2.5} the cells have aggregated to the sides for $n=100$ and $t=30$ and hence created a single large channel in the center of the scaffold, whereas several small channels remain in Figure \ref{sigma_timesPDE}. We note that in Figure \ref{cell_hetero_time_series}, there is a specific time at which spatial heterogeneity emerges (e.g. $E$ is no longer $0$), which corresponds to the onset of cell death in some regions of the scaffold.
\subsection{Onset of Non-Uniform Patterning Due to Shear-Induced Cell Death}\label{instability_time}
As the initial data are approximately uniform, the evolution of the cell density can be determined analytically up to the time that cell death occurs. For $\sigma \lesssim \sigma_c$, the cell growth is purely logistic. This bound is sharp in the limit of $g \to \infty$ where $F_1$ and $F_2$ become step-functions, but it is approximately obeyed for $g=60$ so that for shear stress values below $\sigma_c$, cells will grow logistically. Hence, for a constant initial condition $N_0 \in [0,1]$, we have 
\be\label{pde_logistic}
N(\boldsymbol{x},t) = N_0 \frac{\exp(t)}{1+N_0(\exp(t)-1)}, \quad \boldsymbol{x}\in [0,1]^2, \quad  t < t_s,
\ee
where $t_s$ is the time that it takes for the shear stress to become approximately equal to the threshold parameter, and hence for the cells to no longer undergo uniform logistic growth everywhere.

From Equation \eqref{PDEShear} we see that $t_s$ may be defined implicitly by
\be\label{pde_sigma_instability}
\sigma_c \approx \sigma(t_s) =  \frac{1}{1-\rho N(t_s)}.
\ee
Substituting \eqref{pde_logistic} into \eqref{pde_sigma_instability} gives
\be\label{ts_pde}
t_s = \ln \left(\frac{\left(\sigma_c-1\right)\left(1-N_0\right)}{N_0\left(1-\sigma_c\left(1-\rho\right) \right)}\right).
\ee

Analogously for the lattice model, for $\sigma_i \lesssim \sigma_l$, the cell growth is purely logistic at each node. So for a constant initial condition $N_{i0} = N_0 \in [0,1]$ for each $i$, we have
\be\label{lattice_logistic}
N_i(t) = N_0 \frac{\exp(t)}{1+N_0(\exp(t)-1)}, \quad 1 \leq i \leq n^2, \quad  t < t_s,
\ee
$t_s$ again being the time for the shear stress at some node to become comparable to the shear threshold. From Equation \eqref{LatticeShear} we approximate $t_s$ by
\be\label{lattice_sigma_instability}
\sigma_l \approx \sigma_i(t_s) = \frac{2}{(1-\rho N_i(t_s))^3}.
\ee
We substitute \eqref{lattice_logistic} into \eqref{lattice_sigma_instability} to find
\be t_s = \ln \left(\frac{\left(2^{\frac{2}{3}}\sigma_l^{\frac{1}{3}}-2\right)\left(1-N_0\right)}{N_0\left(2-2^{\frac{2}{3}}\sigma_l^{\frac{1}{3}}\left(1-\rho\right)\right)}\right).\label{ts_lattice}
\ee

\begin{figure}
\centering
\includegraphics[width=.55\textwidth]{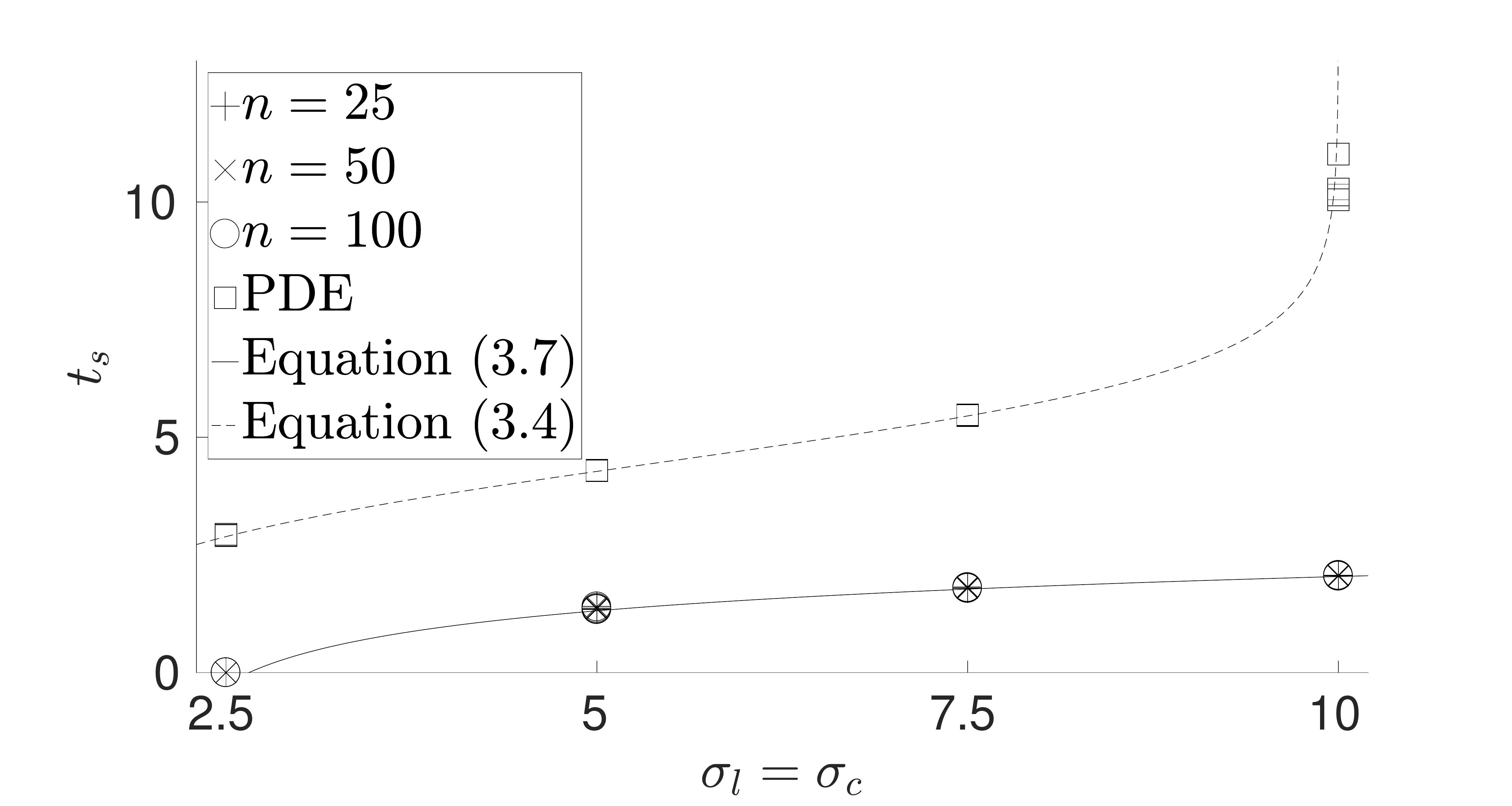}
\caption{Plots of the numerical values of $t_s$ for all parameter combinations of $\delta=10^{-1},10^{-2},10^{-3},10^{-4}$ and $\sigma_c=\sigma_l=2.5,5,7.5$, and $10$ for the PDE and lattice models over $300$ realizations. The symbols $+$, $x$, the circle and the square are for the numerical solutions for lattice sizes $n=25, 50, 100$ and the PDE respectively. The continuous lines are plots of the analytical approximations from \eqref{ts_pde} and \eqref{ts_lattice}. The standard deviation of $t_s$ between 300 realizations is, in each case, less than 1\% of the reported mean value across all realizations.}\label{sigma_instability}
\end{figure}

We compare Equations \eqref{ts_pde} and \eqref{ts_lattice} to the numerically computed values of $t_s$ which correspond to $t_s \equiv \min_t \left(\max_x \left(\sigma(\boldsymbol{x},t)\right) > \sigma_c \right)$ for the PDE simulations, and $t_s = \min_t \left(\max_i (\sigma_i(t)) > \sigma_l \right)$ for the lattice simulations. In Figure \ref{sigma_instability} we plot values of $t_s$ for all lattice and PDE simulations with varying $\delta$, $\sigma_c$, and $\sigma_l$. We see excellent agreement between the analytical and numerical predictions for both the lattice and the PDE. We see that $t_s$ does not depend on the lattice size $n$ or the parameter $\delta$ except for the PDE simulation at $\sigma_c=10$. We note that $\sigma_c=10$ is a limiting case as Equation \eqref{ts_pde} indicates that $t_s \to \infty$ as $\sigma_c \to 10$ if the initial cell density distribution was exactly uniform.

The time $t_s$ can be seen in Figure \ref{sigma_time_series} where the form of the time series changes from a logistic curve, and for all values of $\sigma_l$ and $\sigma_c$ the mean cell density begins decreasing at this time. In most cases this period of decreasing mean cell density is short relative to the simulation timescale. The time $t_s$ can also be seen in Figure \ref{cell_hetero_time_series} as the point where the heterogeneity increases sharply from $E=0$, due to the onset of non-uniform growth.
\subsection{Lattice Oscillations}\label{Oscillations}
\begin{figure}
\begin{subfigure}{.5\textwidth}
\includegraphics[width=1\textwidth]{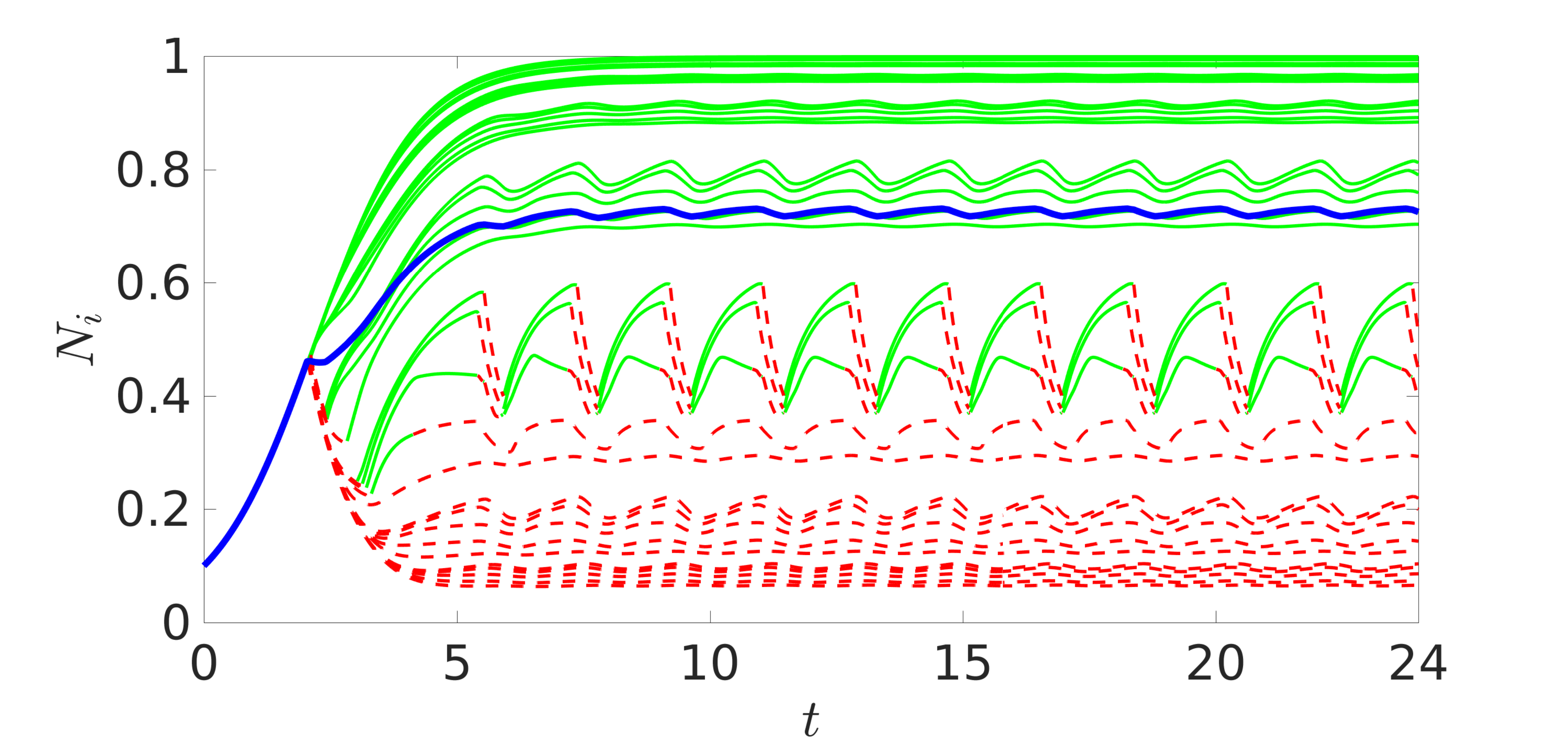}
\caption{$\sigma_l=10$, $n=10$, $\delta=0.01$}
\end{subfigure}
\begin{subfigure}{.5\textwidth}
\includegraphics[width=1\textwidth]{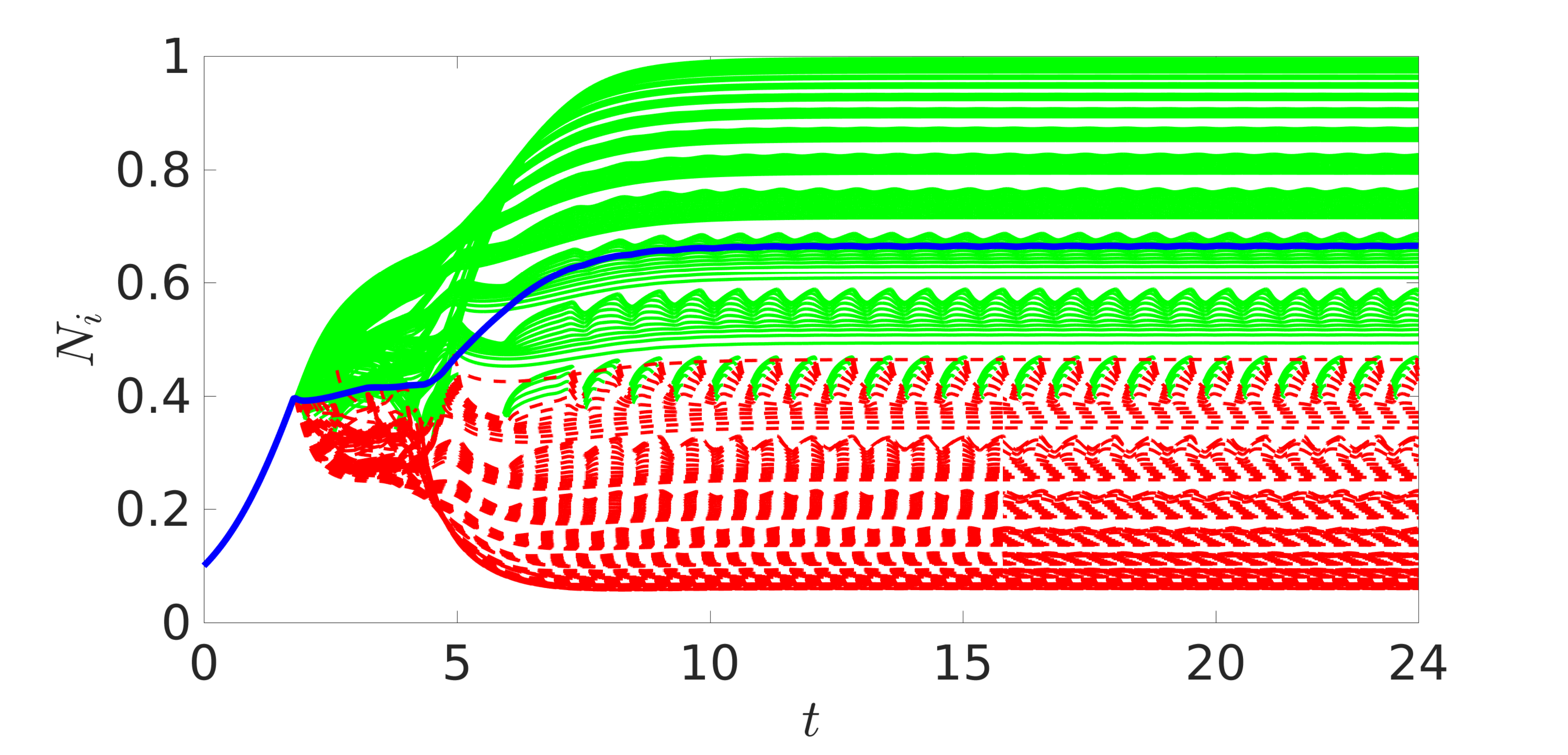}
\caption{$\sigma_l=7.5$, $n=25$, $\delta=0.014$}
\end{subfigure}

\begin{subfigure}{.5\textwidth}
\includegraphics[width=1\textwidth]{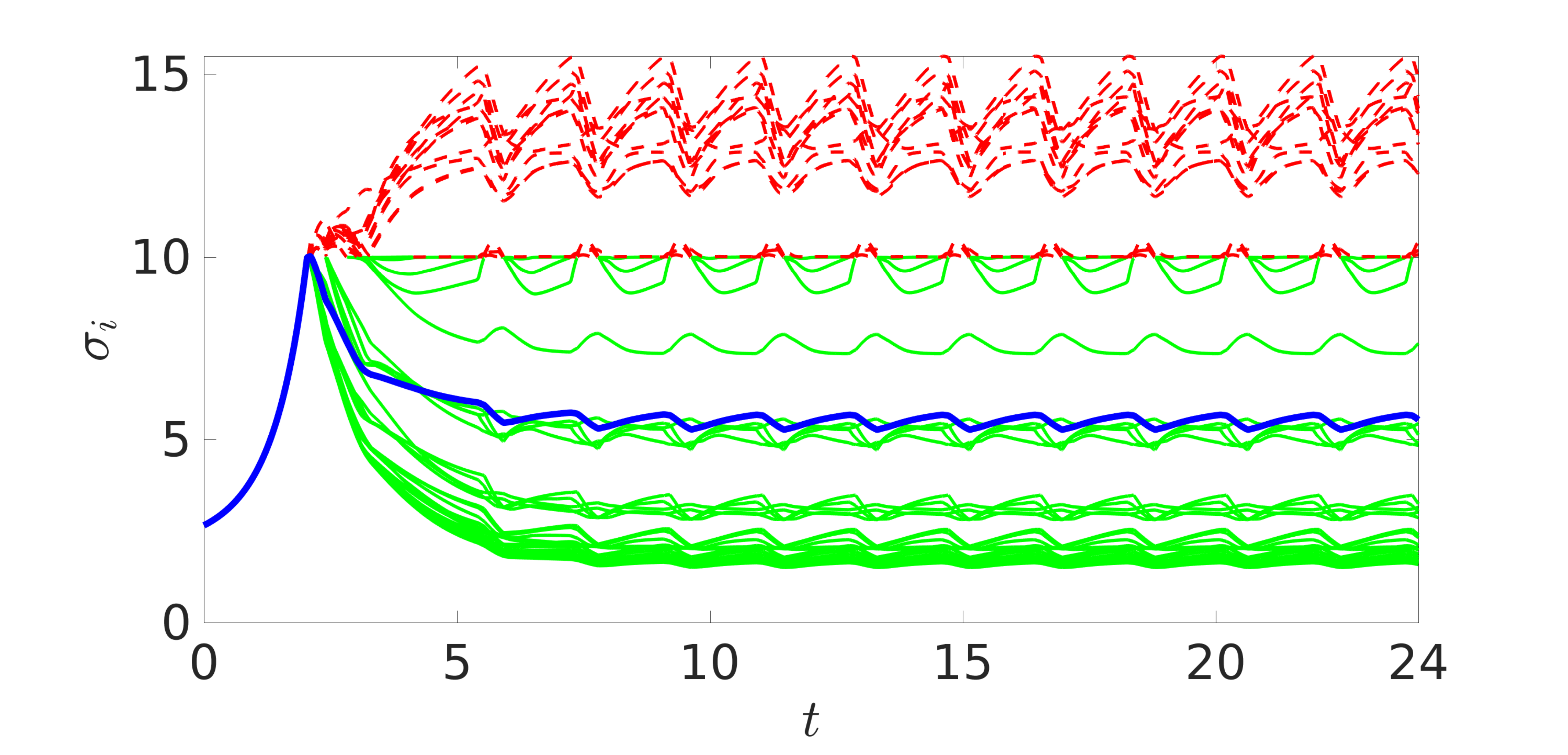}
\caption{$\sigma_l=10$, $n=10$, $\delta=0.01$}
\end{subfigure}
\begin{subfigure}{.5\textwidth}
\includegraphics[width=1\textwidth]{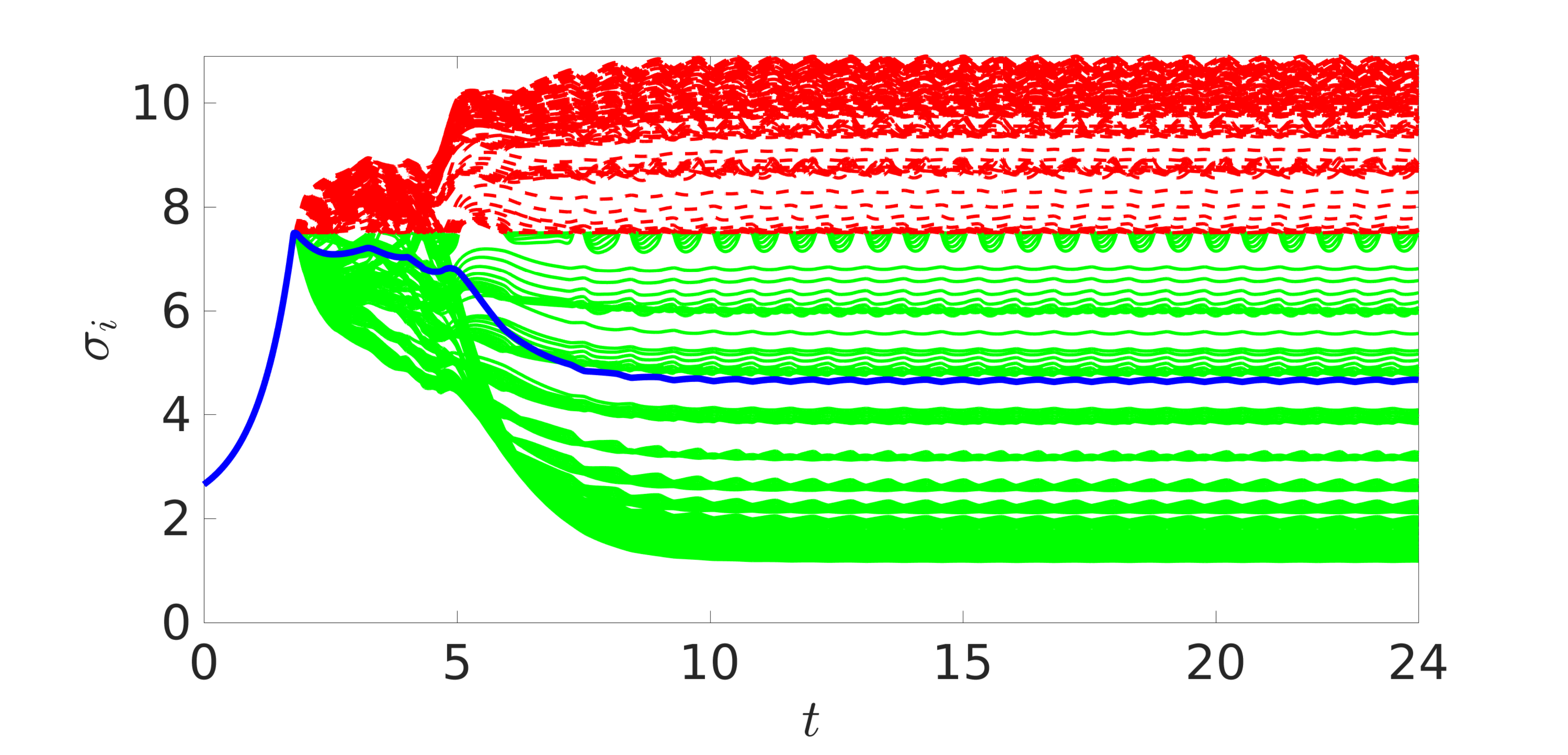}
\caption{$\sigma_l=7.5$, $n=25$, $\delta=0.014$}
\end{subfigure}
\caption{Plots of the cell density at every node for different parameter combinations in (a)-(b), and shear stress at every node in (c)-(d) for one realization of the initial cell densities. Green solid lines correspond to the cells at that node growing logistically ($\sigma_i < \sigma_l$), and red dashed lines correspond to exponential death ($\sigma_i > \sigma_l$). The blue line shows the spatial mean cell density in (a)-(b), and the spatial mean shear stress in (c)-(d).}\label{oscillation_time_series}
\end{figure}

A particularly interesting behaviour we observe in some lattice simulations are oscillations in cell density. Such oscillations are not found the PDE model. Figure \ref{oscillation_time_series} shows a time series plot of the cell density at every lattice node for $n=10$ and $n=25$ in (a)-(b), and corresponding plots of nodal values of the shear stress in (c)-(d). After a period of transient behaviour, the nodal cell densities and nodal values of the shear stress oscillate in phase. Nodes that are growing ($\sigma_i < \sigma_l$) are green solid lines, and nodes that are dying ($\sigma_i > \sigma_l$) are red dashed lines. The spatial mean cell density and shear stress are plotted in blue, which is also oscillating but with a small amplitude. Thus the overall effect of the lattice oscillations on the mean cell density is small. The particular time-series shown in Figure \ref{oscillation_time_series} will have different transient behaviour for different realizations of the initial data, but due to the simple spatial behaviour at large diffusion parameter (e.g. Figure \ref{large_diffusion_plots}), the amplitude and frequency of the oscillations for long times is the same for each realization we simulated.

In regions of the scaffold where the local shear stress exceeds the threshold $\sigma_l$, cells are (exponentially) dying due to high shear stress. These nodes all have relatively low cell density and hence more fluid passes through them, maintaining this high value of shear stress. In other regions of the scaffold with low shear stress, cells grow logistically. Diffusion acts to move cells from regions of high density to low density, and so cells move from regions with low shear stress to regions of high shear stress. In certain parameter regimes, this process finds an equilibrium value where these regions separate, as in Figures \ref{sigma_times2.5} and \ref{sigma_times7.5}, where between regions of growth and death there are some nodes of intermediate cell density. However, for certain values of $n$, $\sigma_l$, and $\delta$, the movement and growth of cells gives rise to a different behaviour from this equilibrium as the growth of cells in regions of high cell density substantially affects fluid flow and shear stress throughout the entire scaffold. As seen in Figure \ref{oscillation_time_series}, diffusion from growing regions can cause regions that are dying to increase in total cell density, and similarly, regions that are growing logistically can decrease in cell density due to diffusion.

%This mechanism is an interplay between cell growth and death, and diffusion. This is a short-range activation long-range inhibition mechanism where cell density in growth regions will increase logistically, but this growth will lead to long-range death as less fluid enters these regions and is forced elsewhere. Additionally, cells will diffuse from regions where they are at a high cell density. This can be seen as red lines in Figure \ref{oscillation_time_series} can be seen to periodically increase due to diffusion from other regions of the scaffold. Spatially between these regions of high and low shear, nodes can be seen to switch between growth and death behaviours. This mechanism is due to fast non-local reaction mediated by the quasi-static description of the shear stress, and slow local diffusion. 

\begin{figure}
\centering
\begin{subfigure}{.47\textwidth}
\includegraphics[width=1\textwidth]{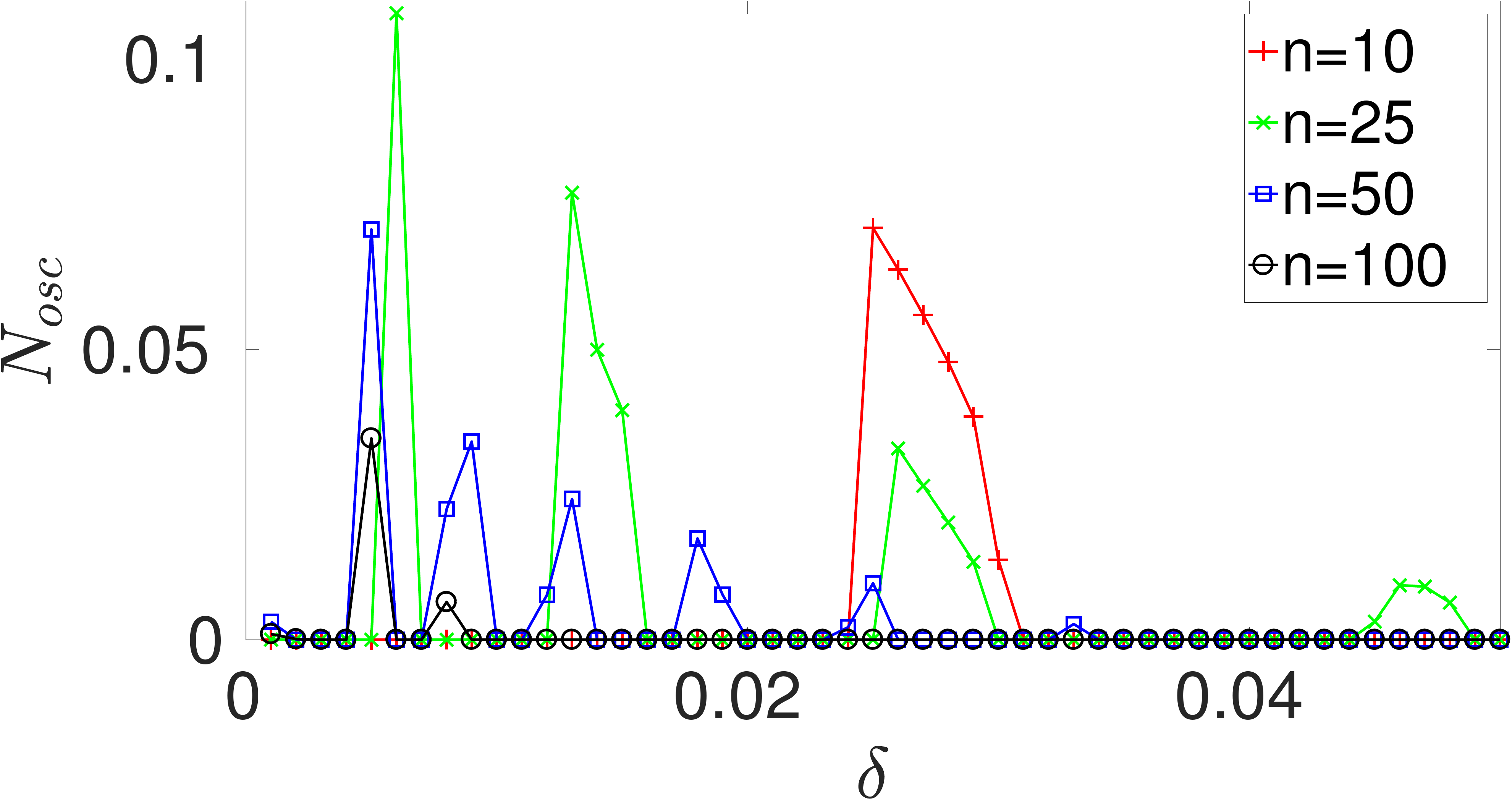}
\caption{$\sigma_l=7.5$}\label{sigma_bifa}
\end{subfigure}
\begin{subfigure}{.47\textwidth}
\includegraphics[width=1\textwidth]{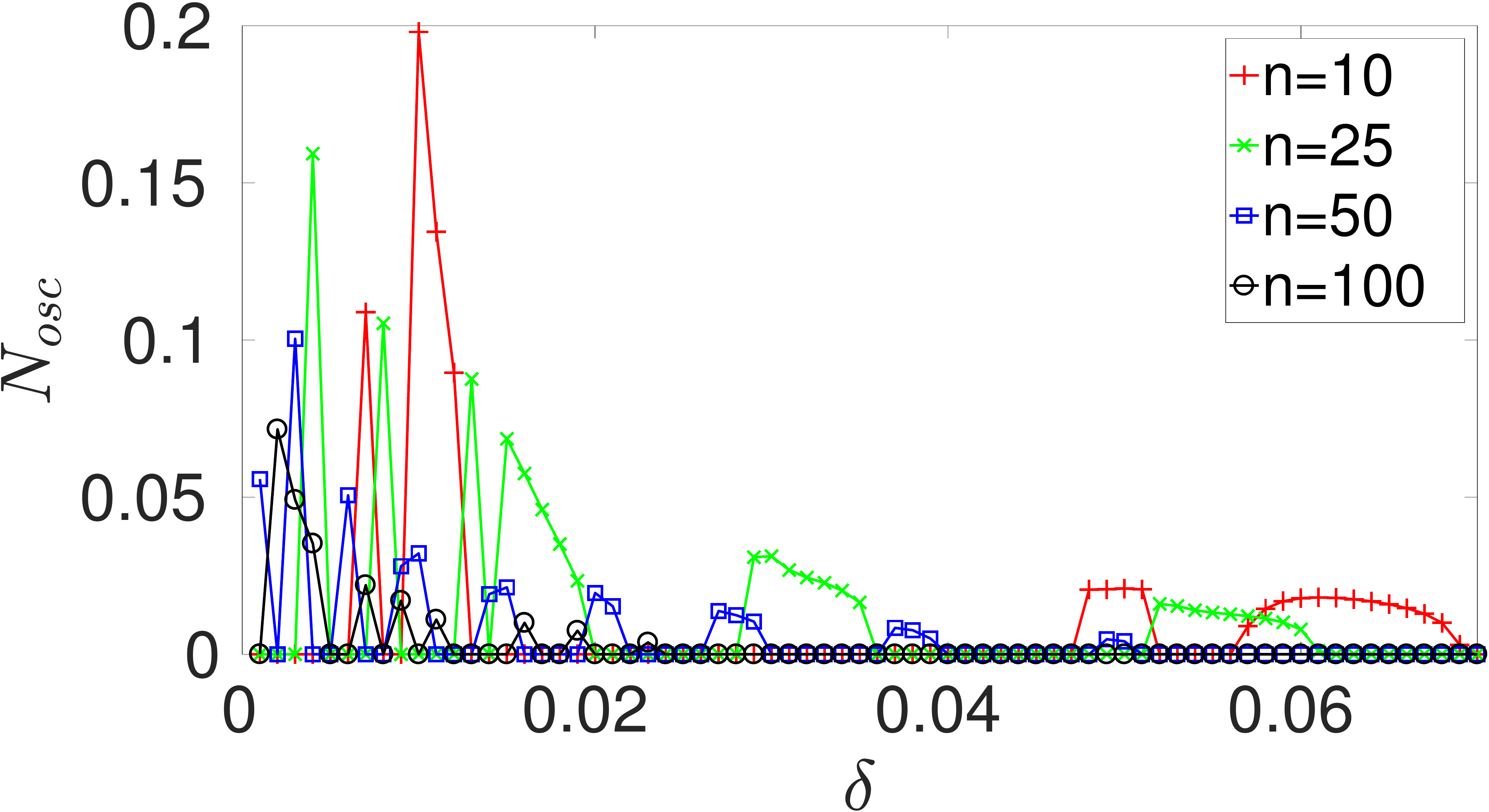}
\caption{$\sigma_l=10$}\label{sigma_bifb}
\end{subfigure}
\\
\begin{subfigure}{.47\textwidth}
\includegraphics[width=1\textwidth]{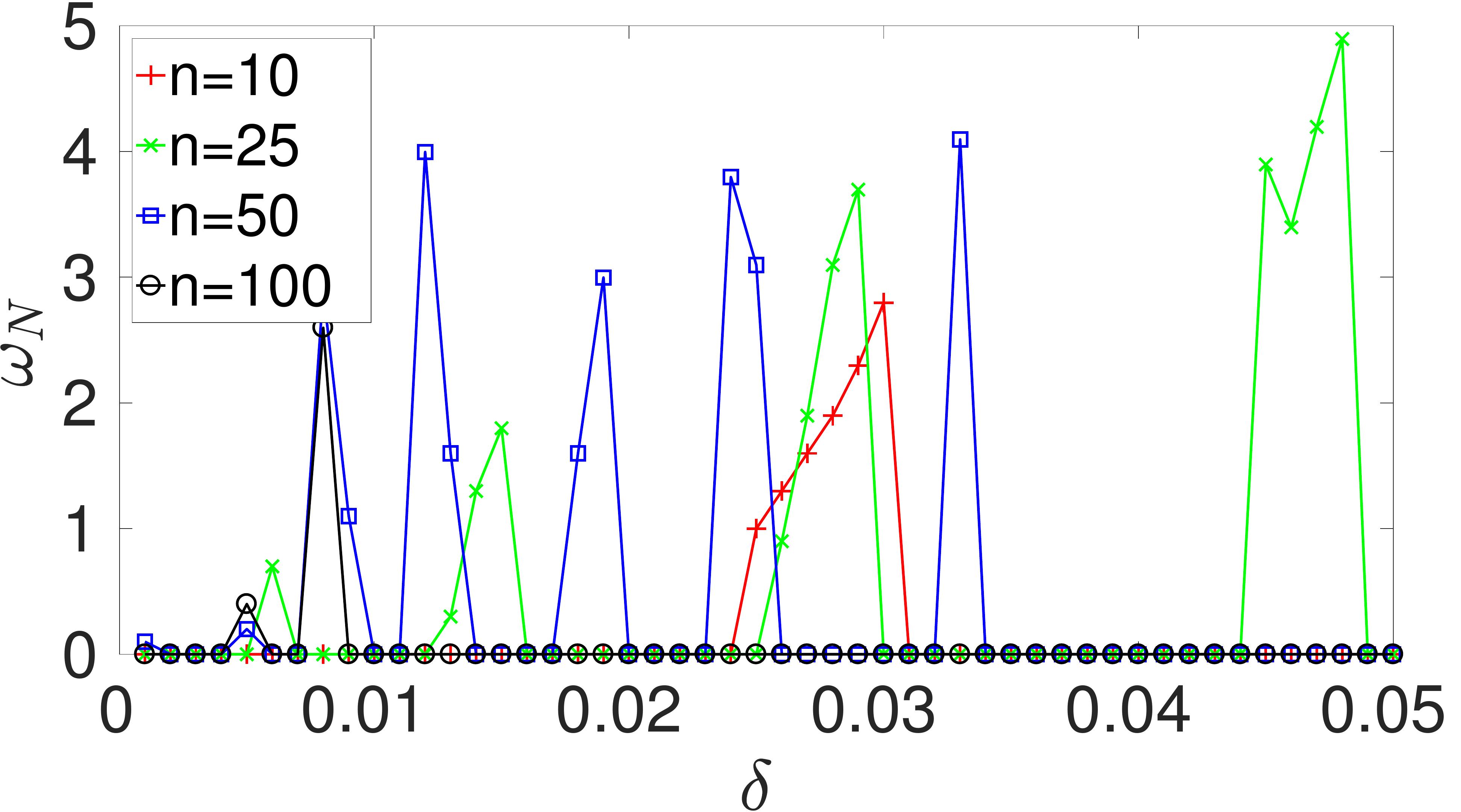}
\caption{$\sigma_l=7.5$}\label{sigma_bifc}
\end{subfigure}
\begin{subfigure}{.47\textwidth}
\includegraphics[width=1\textwidth]{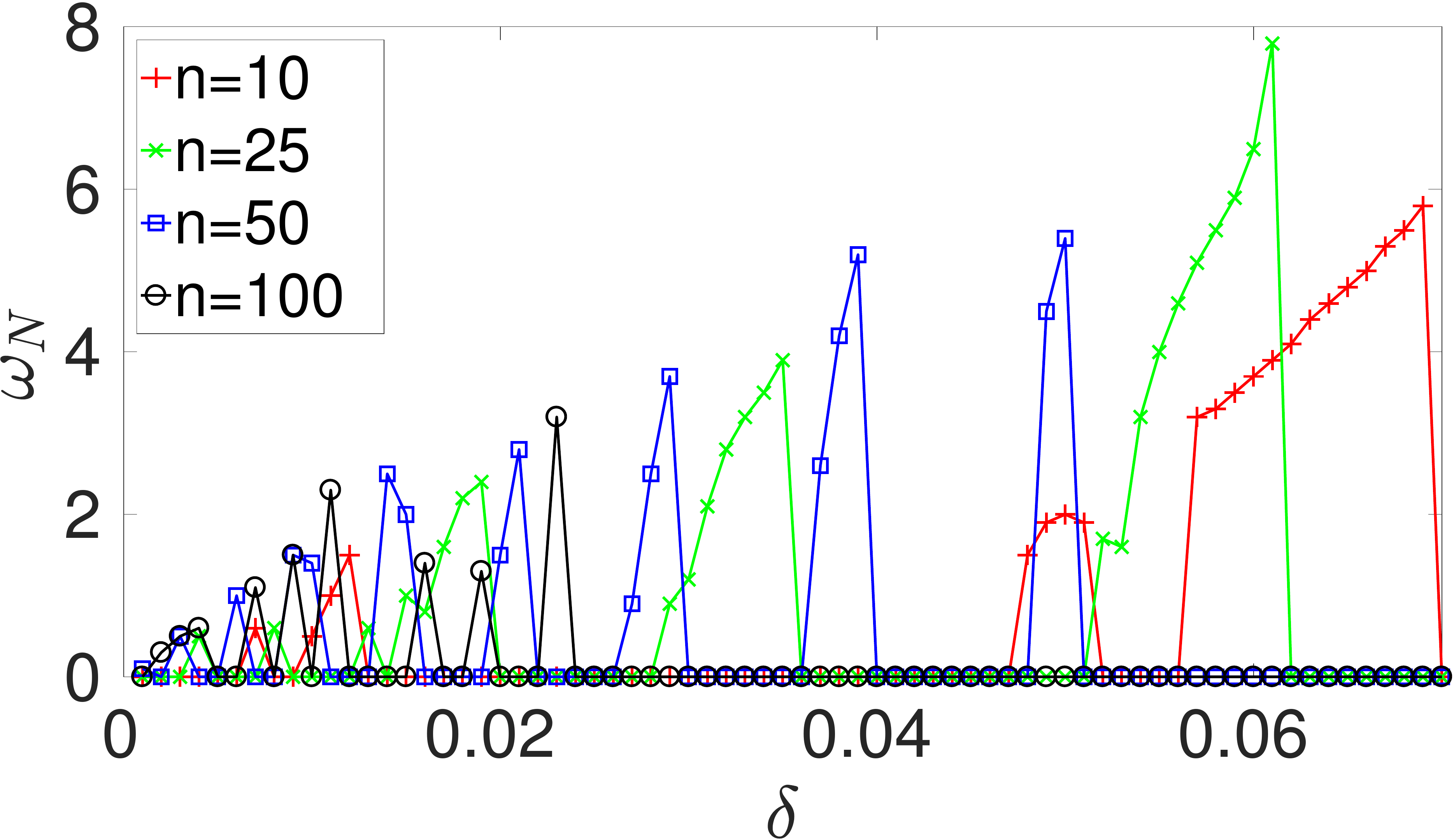}
\caption{$\sigma_l=10$}\label{sigma_bifd}
\end{subfigure}
\caption{Plots (a)-(b) are bifurcation diagrams showing the maximum magnitude of nodal oscillation in the lattice for different lattice sizes. This is plotted in steps of $0.001$ in $\delta$. Plots (c)-(d) are corresponding frequencies for these nodal oscillations. We only show results corresponding to one realization of the initial cell densities, but further simulations had quantitatively identical behaviour.}\label{sigma_bif}
\end{figure}

We plot bifurcation diagrams of the maximal nodal oscillation amplitudes in Figures \ref{sigma_bifa}-\ref{sigma_bifb} for several lattice sizes and two values of the threshold parameter, over a range of the parameter $\delta$. After enough time has passed to ensure we are no longer observing transient dynamics, we compute the maximum nodal amplitude as $N_{osc}=\max_i\{\max_t\{N_i(t)\} - \min_t\{N_i(t)\}\}$ to capture the largest oscillation. We plot the frequency $\omega_N$ of these nodal oscillations computed with the Fast Fourier Transform in Figures \ref{sigma_bifc}-\ref{sigma_bifd}. We note that the frequency is the same for every node in the lattice. 

These bifurcation diagrams show that for each $n$ and $\sigma_l$, the one-dimensional parameter space over $\delta$ is composed of disconnected regions where oscillations are permitted (see behaviour illustrated in Figure \ref{oscillation_time_series}), and regions where cell densities tend to steady states (see Figure \ref{sigma_time_series}). Further simulations within each of these regions (using different realizations of the perturbations of the initial data, for instance) have consistent behaviour, and we conjecture that that these behaviours are generic within each region in parameter space demarcated by the bifurcation between steady state behaviour and oscillations. In Figure \ref{sigma_bif} we see that the number of disconnected regions in $\delta$-space with oscillatory behaviour increases as $n$ increases, and the maximal amplitudes decrease with increasing $n$. We note that for $n=100$, there are many disconnected bifurcation regions for very small $\delta$, but these all have correspondingly small magnitudes in oscillation and so they are not visible in Figure \ref{sigma_bif}. For larger values of $n$, the oscillation amplitudes become so small that they are comparable to numerical truncation and rounding errors, and we conjecture that these oscillations play no role for larger lattices. There is a trend of increasing oscillation frequency and decreasing nodal amplitude for larger values of $\delta$ for a given lattice size. The larger lattices have significantly smaller variations in mean cell density of the scaffold due to these oscillations. This is due to only a small subset of the nodes switching between growth and death behaviours during an oscillation. These oscillations do not occur in any of our PDE simulations. For large $n$, our simulations show that the amplitudes of oscillation in mean cell density decreases, as shown by the blue lines in Figure \ref{oscillation_time_series}, and that the proportion of nodes switching between growth and death behaviours also decreases. Similarly, nodal shear stress oscillations also decrease in amplitude as the lattice size increases. In the other direction, for lattices of size $n=5$ and smaller, no oscillations were observed. We refer to \citep{krause_analysis_2017} for a discussion of Hopf bifurcations in a similar lattice system, where cells die due to high fluid pressure rather than shear stress. There we argue that the oscillations observed in this model are due to a combination of nonlocal effects from the quasi-static fluid equations, coupled with symmetry-breaking bifurcations in the lattice models.
\subsection{Mean Cell Density Predictions}\label{Discussions}
\begin{figure}
\includegraphics[width=.97\textwidth]{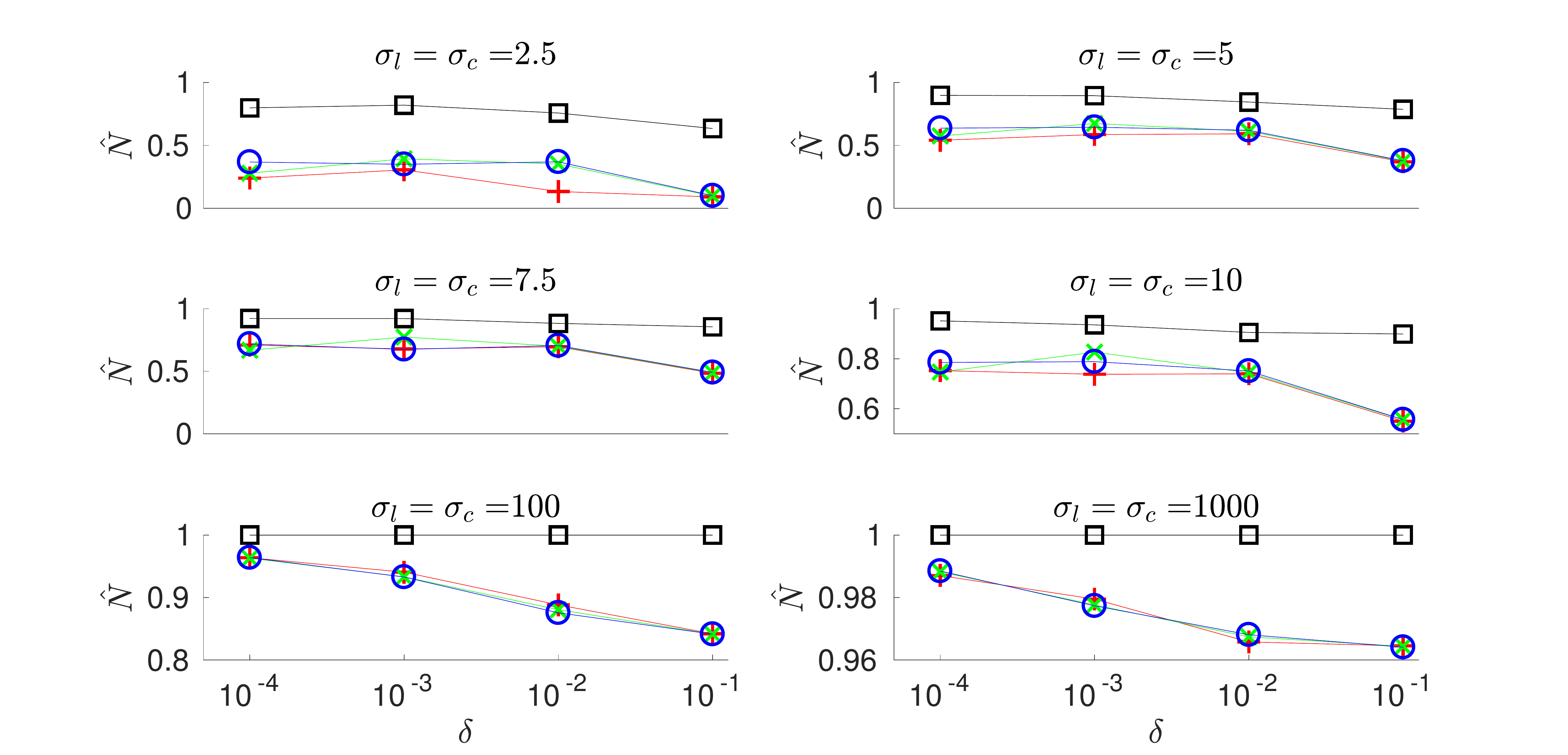}
\caption{Plots of the mean cell density $\hat{N}$ at the final time $t=30$ over the four values of $\delta$ for all parameter combinations for the PDE and lattice models. The symbols $+$, $\times$, $\circ$ and $\square$ are for the lattice sizes $n=25, 50, 100$ and the PDE respectively. Note the variation in the ranges of the mean cell density $\hat{N}$ ($y$-axis). The standard deviation of this spatial mean between 300 realizations is, in each case, less than 1\% of the reported mean value across all realizations.}
\label{sigma_cell1fig}
\end{figure}

We can broadly understand the global behaviour of the numerical simulations by comparing their final mean cell densities computed via Equations \eqref{lattice_mean} and \eqref{PDE_mean}. Figure \ref{sigma_cell1fig} is a plot of the spatial mean of the cell density after $30$ nondimensional units of time. For any given shear stress threshold and $\delta$, the PDE model has a higher mean cell density. This is expected due to the differences between the constitutive assumptions for shear stress in each model (see Figure \ref{ShearNPlot}).

The value of the parameter $\delta$ has a non-monotonic effect for some lattice simulations. In particular, for $\sigma_c, \sigma_l \leq 10$, the maximum mean cell density for the $n=50$ lattice occurs for the intermediate value of the diffusion parameter $\delta=10^{-3}$. For the large value of $\delta=10^{-1}$ in each plot, the lattice size is insignificant in determining the mean cell density. Similarly, for the large shear stress thresholds $\sigma_l=100$, and $1000$ shown in the bottom two plots, cell death occurs only in a very small region of the domain, so the lattice size is almost inconsequential in determining the final mean cell density. For $\sigma_l \leq 10$ and $\delta \leq 10^{-2}$, however, there are differences between the final mean cell density for different lattice sizes. This suggests that for applications where the ratio of proliferation to diffusion timescales is small, and where cells are very sensitive to high shear stress, the topology of the underlying pore network plays a role in determining the final mean cell density.
\subsection{Nonlinear Cell Diffusion}\label{NonlinearDiffusion}
Here we consider the effect of nonlinear cell diffusion. We follow \cite{shakeel_continuum_2013} and use the following (nondimensionalized) functional forms for the diffusion coefficient,
\be\label{nonlineardiff}
D\left(N\right) = \delta \exp\left(\gamma\left(N-1\right)\right), \quad D\left(N_i, N_j\right) = \delta \exp\left(\gamma\left(\frac{N_i+N_j}{2}-1\right)\right),
\ee
for the continuum and lattice models respectively. For values of cell density near the carrying capacity (dimensionless value of 1), diffusion given by \eqref{nonlineardiff} approximates the linear case with the same nondimensional parameter $\delta$. For small values of cell density, the effective diffusion of cells throughout the porous medium is smaller than the corresponding linear value ($\delta$) for all $n$. The dimensionless parameter $\gamma$ determines the strength of this nonlinear effect; \cite{shakeel_continuum_2013} used $\gamma = 2$ to ensure a sufficiently nonlinear behaviour while remaining within a parameter regime where the speed of a proliferating front of cells could be computed. Here we carry out several simulations for varying values of $\gamma$ in the interval $(0, 5]$. For small nonlinearity ($\gamma \lesssim 10^{-2}$), nonlinear cell diffusion simulations have identical mean cell density and oscillatory behaviours as linear cell diffusion simulations. For moderate nonlinearity ($10^{-2} \lesssim \gamma \lesssim 3$), qualitative features such as cell density aggregation and oscillations are retained, but quantitative differences in cell density distributions and oscillation frequency appear. For larger values of $\gamma$, we no longer observe oscillations in cell density.

\begin{figure}
\centering
$n=25$ \hspace{3cm} $n=50$ \hspace{3cm} $n=100$

\includegraphics[width=.27\textwidth]{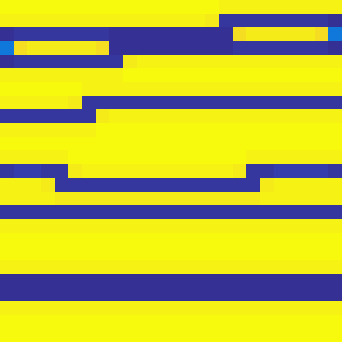} \quad
\includegraphics[width=.27\textwidth]{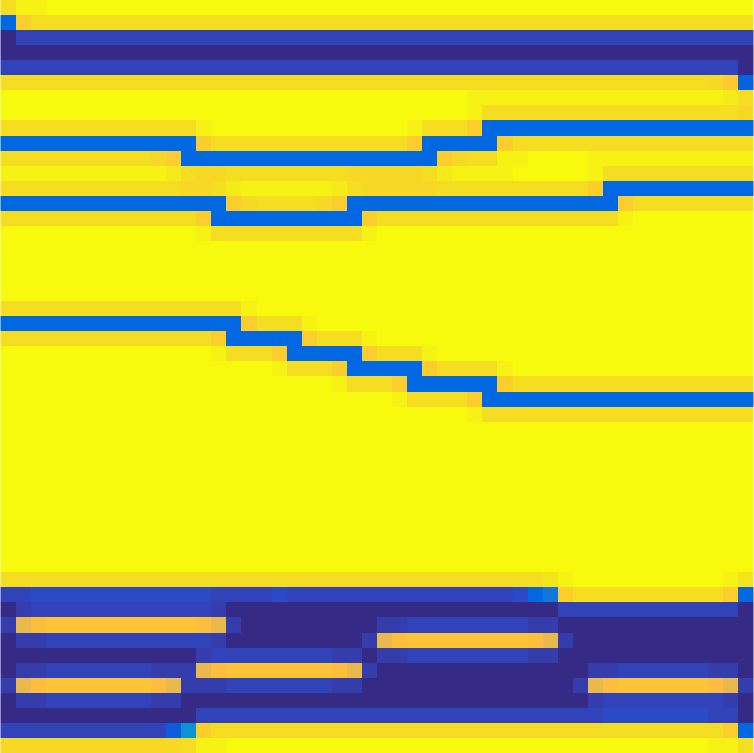} \quad
\includegraphics[width=.27\textwidth]{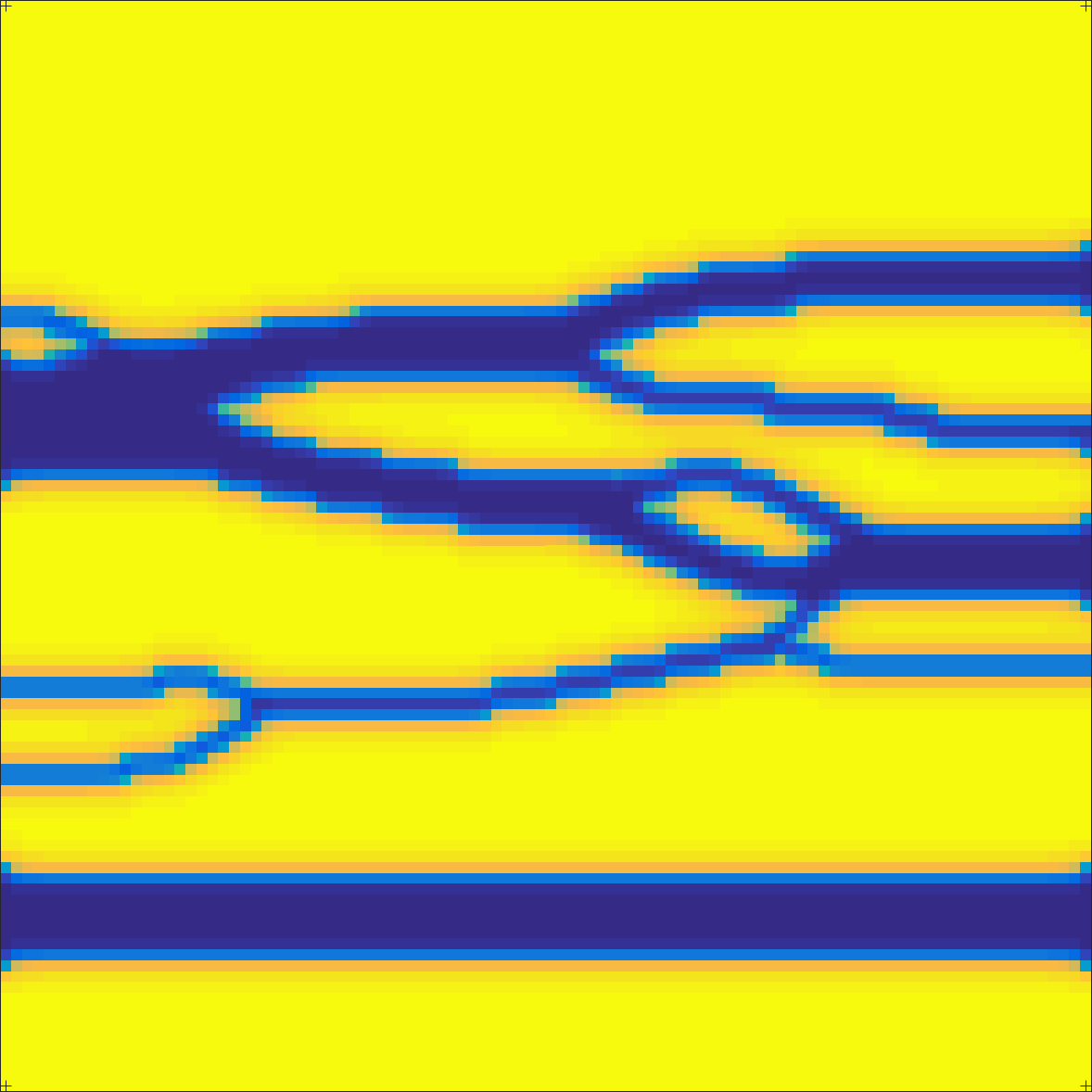}

\hspace{-.7cm}\includegraphics[clip, trim={0 23cm 0 0}, width=.96\textwidth]{latticeshearcolorbar.png}
\caption{Cell density plots with $\delta=10^{-4}$, $\gamma=2$, and $\sigma_c=7.5$, for $n=25, 50$, and $100$ at $t=30$ with the same realization of the initial data as used in Figure \ref{sigma_times7.5}.}
\label{sigma_plots_nonlinear}
\end{figure}

We first consider the spatial cell density distributions with nonlinear cell diffusion. We take $\gamma=2$ in Equations \eqref{nonlineardiff} and repeat the simulations (using the same realization of the initial data) from Figure \ref{sigma_times7.5}, and display these in Figure \ref{sigma_plots_nonlinear}. While the steady state cell density distributions have changed, the organization of cells into high density aggregates occurs in both the linear and nonlinear cell diffusion simulations. Additional simulations (with different realizations of the initial cell density, and different values of $\sigma_l$) show qualitatively similar coarsening behaviour in the $n=100$ simulations with nonlinear cell diffusion compared to the linear case (see Figures \ref{sigma_times2.5}-\ref{sigma_times7.5}). Similarly, $n=25$ simulations including nonlinear cell diffusion quickly settle into a steady state as in the linear case. Quantitatively, the change in the spatial mean cell density for $\gamma \leq 1$ is less than 5\% in each case shown in Figure \ref{sigma_cell1fig}. Larger values of the nonlinearity parameter $\gamma$ leads to larger changes to the spatial mean cell densities reported in Figure \ref{sigma_cell1fig} for linear cell diffusion, but we leave further quantification of this to future work.

\begin{figure}
\begin{subfigure}{.49\textwidth}
\includegraphics[width=1\textwidth]{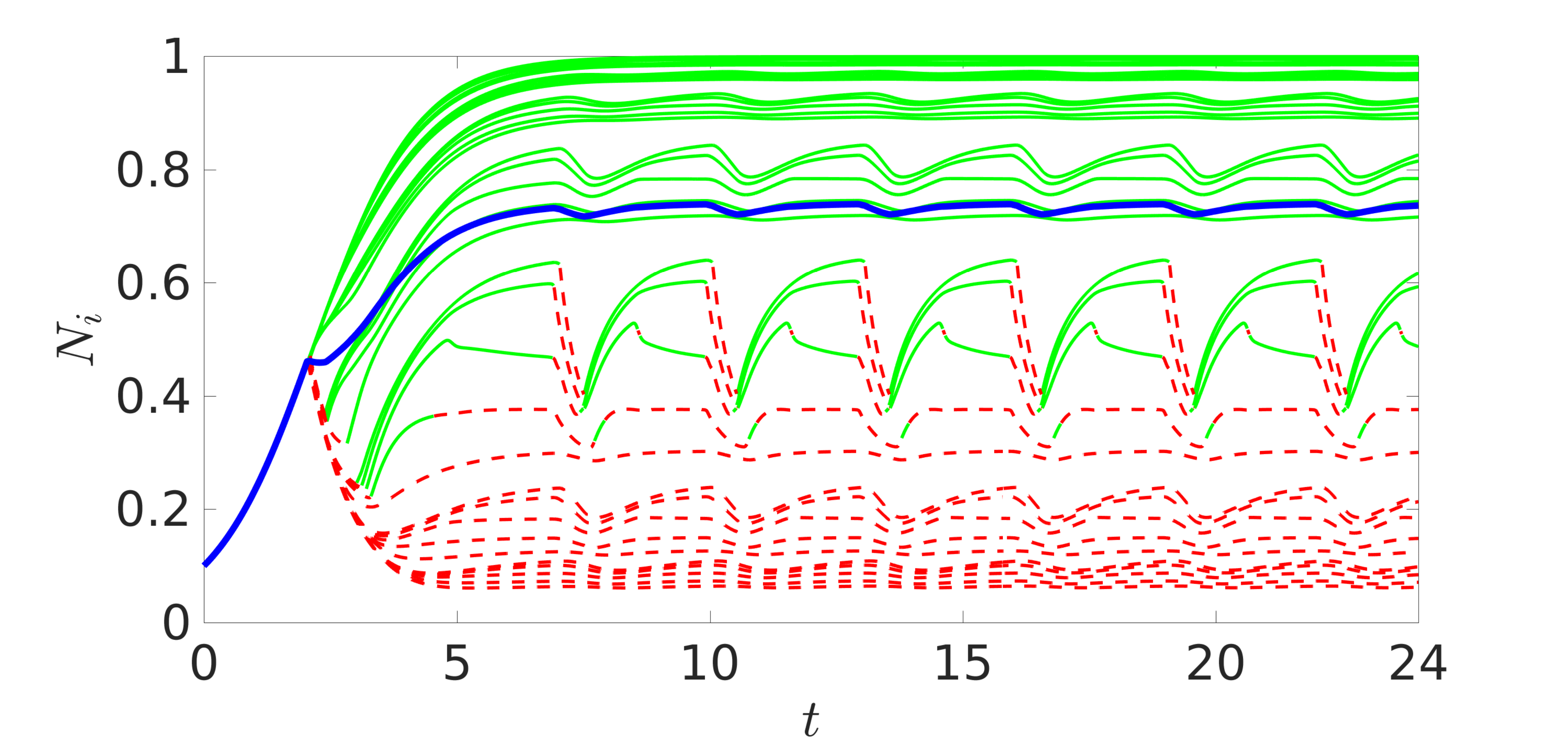}
\caption{$\sigma_l=10$, $n=10$, $\delta=0.01$,$\gamma=0.1$}
\end{subfigure}
\begin{subfigure}{.49\textwidth}
\includegraphics[width=1\textwidth]{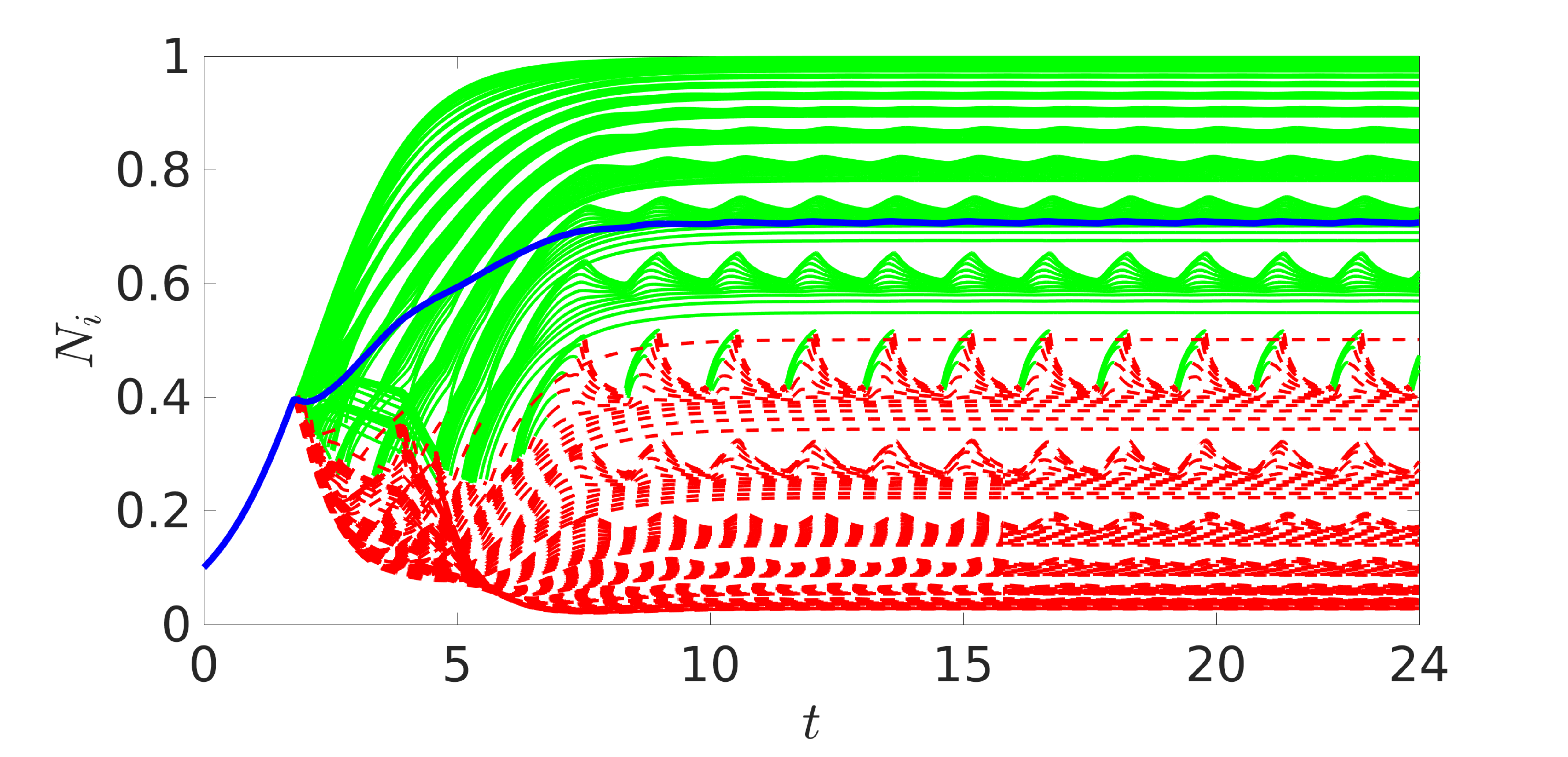}
\caption{$\sigma_l=7.5$, $n=25$, $\delta=0.014$, $\gamma=1$}
\end{subfigure}
\caption{Plots of the cell density of every node for the same parameter combinations shown in Figure \ref{oscillation_time_series} with nonlinear cell diffusion given by \eqref{nonlineardiff}. Green solid lines correspond to the cells at that node growing logistically ($\sigma_i < \sigma_l$), and red dashed lines correspond to exponential death ($\sigma_i > \sigma_l$). The blue line shows the spatial mean cell density.}\label{nonlinear_oscillation_time_series}
\end{figure}

Next we consider the effect of nonlinear cell diffusion on the oscillations in cell density found in Section \ref{Oscillations}. For small values of the nonlinearity, e.g. $\gamma \leq 10^{-2}$, the oscillatory behaviour of solutions is quantitatively the same up to a slight phase shift due to differences in transient dynamics. We plot two example time series in Figure \ref{nonlinear_oscillation_time_series} that use the same parameter sets as those in Figure \ref{oscillation_time_series}, but now including nonlinear cell diffusion with $\gamma=0.1$ and $1$. We see that the period of the oscillation has approximately doubled in both cases compared with the linear diffusion simulations shown in Figure \ref{oscillation_time_series}. For increasing values of $\gamma$, many of the regions (in $\delta$) of oscillatory behaviour shown in Figure \ref{sigma_bif} become smaller, and completely disappear for $\gamma \geq 3$. We conclude that  large values of nonlinear diffusion can be stabilizing in that oscillatory regions of parameter space shrink in size, and all oscillatory behaviour disappears for large enough values of $\gamma$. As before, we do not observe any oscillatory behaviour in the continuum model with nonlinear cell diffusion.
\section{Conclusions}\label{Conclusions_M}
In this paper we have explored complementary lattice and continuum models for a bioactive porous tissue scaffold, and demonstrated several important differences in model behaviours. The lattice simulations show that considering finite pore networks when modelling cell growth within a porous scaffold will lead to qualitative differences compared with PDE models. Varying degrees of spatial heterogeneity in the cell density distribution are displayed in the lattice model, whereas for most of the parameter space the PDE solutions exhibit simpler spatial structures, such as one or two horizontal channels. The significant variation between different lattice sizes indicates that considering the spatial network explicitly has a nontrivial impact on pattern formation. The smoothing behaviour observed in both the PDE and large $n$ lattice models is reminiscent of models of coarsening in condensed matter physics, where initial spatial irregularities evolve into larger spatial structures. That this coarsening occurs at different timescales and to different degrees due to the size of the lattice is an unexpected result. We conjecture that, modulo constitutive differences, a PDE and a very large lattice (e.g. $n > 100$) should have comparable behaviour, but for smaller pore networks the finite structure of the lattice becomes important to the overall growth process within the scaffold.

We observe small amplitude oscillations in cell density exclusively in the lattice model. While the effects of these oscillations on the total mean cell density of the scaffold are small, and would likely not be physically observed given experimental noise, their presence is an interesting effect of finite lattice size. Additionally, the larger variations in nodal values of shear stress (see Figure \ref{oscillation_time_series}(c)-(d)) has important implications for shear-stress mediated tissue growth. In particular, shear-stress can play a role in both cell proliferation and cell differentiation \citep{iskratsch_appreciating_2014}. These spatiotemporal oscillations in shear stress could potentially lead to the formation of heterogeneity in the cell populations within the scaffold, and hence influence the integrity and uniformity of artificial tissue \citep{yin2016engineering}. 

Nonlinear interactions could amplify these oscillations. For instance, some tissue scaffolds are cyclically loaded to facilitate cell growth via mechano-transduction \citep{angele2004cyclic, pohlmeyer_cyclic_2013, nessler_influence_2016}. In certain parameter regimes, this oscillatory forcing could interact in non-trivial ways with these lattice induced oscillations, for instance inducing resonance in the cell density oscillations. This is especially true given that small amplitude oscillations in cell density give rise to larger shear stress variations, which suggests feedback between cell proliferation and shear stress. We leave the investigation of such possibilities to future work.

Some strategies for quantifying experimental cell distributions have been proposed which could validate the steady-state distributions of cell density we demonstrate in Figures \ref{sigma_times2.5}-\ref{sigma_timesPDE} \citep{thevenot2008method}, though we note that these approaches involve sacrificing tissue scaffolds after cells have stopped proliferating. Observing oscillations in cell density experiments would be difficult given the small oscillation amplitude, and the difficulty in producing temporal datasets from tissue scaffolds due, e.g., to requiring the scaffolds to be sacrificed in order to ascertain cell density. Models can also be validated by comparing other key outputs, such as the mean cell densities given in Figure \ref{sigma_cell1fig}, against experimentally accessible data \citep{odea_continuum_2012}. While validating these models is difficult, once validated such theoretical models provide predictions for the spatiotemporal evolution of important quantities (e.g. shear stress and cell density) throughout the scaffold, without the need for large numbers of costly and time-consuming experiments.

While we observed quantitative differences between these paradigms (see Figure \ref{sigma_cell1fig}), we attribute some of these differences to the constitutive assumptions as shown in Equations \eqref{PDE_Vel}-\eqref{LatticeShear}, and these can vary between particular applications of these models. We have explored simulations using different functional forms of $\phi$, $k$, and $R_{ij}$, as well as different values of $\rho$. These preliminary simulations were consistent with the qualitative differences reported here between lattice and continuum models. The constitutive choices we have made in this paper do give results that are consistent with \cite{nava_multiphysics_2013} in that our 2-D fluid model in the PDE over-predicts cell density compared with the 3-D fluid (`tube') model of the lattice. We also note that despite the simplicity of our models, the qualitative differences between lattice and continuum formulations exist even when we introduce nonlinear diffusion into the model. We conjecture that these differences are inherent to the forms of the models used here suggesting that discrete spatial geometry should be carefully incorporated into macroscopic tissue engineering models. This is in line with the study of discrete and continuum cell monolayers \citep{byrne2009individual}, where insights from individual and lattice cellular-automata models were necessary to inform the continuum formulation. 

These differences in the type of behaviours displayed by continuum and lattice models exist even though the network geometry studied here, a square grid lattice, is simple. More complicated hexagonal and face centred cubic lattices \citep{welter2013interstitial}, off-lattice models \citep{vilanova2017mathematical}, and other complex microscale models have been used in hybdrid discrete-continuum settings. In addition to a more complex network geometry, there are many components that could be included in this lattice model, such as a model of nutrient transport, or a more detailed microscale relationship between cell density and fluid flow. Nevertheless, these simple models give insight into the kinds of differences one can expect from each modelling paradigm. 

For experiments involving small pore sizes or large constructs, the pore network is relatively dense and we expect continuum and lattice models to be comparable in terms of predicting cell proliferation and global fluid properties, such as scaffold permeability. For small scaffolds or scaffolds with large pores, however, we expect lattice models to significantly differ from continuum models in their predictions. The variations between the lattices of different size implies that a good model of the finite network geometry of a scaffold is important in understanding the tissue growth process for such scaffolds. Given a realistic description of the evolution of cells and fluid flow at the pore scale, our approach gives an alternative to continuum approaches such as mathematical homogenization in upscaling these microscale processes. Numerical solution of the lattice model is no more difficult than an equivalent discretization of a continuum model, and so the only practical disadvantage of using a lattice model would be the requirement of specifying the network topology, and determining appropriate fluid and cell properties at the scale of nodes in the network. We believe that pursuing these kinds of models can lead to novel insights in understanding the growth of artificial tissue, and eventually in developing clinically successful technologies.
\section{Acknowledgements}
D. Beliaev was partially funded by Engineering \& Physical Sciences Research Council (EPSRC) Fellowship ref. EP/M002896/1. We gratefully acknowledge the anonymous referees for providing critical analysis that significantly improved an earlier version of this manuscript.

\bibliographystyle{apalike}
\bibliography{LitReview}
\end{document}